\documentclass[%
reprint,
superscriptaddress,
showpacs,preprintnumbers,
 amsmath,amssymb,
 aps,
prd,
]{revtex4-1}

\usepackage{graphicx}
\usepackage{dcolumn}
\usepackage{bm}
\usepackage{bbold}
\usepackage{amssymb,amsmath}
\usepackage{hyperref}

\usepackage{color}
\usepackage{graphicx}
\usepackage{amsmath, amssymb}
\usepackage{amsfonts}
\usepackage{dcolumn}
\usepackage{subfigure}
\usepackage{hyperref}

\newcommand{\Tr}{\ensuremath{\operatorname{Tr}}}

\newcommand{\overbar}[1]{\mkern 1.5mu\overline{\mkern-1.5mu#1\mkern-1.5mu}\mkern 1.5mu}

\def\Fig#1{Fig.~\ref{#1}}

\def\Eq#1{Eq.~(\ref{#1})}
\def\eq#1{(\ref{#1})}
\def\eqref#1{(\ref{#1})}

\def\lA0{{\langle A_0 \rangle}}
\def\bA0{{\bar{A}_0}}

\def\0#1#2{\frac{#1}{#2}}

\graphicspath{{./figures/}{./}}

\begin{document}

\preprint{}

\title{On the relevance of matter and glue dynamics for baryon
  number fluctuations}

\author{Wei-jie Fu}
\affiliation{Institut f\"{u}r Theoretische Physik, Universit\"{a}t
  Heidelberg, Philosophenweg 16, 69120 Heidelberg, Germany}

\author{Jan M. Pawlowski}
\affiliation{Institut
 f\"{u}r Theoretische Physik, Universit\"{a}t Heidelberg,
 Philosophenweg 16, 69120 Heidelberg, Germany}
\affiliation{ExtreMe Matter Institute EMMI, GSI, Planckstr. 1, 64291
  Darmstadt, Germany}


\begin{abstract}
  We investigate the impact of the matter and glue dynamics on baryon
  number fluctuations and the kurtosis of baryon number
  distribution. This is done within the framework of QCD-improved low
  energy effective models. In particular we include the momentum scale
  dependence of the quark-meson scattering and the non-trivial
  dispersions of both, quarks and mesons. On the gluonic side we take
  into account the backreaction of the matter sector on the glue
  dynamics. It is shown that the above fluctuations lead to a more
  rapid change of the baryon number fluctuations as well as the
  kurtosis of with the chiral crossover.

  We also study the signatures of quark confinement in low energy QCD.
  It is shown that contrary to the common picture the effective
  thermal distribution in the presence of confining glue backgrounds
  does not tend towards the colourless baryonic one. Instead, the
  dominance of colourless hadronic states is obtained in a subtle
  interplay of quark and glue contributions to the canonical potential.
\end{abstract}

\maketitle

\section{\label{sec:intr}Introduction}

Studies of the QCD phase structure and the QCD thermodynamics have
attracted lots of attentions in recent years. QCD matter composed of
deconfined, coloured quarks and gluons, i.e. the Quark-Gluon Plasma,
produced before the quark-hadron transition at $\sim
10\,\mu\mathrm{s}$ in the evolution of early
Universe~\cite{Schwarz:2003du}, is believed to be recreated in
heavy-ion collisions at Relativistic Heavy-Ion Colliders such as
(RHIC)~\cite{Adams:2005dq,Adcox:2004mh} and LHC~\cite{Aamodt:2010pa}.

The chiral and confinement-deconfinement crossovers for QCD with
physical quark masses at vanishing density may turn into first order
transitions at high density. One of the key challenges concerning the
phase structure of QCD is to get hold of the existence and location of
the corresponding critical end point (CEP),
\cite{Stephanov:2007fk}. The Beam Energy Scan (BES) program at RHIC is
directly aimed at this task, where the beam energy or collision
centrality dependence of the fluctuations of conserved charges, such
as moments of net-proton multiplicity distributions, is employed to
locate the QCD critical point.

In turn, the experimental programme should be accompanied by reliable
theoretical predictions for the above observables and their relation
to the CEP are highly demanded. A particularly promising direction is
the computation of the fluctuation of conserved charges at vanishing
and finite density, see e.g. \cite{Borsanyi:2013hza,Ding:2014kva,
Bellwied:2015lba,Ding:2015fca}. Moreover, it has been
suggested that the values of these observables at small density
already give access to the existence and location of the CEP, see e.g.\ 
\cite{Karsch:2010ck}. 

In the present work we investigate the QCD phase structure and
thermodynamics as well as baryon number fluctuations within the
quark-meson (QM) model and the QCD-enhanced Polyakov--quark-meson
(PQM) model. The thermal and quantum dynamics of the models is
accessed with the functional renormalisation group (FRG), for
QCD-related reviews see \cite{Litim:1998nf, Berges:2000ew,
  Pawlowski:2005xe, Gies:2006wv,
  Schaefer:2006sr,Pawlowski:2010ht,Braun:2011pp,
  vonSmekal:2012vx}. Recent developments in FRG applications to QCD,
and in particular the embedding of low energy effective models in
first principle QCD can be found in
\cite{Braun:2009gm,Braun:2009si,Pawlowski:2010ht,Haas:2013qwp,Herbst:2013ufa,Pawlowski:2014zaa,%
  Helmboldt:2014iya,Mitter:2014wpa,Braun:2014ata}, for a recent survey
see \cite{Pawlowski:2014aha}. Baryon number fluctuations and general
higher moments have been studied with the FRG in
\cite{Skokov:2010wb,Skokov:2010uh,Friman:2011pf,Skokov:2011rq,Skokov:2012ds,%
  Morita:2012kt,Morita:2013tu,Morita:2014fda,Morita:2014nra}, see
\cite{Fu:2009wy,Fu:2010ay,Skokov:2010sf,Karsch:2010hm,%
  Schaefer:2011ex,Wagner:2009pm} for corresponding mean field results
and interesting algorithmic developments.

Here we build on the progress made in QCD-embedded low energy
effective model, see 
\cite{Pawlowski:2010ht,Haas:2013qwp,Herbst:2013ufa,Pawlowski:2014zaa,Helmboldt:2014iya},
to improve on the existing fluctuations studies within an advanced FRG
study of these observables: the momentum scale dependence of the
quark-meson scattering is taken into account as well as the
non-trivial dispersions of both quarks and mesons. Furthermore the
backreaction of the matter sector on the glue dynamics is included,
leading to a temperature-dependence modification of the Polyakov loop
potential.

The paper is organized as follows. In Section~\ref{sec:QCDlow} the
embedding of low energy effective models in QCD within the FRG
framework is recalled. In Section~\ref{sec:QMmodel} the QM model is
discussed. The significance of the additional matter fluctuations
considered in the present work is most cleanly seen without the
Polyakov loop potential. We derive the flow equations for the
couplings, and present numerical results on the phase structure,
thermodynamics and baryon number fluctuations. In
Section~\ref{sec:PQM} we discuss the QCD-enhanced PQM model. This
includes a discription of the QCD-enhancement of the Polyakov loop
potential, as well as an discussion of the intricacies of the relation
between color confinement and hadronic properties. At the end of this
Section our final results are presented for the order parameter, the
thermodynamics and the higher moments including a 2+1 flavour estimate
for the kurtosis in comparison to lattice data. Some useful formulae,
such as threshold functions, flow equations, and further discussions
can be found in the Appendix. Notably, in Appendix~\ref{app:Yukawa} we
discuss the frequency and chemical potential dependence of fermionic
couplings. This is done in view of the silver blaze property, and the
thermal decay properties specifically relevant for the thermodynamics at low
temperatures.

\section{From QCD to low energy effective models}  
\label{sec:QCDlow}           
%
\begin{figure}[t]
\centering
\includegraphics[width=8.5cm]{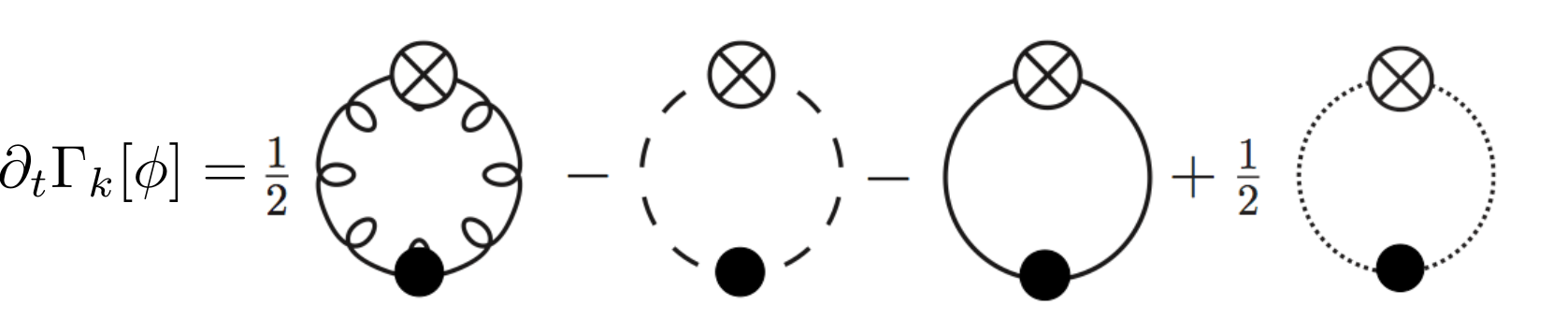}
\caption{Flow equation for the effective action or grand potential
  $\Gamma$ or $\Omega$ in full QCD. The first two loops with gluons
  and ghosts give the glue contribution $\Omega_{\text{\tiny glue}}$, the quark
  and mesonic loops give the matter contributions $\Omega_{\text{\tiny matt}}$.}
\label{fig:FRG-QCD_rebosonised}
\end{figure} 
%
It has been argued, see e.g.\ \cite{Pawlowski:2014aha}, that low
energy effective models can be derived systematically from QCD
allowing for a precise determination of the fundamental model
parameters. Within the functional renormalisation group framework this
can be discussed already on the level of the flow equation for the
scale-dependent effective action, $\Gamma_k[\Phi]$, with the
super-field $\Phi=(A_\mu,c,\bar c,q,\bar q,\phi,...)$ with
$\phi=(\sigma,\vec \pi)$ and possible furthe effective hadronic
fields.  The flow equation reads schematically
\begin{align}
  \label{eq:dtGam}
  \partial_{t}\Gamma_{k}[\Phi]=\frac{1}{2}\mathrm{Tr}\,G_{\Phi
    \Phi}[\Phi]\partial_{t} R^{\Phi}_{k}\,, \quad t=\ln (k/\Lambda)\,,
\end{align}
where 
\begin{align}\label{eq:GPhi} 
  G_{\Phi_i \Phi_j}[\Phi] =
  \left(\0{1}{\0{\delta^2\Gamma_k[\Phi]}{\delta\Phi^2}+R_k^\Phi}\right)_{ij}\,,
\end{align}
is the full field-dependent propagator, $k$ is the infrared cutoff
scale, and $\Lambda$ is some reference scale. The flow equation
\eq{eq:dtGam} is depicted in \Fig{fig:FRG-QCD_rebosonised}, for more
details and QCD results, see e.g.\
\cite{Pawlowski:2014aha,Braun:2009si,Mitter:2014wpa,Braun:2014ata}. All
loops in the flow carry only momenta $q^2\lesssim k^2$ with the
infrared cut-off scale $k$, and \eq{eq:dtGam} implements a successive
integration of momentum modes. The first two terms in
\Fig{fig:FRG-QCD_rebosonised} constitute the contributions of the glue
system, while the third term entails the quark fluctuations and the
last loop stands for the fluctuations of the effective hadronic
degrees of freedom. Hence, the effective action can be written as
\begin{align}\label{eq:Gasplit}
  \Gamma_k[\Phi]=
  \Gamma_{\text{\tiny{glue}},k}[\Phi]+\Gamma_{\text{\tiny{matt}},k}[\Phi]\,,
  \quad
  \Gamma_{\text{\tiny{matt}},k}=\Gamma_{q,k}+
\Gamma_{\phi,k}\,,
\end{align}
where $\Gamma_{\text{\tiny{glue}},k}$ includes the integrated flow of
gluon and ghost loop, $\Gamma_{q,k}[\Phi]$ that of the quark loop, and
$\Gamma_{\phi,k}[\Phi]$ that of the hadronic degrees of freedom. Note,
that the split in quark and hadronic contributions does not reflect an
effective theory setup. Within the framework of dynamical
hadronisation, introduced in
\cite{Gies:2001nw,Gies:2002hq,Pawlowski:2005xe,Floerchinger:2009uf},
it has turned out to be a very effective and powerful parameterisation
of matter fluctuations in ab initio QCD in terms of genuine quark
scatterings and resonant momentum channels with hadronic quantum
numbers, for applications to QCD see e.g.\
\cite{Mitter:2014wpa,Braun:2014ata}.

\begin{figure}[t]
\centering
\includegraphics[scale=0.7]{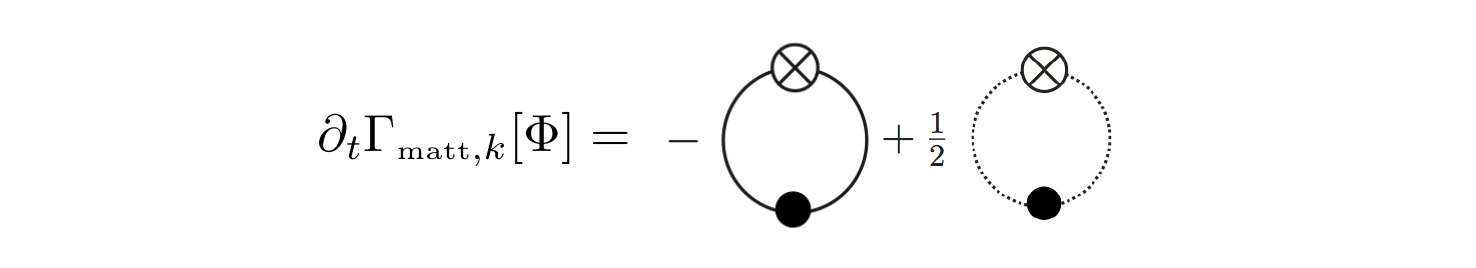}
\caption{Flow equation for the effective action or grand potential
  $\Gamma$ or $\Omega$ in the low energy regime. The gluon and quark loops in
  \Fig{fig:FRG-QCD_rebosonised} are suppressed and the quark and
  mesonic loops give the matter contributions $\Omega_{\text{\tiny
      matt}}$.}
\label{fig:FRG-EFT_rebosonised}
\end{figure}

Moreover, it facilitates the embedding of low energy effective
theories of QCD such as NJL-type and QM-type models with and without
confining Polyakov loop potential in first principle QCD: For the sake
of definiteness we restrict ourselves to Landau gauge QCD. There, the
physical mass gap of QCD is reflected in a mass gap in the gluon
propagator of about 1 GeV. Accordingly, glue fluctuations decouple
below this scale. The ghost has no mass gap but only couples to QCD
matter via the gluon. Hence, it effectively shares the gluon
decoupling. In summary, the fluctuations of the glue sector decouple
from the matter sector in the infrared. Still, the effective potential
of the glue sector,
$\Omega_{\text{\tiny{glue}}}=\Gamma_{\text{\tiny{glue}}}$, is required
for solving the quantum equations of motion for low energies. There,
it is relevant for determining the physical glue background. In
summary, integrating the QCD flow down to scales $k=\Lambda$ with
$\Lambda\lesssim 1$ GeV leaves us with an effective matter theory with
glue background. The flow equation for this QCD-embedded low energy
effective theories is given by the two matter loops in
\Fig{fig:FRG-QCD_rebosonised}, see \Fig{fig:FRG-EFT_rebosonised}.
%
%
The total effective action is given by \eq{eq:Gasplit} where
$\Gamma_{\text{\tiny{glue}},k}$ is an external input. Since it does
not enter the flow equation of the matter part, we only need it at
$k=0$. 

In a flow equation approach to QCD, $\Omega_{\text{\tiny{glue}}}$
comprises the contributions from the gluonic and ghost loop in
\Fig{fig:FRG-QCD_rebosonised}, while
$\Omega_{\text{\tiny{matt}}}=\Gamma_{\text{\tiny{matt}}}$ comprises
that from the quark and mesonic loop in
\Fig{fig:FRG-QCD_rebosonised}. The representation in terms of a flow
equation facilitates the discussion of the repsective scales. Dropping
$\Omega_{\text{\tiny matt}} $ the grand potential reduces to the pure
glue potential with a transition temperature $T_{\text{\tiny glue}}
\approx 200$\,MeV, see
\cite{Pawlowski:2010ht,Haas:2013qwp,Herbst:2013ufa}. The glue
potential (in the current approximation) features a first order phase
transition, and below $T_{\text{\tiny glue}}$ the Polyakov loop
expectation vanishes. The possible minor center-breaking corrections
of the glue propagation due to the vacuum polarisation are neglected.
We shall see, that for the present argument we can safely drop these
contributions as they only increase the expectation value of the
Polyakov loop.
 
We conclude that the non-vanishing expectation values of $L$ and $\bar
L$ in QCD are dominantly triggered by the quark loop in
$\Omega_{\text{\tiny matt}}$. The influence of the non-trivial $A_0$-
or $L,\bar L$-background will be discussed at the end of this
work. For the time being we concentrate on the impact of improved
matter fluctuations on the higher moments. This concerns in particular
the influence of the deformed dispersion that relates to non-trivial
wave function renormalisations of quarks and mesons.

\section{Quark-meson model}
\label{sec:QMmodel}
In the spirit of the arguments of the last section we first
concentrate on the low energy matter quantum, thermal and density
fluctuations for two-flavour QCD. These fluctuations are well-captured
within a quark meson model at scales well below 1 GeV, see the reviews
\cite{Schaefer:2006sr,Braun:2011pp} and references therein.  Moreover, 
full QCD calculations indicate that the width of the transition region
from the quark-gluon regime to the hadronic one is small. This
guarantees that the effective model is applicable and reliable at low
energy,
see~\cite{Pawlowski:2014aha,Braun:2009si,Mitter:2014wpa,Braun:2014ata}. The
scale-dependent effective action of the quark-meson model is given by
\begin{align}\nonumber 
  \Gamma_{k}&=\int_{x}\Big\{Z_{q,k}\bar{q}(
  \gamma_{\mu}\partial_{\mu}-\gamma_{0}\mu)q +
  \frac{1}{2}Z_{\phi,k}(\partial_{\mu}\phi)^2\\[2ex]
  &\quad+h_{k} \,\bar{q}\left( T^{0}\sigma+i\gamma_{5}
  \vec{T}\cdot\vec{\pi}\right) q+V_{k}(\rho)-c\sigma\Big\}+\cdots
  \,,
 \label{eq:action}
\end{align}
with $\int_{x}=\int_0^{1/T}d x_0 \int d^3 x$, and the dots indicate
higher order terms in derivatives and fields. Accordingly,
\eq{eq:action} has to be understood in the spirit of both, a
derivative expansion at low energies as well as a vertex expansion in
terms of the fluctuation physics. The validity of the latter is
important for having a good grip on the fluctuation physics at low
energies, and is crucial for a good access to higher moments. Note
that the validity of the expansion in terms of the fluctuation physics
is based on the relevance ordering of fluctuating modes that propagate
off-shell in the loops.  This should not be confused with the relevance
of higher asymptotic states in the effective action. This important
distinction is at the root of the quantitative agreement for the QCD
thermodynamics computed in Polyakov quark-meson models including the
trace anomaly, see \cite{Herbst:2013ufa}. In turn, in approaches based
on an expansion in asymptotic states a large number of the latter is
necessary to have quantitative access to the full fluctuation physics.

The meson field $\phi=(\sigma,\vec{\pi})$ introduced in \eq{eq:action}
is in the $O(4)$ representation, with $\rho=\phi^2/2$. Here $k$ is the
infrared (IR) cutoff scale of FRG, see Sec.\ref{sec:FRG} for more
details; $\mu$ is the quark chemical potential. $\vec{T}$ are the
$SU(N_{f})$ generators with
$\mathrm{Tr}(T^{i}T^{j})=\frac{1}{2}\delta^{ij}$ and
$T^{0}=\frac{1}{\sqrt{2N_{f}}}\mathbb{1}_{N_{f}\times N_{f}}$. The
field-dependent effective potential $V_{k}(\rho)$ is $O(4)$ invariant,
and the chiral symmetry is explicitly broken by the linear term
$-c\sigma$. Thus, the mass of the Pions $\vec \pi$ is proportional to
the linear breaking parameter $c$.  

The quark and meson wave function renormalisations, i.e. $Z_{q,k}$ and
$Z_{\phi,k}$, are scale-, momentum-, and field-dependent. Moreover, at
finite temperature $Z$'s split into $Z^{\parallel}$ and $Z^{\perp}$,
corresponding to those longitudinal and transversal to the heat bath,
respectively. The transveral ones dominate the fluctuations
physics. It has also been shown, that the approximation with
scale-dependent $Z$'s already captures the effects of the non-trivial
momentum and frequency dependence for the propagators quantitatively,
see \cite{Helmboldt:2014iya}. Hence, as a good approximation we only
keep the scale-dependence for the wave function renormalizations, and
assume $Z^{\parallel}=Z^{\perp}$ throughout this work.  Quark and
meson fields interact with each other via a scale-dependent Yukawa
coupling $h_{k}$ in Eq.~(\ref{eq:action}). For the fermionic
parameters $Z_q$ and $h$ a consistent treatment of the frequency- and
$\mu$-dependence is crucial for the correct low temperature
physics. This is discussed in detail in
Appendix~\ref{app:Yukawa}. Higher order quark-mesonic interactions can
be included through the dependence of $h_{k}(\rho)$ on the meson field
\cite{Pawlowski:2014zaa}, which will be discussed elsewhere.

We investigate the thermodynamics of the QM model within various 
truncations. This sheds light into the roles played by the different
fluctuations. Here, we denote the truncation with a full effective
potential and both running $Z_k$'s and $h_k$, as 'full'. If the flows
of $Z$'s and $h$ are turned off, i.e.
\begin{equation}
\partial_{t}Z_{\phi/q}=0,\quad\partial_{t}h=0, \label{}
\end{equation}
with $t$ in (\ref{eq:dtGam}) and the reference scale $\Lambda$ being
the ultraviolet (UV) cutoff, the effective action in
Eq.~(\ref{eq:action}) is that of the local potential approximation
(LPA). Apart from LPA we also investigate the $\mathrm{LPA}^{\prime}$
approximation, i.e. the LPA with running $Z$'s, but with a constant
$h$. Here we choose a constant renormalised Yukawa coupling,
$\partial_{t} \bar h=0$ with $\bar h=h/Z^{1/2}_\phi Z_q$, for more
details see the next section.  A vanishing flow of $\bar h$ provides a
good approximation to the small scale-dependence of $\bar h$ in full
QCD, see \cite{Mitter:2014wpa,Braun:2014ata}. In turn,
$\partial_{t}h=0$ does not work well due to the large anomalous
dimension of the mesons. Another approximation is characterized by a
running $h_k$ but constant $Z$'s, which is denoted as
$\mathrm{LPA}+h_{k}$ in our work. In Table~\ref{tab:trun} we summarise
the truncations studied in this work.

\begin{table}[t]
  \centering
  \begin{tabular}[c]{c|c}
   \hline \hline
    Label & Truncations\\ \hline
    LPA & $\partial_{t}Z_{\phi/q}=0,\;\partial_{t}h=0$\\
    $\mathrm{LPA}^{\prime}$ & $\partial_{t}Z_{\phi/q}\neq 0,\;\partial_{t}\bar{h}=0$\\
    $\mathrm{LPA}+h_{k}$ & $\partial_{t}Z_{\phi/q}=0,\;\partial_{t}h\neq 0$\\
    full & $\partial_{t}Z_{\phi/q}\neq 0,\;\partial_{t}h\neq 0$\\ \hline
  \end{tabular}
  \caption{Four different truncations and their labels used in this work.}
  \label{tab:trun}
\end{table}
\subsection{Flow equations}
\label{sec:FRG}
As discussed in detail in Section~\ref{sec:QCDlow}, the flow equation
for the quark meson model follows from that of first principle QCD for
low cutoff scales $k\lesssim$ 1GeV, where the gluonic fluctuations
decouple. Then the matter loops carry the low energy fluctuations in
the presence of a glue background. The glue sector only gives rise to
a background potential for $A_0$ or the Polyakov loop $L,\bar L$
respectively. Dropping (the flow of) the background potential and
restricting ourselves for the time being to trivial backgrounds
$A_0=0$ or $L,\bar L=1$ we arrive at the flow equation for the quark
meson model, see \Fig{fig:FRG-EFT_rebosonised}, 
\begin{align}
  \label{eq:QMflow}
  \partial_t\Gamma_{k}[\Phi]=\frac{1}{2}\mathrm{Tr}\,
  G_{\phi\phi}[\Phi]\partial_{t}R^{\phi}_{k}
  -\mathrm{Tr}\,G_{q\bar{q}}[\Phi]\partial_{t}R^{q}_{k}\,,
\end{align}
where the super field now only includes the fluctuating quark and
meson fields, $\Phi=(q,\bar{q},\phi)$, and the regulator
$R^\Phi=(R^{q}_k\,,\, R^{\phi}_{k})$ suppresses infrared fluctuations
of quark and meson fields respectively. The traces in
Eq.~(\ref{eq:QMflow}) sum over momenta and internal quark (Dirac and
color) and meson (flavour) indices. The full, field-dependent
propagator, \eq{eq:GPhi} reduces to the coupled one of quarks and
mesons, to wit,
\begin{align}
  \label{eq:prop}
 G_{\phi\phi/q\bar q}[\Phi]=\left(
 \0{1}{ \0{\delta^2
        \Gamma_k[\Phi]}{\delta \Phi^2}
      +R^\Phi_k }
\right)_{\phi\phi/q\bar q}\,,
\end{align}
for the diagonal parts that show up in \eq{eq:QMflow}.

\subsection{Flow equations for the effective
  potential} \label{sec:dtV} In this work we use $3d$ flat
regulators, \cite{Litim:2000ci,Litim:2001up}, for quarks and mesons,
see also Appendix~\ref{app:threshold}. The flow of the effective potential $V_{k}(\rho)$ 
is obtained by substituting Eqs.~(\ref{eq:action}),(\ref{eq:prop}) in (\ref{eq:QMflow}) and 
evaluating the flow at constant mesonic fields, 
\begin{align}\nonumber 
    \partial_{t}V_{k}(\rho)&=\frac{k^{4}}{4\pi^{2}}\big[(N_{f}^{2}-
    1)l_{0}^{(B,4)}(\bar{m}_{\pi,k}^{2},\eta_{\phi,k};T)\\[2ex]
\nonumber 
    &\quad+l_{0}^{(B,4)}(\bar{m}_{\sigma,k}^{2},\eta_{\phi,k};T)\\[2ex]
    &\quad-4N_{c}N_{f}l_{0}^{(F,4)}(\bar{m}_{q,k}^{2},\eta_{q,k};T,\mu)\big]\,,
 \label{eq:Vflow}\end{align}
where $l_{0}^{(B/F,4)}$, e.g.\ \cite{Braun:2009si,Pawlowski:2014zaa},
are the threshold functions, see Appendix~\ref{app:threshold}. The
renormalised dimensionless meson and quark masses are given by
  \begin{align}\nonumber 
    \bar{m}_{\pi,k}^{2}&=\frac{V_{k}^{\prime}(\rho)}{k^{2}Z_{\phi,k}}\,,\\[2ex]
\nonumber 
    \bar{m}_{\sigma,k}^{2}&=\frac{V_{k}^{\prime}(\rho)+2\rho 
V_{k}^{\prime\prime}(\rho)}{k^{2}Z_{\phi,k}}\,,\\[2ex]
    \bar{m}_{q,k}^{2}&=\frac{h_{k}^{2}\rho}{2k^{2}Z_{q,k}^{2}}\,,
 \label{eq:mass} \end{align}
and the anomalous dimensions are defined by
\begin{equation}
  \label{eq:anom}
  \eta_{\phi,k}=-\frac{\partial_{t}Z_{\phi,k}}{Z_{\phi,k}}\,,\quad \quad \eta_{q,k}=
-\frac{\partial_{t}Z_{q,k}}{Z_{q,k}}\,.
\end{equation}
Here we consider frequency and spatial momentum-independent anomalous
dimension, the anomalous dimensions are evaluated at low frequencies
and spatial momenta. The related depencence if covered by the
$k$-dependence. The precise definition of the anomalous dimensions in
\eq{eq:anom} and a detailed discussion of this approximation is found
in Appendix~\ref{app:mesons} for $\eta_\phi$, and in
Appendix~\ref{app:Yukawa} for $\eta_q$. The latter also contains an
evaluation of the $\mu$ and $T$ dependence important for the silver
blaze property of QCD and the thermodynamics at low temperatures. 

In general it is more convenient to work with renormalized fields and
renormalization group (RG)-invariant quantities. We denote them by
symbols with bar, as the renormalized masses shown in
Eqs.~(\ref{eq:mass}). The relations between renormalised quantities
and original ones read
\begin{equation}
  \label{}
  \bar{\phi}=Z_{\phi,k}^{\frac{1}{2}}\phi\,,\quad \bar{h}_{k}=
  \frac{h_{k}}{Z_{q,k}Z_{\phi,k}^{\frac{1}{2}}}\,,
\end{equation}
taking the meson field and Yukawa coupling as examples. Thus we have
$\bar{\rho}=Z_{\phi,k}\rho$ and the effective potential
$\bar{V}_{k}(\bar{\rho})=V_{k}(\rho)$. Note that the chiral symmetry
breaking term $-c\sigma$ in Eq.~(\ref{eq:action}) is linear in the
fields. Hence it neither contributes to the right hand side of any
flow, nor does it flow. Considering $\bar{c}_k=c/Z_{\phi,k}^{1/2}$,
this leads to 
\begin{equation}
  \label{eq:cflow}
  \partial_t\bar{c}_k=\frac{1}{2}\eta_{\phi,k}\bar{c}_k\,.
\end{equation}
In the present work, we solve the flow equation for the effective
potential, Eq.~(\ref{eq:Vflow}), within a Taylor expansion about a
fixed unrenormalised field value $\kappa$, to be contrasted to one
about the scaling minimum $\rho_{0,k}$, see e.g.\ \cite{Papp:1999he}.
It has been shown in \cite{Pawlowski:2014zaa} that such a Taylor
expansion about a fixed $\kappa$ has the most rapid convergence. Expanded in
renormalised fields, the effective potential then reads
\begin{equation}
  \label{eq:Vexpa}
  \bar{V}_{k}(\bar{\rho})=\sum_{n=0}^{N}\frac{\bar{\lambda}_{n,k}}{n!}(
\bar{\rho}-\bar{\kappa}_k)^n\,,
\end{equation}
with $\bar{\lambda}_{n,k}=\lambda_{n,k}/Z_{\phi,k}^n$ and $\bar
\kappa_k= Z_\phi \kappa$ with 
\begin{equation}
  \label{eq:kapflow}
  \partial_t \bar{\kappa}_k=-\eta_{\phi,k}\bar{\kappa}_k\,.
\end{equation}
In accordance with the convergence
discussion in \cite{Pawlowski:2014zaa} we already find a good
convergence for $N=5$. For example, the difference for $f_\pi=\bar
\sigma$ between $N=5$ and $N=6,7$ is less than 1\% for all
temperatures. Consequently we use $N=5$ for all computations. 
%
\begin{figure*}[t]
\includegraphics[scale=0.6]{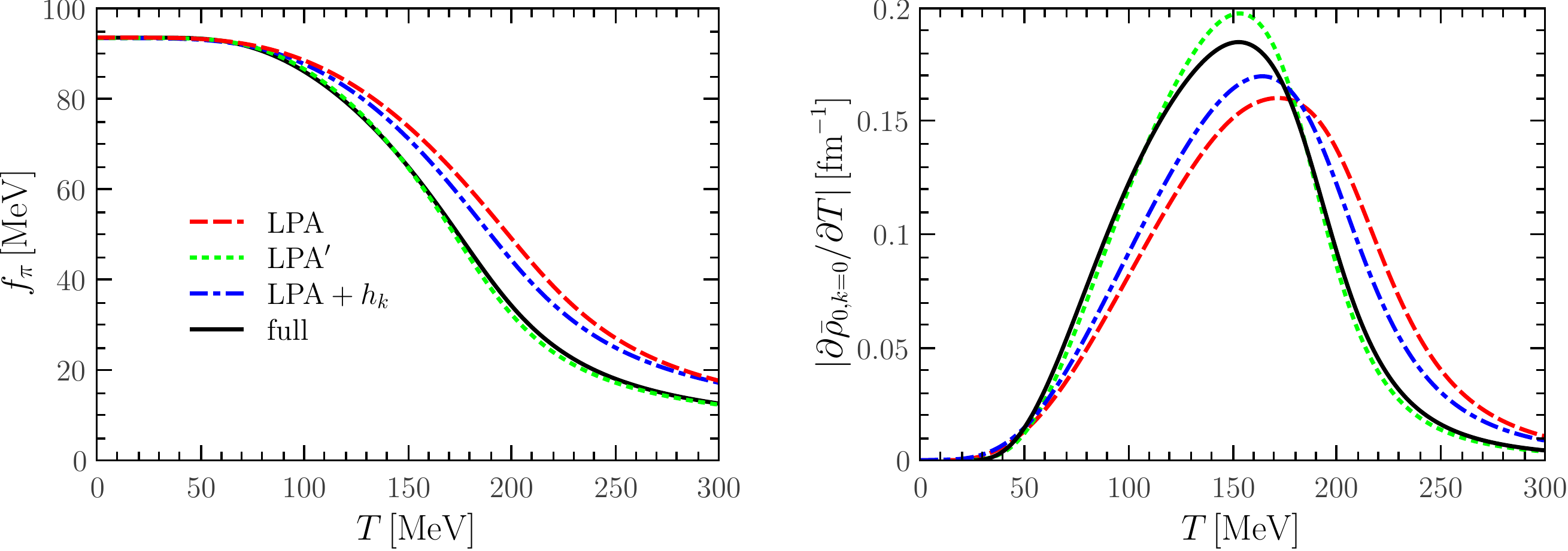}
\caption{Order parameter of the chiral phase
  transition: $f_{\pi}=(2\bar{\rho}_{0,k=0})^{1/2}$ (left panel), and
  $|\partial \bar{\rho}_{0,k=0}/\partial T|$ (right panel) as
  functions of the temperature in the different truncations of 
  Table~\ref{tab:trun}.}\label{fig:fpi}
\end{figure*}
%
Inserting Eq.~(\ref{eq:Vexpa}) into the left hand side of
Eq.~(\ref{eq:Vflow}), one arrives at
\begin{align}\nonumber 
    &\partial_{\bar{\rho}}^{n}\Big(\partial_t\big|_{\rho}
    \bar{V}_{k}(\bar{\rho})\Big)\Big|_{\bar{\rho}=\bar{\kappa}_k}\\[2ex] 
    =&(\partial_t\bar{\lambda}_{n,k}-n\eta_{\phi,k}\bar{\lambda}_{n,k})-(\partial_t
    \bar{\kappa}_k+\eta_{\phi,k}\bar{\kappa}_k)\bar{\lambda}_{n+1,k}\,,
   \label{eq:lamflow}\end{align}
 Note that the term proportional to the higher order coupling
 $\bar{\lambda}_{n+1,k}$ vanishes due to Eq.~(\ref{eq:kapflow}). This
 is at the root of the rapid convergence of the present expansion
 scheme, see \cite{Pawlowski:2014zaa}. In our computation the
 expansion point is chosen such that $\bar{\kappa}_{k=0}=\bar
 \rho_{0,k=0}$. Hence it is an expansion at the physical minimum, and
 we are well within the convergence radius of such an expansion.

Apart from the flow equation for the effective potential, we also need
the flow equations of the Yukawa coupling, and the anomalous
dimensions.  They are deferred to the Appendix~\ref{app:Yukawa}.

\subsection{Initial conditions} 
\label{sec:thermo}
It is left to specify the input parameters of the effective
model. Within the FRG approach, this is done by fixing the initial
effective action $\Gamma_\Lambda$ in the vacuum. The UV-cutoff scale
is chosen to be $\Lambda=700\,\mathrm{MeV}$ throughout this
work. 

Choosing a larger $\Lambda$ requires the inclusion of dynamical
hadronisation as well as that of gluonic fluctuations: the validity
range of the current model can be accessed in the vacuum by using the
results of the QCD flows in \cite{Mitter:2014wpa,Braun:2014ata} at a
given cutoff scale $k=\Lambda$ as initial effective action of the
model at hand. The results of such a computation can then be compared
with the results of the QCD flows. The larger the initial cutoff scale
$\Lambda$ is, the larger the deviations grow. This limits the
reliability of the model predictions. 

In turn, choosing a smaller $\Lambda$ limits our thermal range, as the
physics of temperatures with $T/\Lambda \gtrsim 1/7 - 1/5$ is severely
contaminated by cutoff effects, see \cite{Helmboldt:2014iya}. This
entails that the maximal temperature accessible within the current
model is $T_{\text{\tiny}{max}} = 100-140$ MeV. In the present work we
also amend the model with temperature-dependent initial conditions
which gives access to higher temperatures. This is detailed in the
next Section~\ref{sec:pres}.
\begin{table}[b]
  \centering
  \begin{tabular}[c]{c|c|c|c|c}
    \hline \hline 
    Truncations & $\bar{\lambda}_{\Lambda}$ & $\bar{h}_{\Lambda}$ & $
    \bar{c}_{\Lambda}$ ($\times 10^{-3}\mathrm{GeV}^3$)& $m_{\sigma}$ ($\mathrm{MeV}$)\\ \hline
    LPA & 43 & 6.5 & 1.7 & 585.3\\
    $\mathrm{LPA}^{\prime}$ & 72 & 6.5 & 1.98 & 584.7\\
    $\mathrm{LPA}+h_{k}$ & 36.8 & 5.5 & 1.7 & 551.2\\
    full & 77 & 7.1 & 2.0 & 578.9\\ \hline
  \end{tabular}
  \caption{Input parameters and the predicted $\sigma$-meson mass with different truncations.}
  \label{tab:para}
\end{table}

At the initial scale the effective potential is well approximated by a
classical potential, to wit
\begin{align}\label{eq:VLambda}
\bar{V}_{\Lambda}(\bar{\rho})=\frac{\bar{\lambda}_{\Lambda}}{2}\bar{\rho}^2\,,
\end{align}
with the relevant coupling
$\bar{\lambda}_{\Lambda}=\bar\lambda_{2,\Lambda}$. For the sake of
simplicity we have chosen the mass $m_{\phi,\Lambda}^2=0$. This fixes 
the $\sigma$-mass and is respsonsible for its minor variation for the different 
truncations as well as a relatively large meson scattering coupling
$\bar \lambda_\Lambda$ at the initial scale, see
Table~\ref{tab:para}. Heuristically, if aiming at maximising the
effective thermal and chemical potential range one has to utilise the
meson mass parameter in order to minimise $\bar \lambda_\Lambda$. Such a choice leads
to small mesonic contributions at large cutoff scales $k\to \Lambda$
and hence at large $T$ and $\mu$. This simulates the
melting of the mesons at large scales. Still, it rather is a feature
of the model instead of built-in physics. In the present framework the
melting of mesons is naturally induced by dynamical hadronisation and
will be considered elsewhere. Here, we stick to \eq{eq:VLambda}.

The other two relevant couplings are the Yukawa coupling
$\bar{h}_{\Lambda}$ and the coefficient of the linear breaking term
$\bar{c}_{\Lambda}$. The three relevant couplings of the model are
determined by fitting hadronic observables in the vaccum, the $\pi$
decay constant $ f_{\pi}=\bar\sigma$ with $f_\pi =93.5\,\mathrm{MeV}$,
the $\pi$-meson mass $m_{\pi}=135\,\mathrm{MeV}$, and the quark mass
$m_q= 1/2\, \bar h\, \bar \sigma$ with
$m_{q}=303.5\,\mathrm{MeV}$. The three couplings at the initial scale
$k=\Lambda$ as well as the predicted $\sigma$-meson mass at vanishing
cutoff $k=0$ are summarised in Table~\ref{tab:para} for the different
truncations discussed in Section~\ref{sec:QMmodel}.

\subsection{Pressure and entropy}
\label{sec:pres}
\begin{figure*}[t]
\includegraphics[scale=0.6]{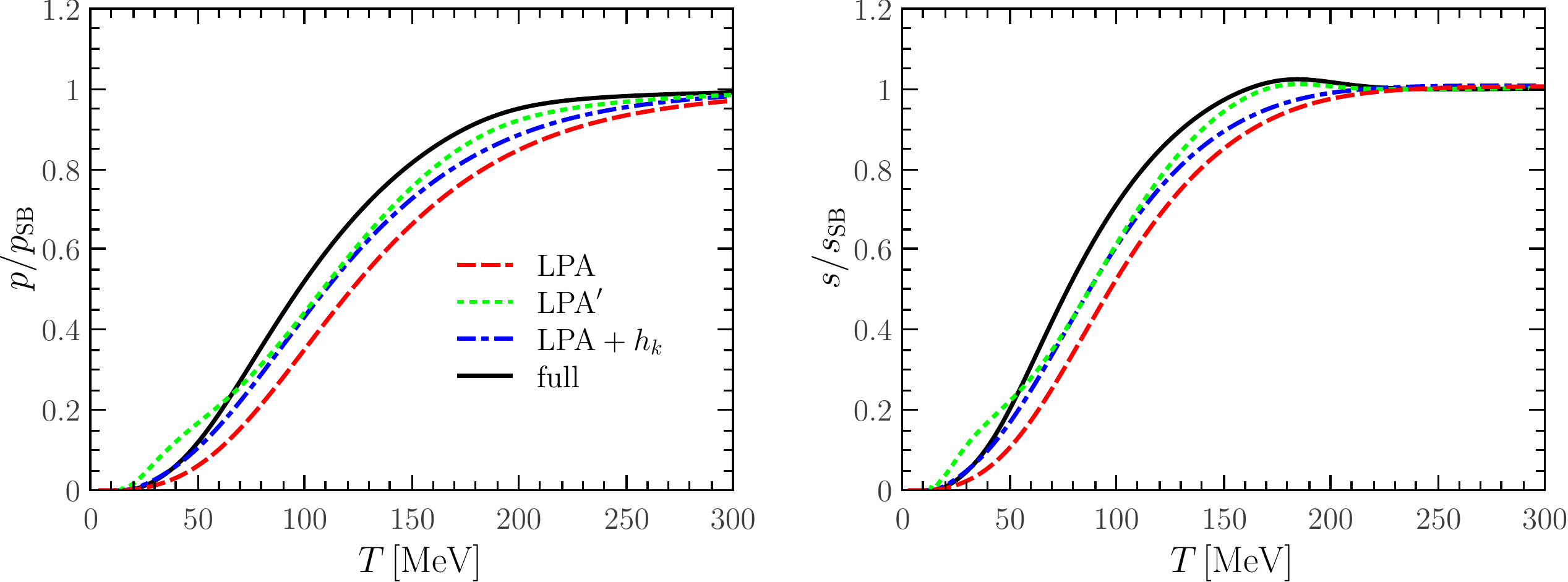}
\caption{Pressure (left panel) and entropy density
  (right panel), normalised to their Stefan-Boltzmann values, as
  functions of $T$.}\label{fig:ps}
\end{figure*}

The thermodynamical potential density, i.e. grand potential
density, $\Omega_{k}(\bar{\rho})$ is related to the effective
potential through
\begin{equation}\label{eq:grandpot}
  \Omega_{k}(\bar{\rho})=\Gamma_{k,T}[\bar\rho]-\Gamma_{k,T=0}[\bar\rho]\,,
\end{equation}
which is normalised to zero at vanishing temperature. It has been
argued above that the thermal range is given by
$\Lambda/T_{\text\tiny{max}}\approx 5-7$, \cite{Helmboldt:2014iya}.
The ultraviolet cutoff $\Lambda$ used in the present work has been
chosen as $\Lambda= 700$ MeV in order to minimise the systematic
errors of missing glue fluctuations and additional four-fermi and
other higher order couplings that can be described within dynamical
hadronisation. However, then temperature-dependent initial conditions
are required for $T\gtrsim 100-140$ MeV: Already the flow at the
initial scale is sensitive to thermal fluctuations for these
temperatures, and hence the initial action as well as thermodynamical
quantities, such as the pressure, entropy, fluctuations etc. are. This
has been discussed at length in \cite{Herbst:2013ufa}, where it has
been used for a systematic error estimate. In the present case the
thermodynamical potential density at $\Lambda$ can be estimated by the
integrated flow of the difference in \eq{eq:grandpot} without feeding
back the flow of the parameters. This leads to
\begin{align}\nonumber 
  \Omega_{\Lambda}\approx &-\frac{4N_c
    N_f}{12\pi^2}\int_{\Lambda}^{\infty}dk\, k^3
  \bigg[\frac{1}{\exp\big(\frac{k-\mu}{T}\big)+1}\\[2ex]
  &\hspace{3cm}+\frac{1}{\exp\big(\frac{k+\mu}{T}\big)+1}\bigg]\,,
\label{eq:UVgrandpot}\end{align}
where we have dropped the negligible contribution of the mesonic
degrees of freedom for cutoff scales above the chiral symmetry
breaking scale $k_\chi\ll \Lambda$. We estimate that this procedure
effectively extends the thermal range to $T\lesssim \Lambda/5 -
\Lambda/3$, as it mimics the thermal change of the initial condition
at $k=\Lambda$. For the ultraviolet cutoff $\Lambda= 700$ MeV used
here this gives us access to temperatures with $T_{\text{\tiny{max}}}
\approx 140 - 230$ MeV. The corresponding systematic error estimates
from \cite{Herbst:2013ufa} will be used later for an error estimate of
the bayronic fluctuations.

The pressure and entropy density follow readily as 
\begin{align}\label{eq:p+s}
  p=-\Omega_{k=0}(\bar{\rho}_{0,k=0})\,,\quad {\rm and}\quad
  s=\frac{\partial p}{\partial T}\,, 
\end{align}
from the grand potential $\Omega$. \Fig{fig:fpi} (left panel) compares
the temperature dependence of the order parameter
$f_{\pi}=\bar{\sigma}$ on $T$ in different truncations. One observes
that the truncations cluster into two groups: full and
$\mathrm{LPA}^{\prime}$ in one group, LPA and $\mathrm{LPA}+h_{k}$ in
the other. The crossovers in LPA and $\mathrm{LPA}+h_{k}$ are broader than 
that of full approximation and $\mathrm{LPA}^{\prime}$. This entails that the
non-trivial momentum dependence of the propagators, encoded in the
wave function renormalization factors, increases the strength of the
crossover and lowers the pseudo-critical temperature. This is seen very
clearly in the right panel of Fig.~\ref{fig:fpi}, where $|\partial
\bar{\rho}_{0,k=0}/\partial T|$ as a function of temperature is
depicted. We also expect larger changes in $\mathrm{LPA}^{\prime}$ and
full approximation for the high-order fluctuations, see
Section~\ref{sec:barfluc}.

In Fig.~\ref{fig:ps} we show the pressure and entropy density as
functions of the temperature in different truncations. We have 
normalised them with those of massless ideal gas of quarks, i.e. the
Stefan-Boltzmann (SB) limit,
\begin{align}
  \frac{p_{\mathrm{SB}}}{T^4}=2N_cN_f\bigg[\frac{7\pi^2}{360}+\frac{1}{12}
\Big(\frac{\mu}{T}\Big)^2+\frac{1}{24\pi^2}\Big(\frac{\mu}{T}\Big)^4\bigg]\,.
\end{align}
One observes that the pressures calculated with truncations full 
and $\mathrm{LPA}^{\prime}$ increase with $T$ more rapidly than those
with the other two truncations, and their values are also
larger. These findings indicate that propagators with improved
momentum-dependence stiffen the equation of state. A similar behavior
is found in the calculations of entropy density, as shown in the right
panel of Fig.~\ref{fig:ps}. The entropy obtained with wave function
renormalization factors rises more rapidly.
\begin{figure*}[t]
\includegraphics[scale=0.6]{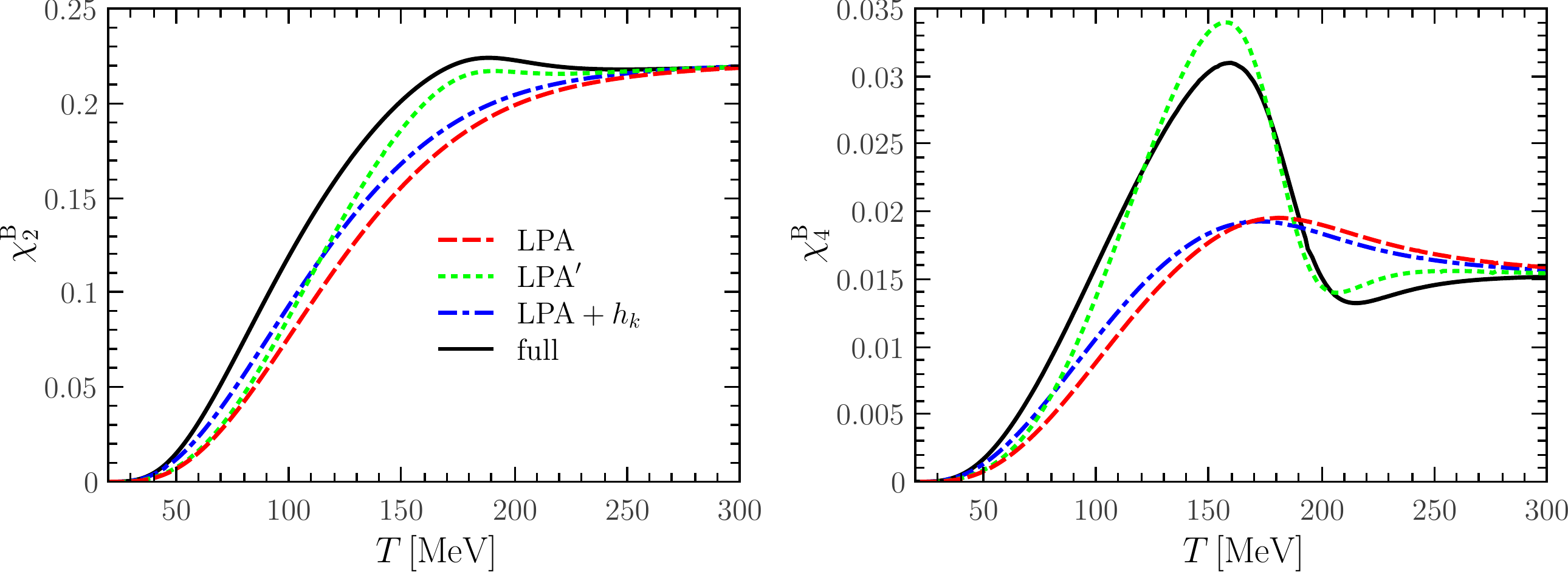}
\caption{Quadratic (left panel) and quartic (right
  panel) baryon number fluctuations as functions of $T$ in the QM
  model for the different truncations, see Table~\ref{tab:trun}.}\label{fig:chi2B}
\end{figure*}
\begin{figure}[t!]
  \includegraphics[scale=0.6]{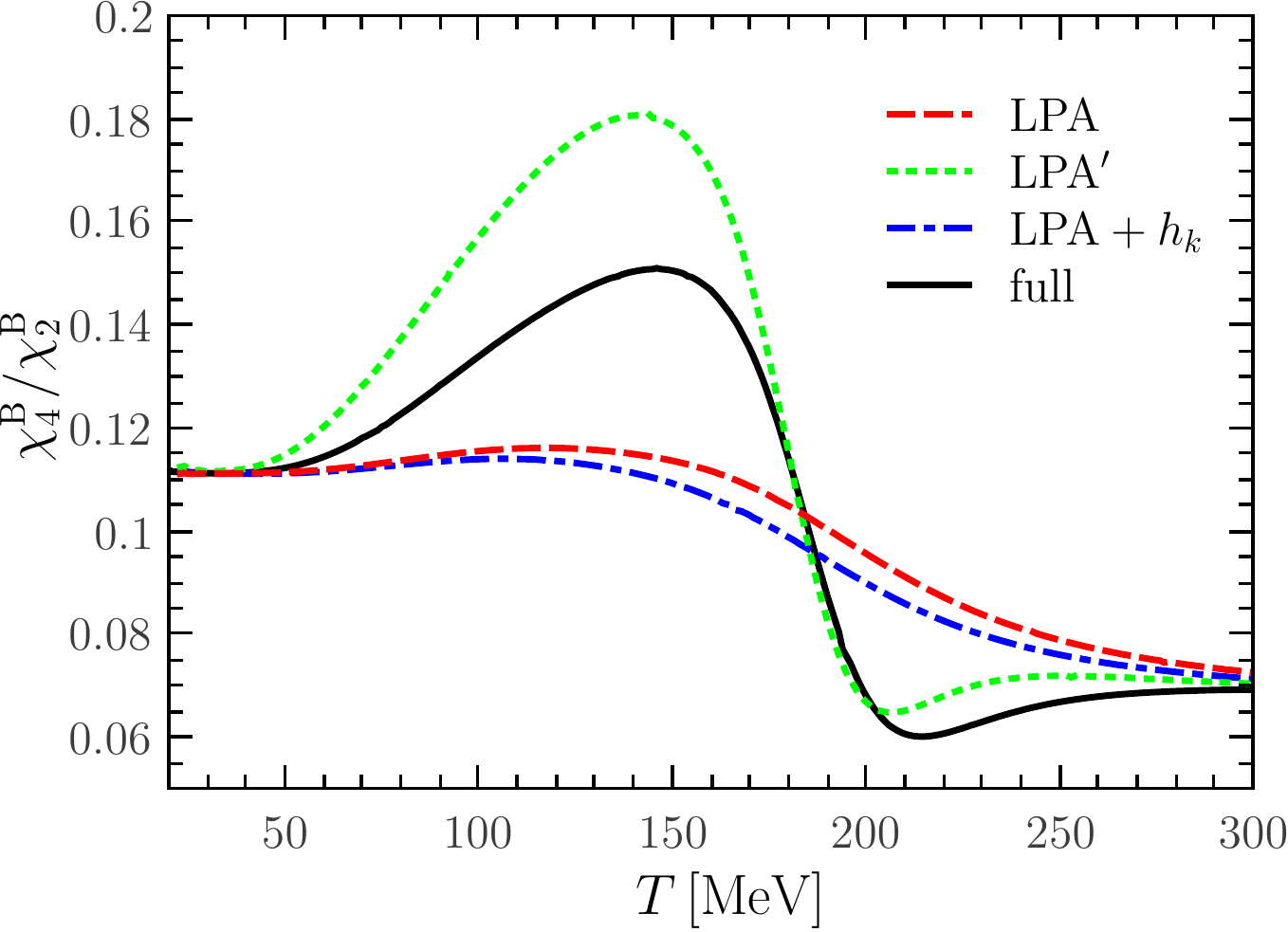}
  \caption{Kurtosis of the baryon number distribution
    as a function of the temperature in the QM
    model.}\label{fig:R42QM}
\end{figure}

In approximations with scale-dependent wave function renormalisations
both the entropy as well as quadratic baryonic fluctuation show a peak
at about 170 -180 MeV. This is an artefact of the current
approximation, and can be traced back to the systematic errors
discussed before. First of all these temperatures are above the
estimate from \cite{Helmboldt:2014iya} for the maximal temperature
$T_{\text{\tiny{max}}} \approx 100 - 140$ MeV as well as in the border
regime of the enhanced estimate with $T_{\text{\tiny{max}}} \approx
140 - 230$ MeV. Hence they are subject to cutoff effects. Moreover, we
already know from QCD flows at vanishing temperature that the present
local approximation with scale-dependent wave function
renormalisations without full momentum dependence overestimates the
anomalous dimensions in the scale regime slightly above the chiral
scale, see \cite{Mitter:2014wpa} and \cite{Braun:2014ata} for a
comparison of the scale and momentum-dependence. A related comparison
concerning the scale dependent running strong coupling with and
without momentum dependence can also be found in
\cite{Mueller:2015fka}, see Fig.~12. This calls for a refined analysis
with momentum and frequency dependent propagators which will be
considered elsewhere.

Finally, we remark that the $\mathrm{LPA}^{\prime}$ and full
approximation leads to a very small negative pressure for small
temperatures $T\lesssim 80$ MeV, see the inlay plot in
Fig.~\ref{fig:pre1}. Its origin can be traced back to the anomalous
dimension of the qaurk, and hints at a -minor- overestimation of
thermal flucutation effects of the quark propagation for constant
anomalous dimensions.  We expect this effect to disappear once the
full frequency and momentum dependence of the propagator is taken into
account. In any case, it is a very minor effect, as shown in
Fig.~\ref{fig:pre1}, and only exists the low-temperature regime far
below the pseudo-critical temperature.

\subsection{Baryon number fluctuations}
\label{sec:barfluc}

The baryon number fluctuations are usually described by the
generalized susceptibilities, 
\begin{equation}\label{eq:suscept}
  \chi_n^{\mathrm{B}}=\frac{\partial^n}{\partial (\mu_{\mathrm{B}}/T)^n}\frac{p}{T^4}\,,
\end{equation}
with baryon number chemical potential
$\mu_{\mathrm{B}}=3\mu$. $\chi_n^{\mathrm{B}}$'s are related with the
cumulants of baryon multiplicity distributions. Denoting the ensemble
average of the baryon number with $\langle N_{\mathrm{B}}\rangle$, the
quadratic and quartic fluctuations are given by
\begin{align}\nonumber 
  \chi_2^{\mathrm{B}}&=\frac{1}{VT^3}\langle{\delta N_{\mathrm{B}}}^2\rangle\,,\\
  \chi_4^{\mathrm{B}}&=\frac{1}{VT^3}\Big(\langle{\delta
    N_{\mathrm{B}}}^4\rangle-3\langle{\delta
    N_{\mathrm{B}}}^2\rangle^2\Big)\,,
\end{align}
where $V$ is the volume of the system, and $\delta N_{\mathrm{B}}:=
N_{\mathrm{B}}-\langle N_{\mathrm{B}}\rangle$. In fact,
$\langle{\delta N_{\mathrm{B}}}^2\rangle$ is just the width of the
Gaussian distribution, denoted specifically by $\sigma^2$. The
kurtosis of non-Gaussian distributions $\kappa$ (the skewness is
vanishing at $\mu_{\mathrm{B}}=0$) is defined by
\begin{equation}
\kappa\sigma^2=\frac{\chi_4^{\mathrm{B}}}{\chi_2^{\mathrm{B}}}\,.
\end{equation}
The quadratic baryon number fluctuation $\chi_2^{\mathrm{B}}$ as a
function of $T$ with different truncations is presented in left panel of
Fig.~\ref{fig:chi2B}. As expected, $\chi_2^{\mathrm{B}}$ in LPA and
$\mathrm{LPA}+h_{k}$ signal a broad crossover, while in the other two
truncations show a steeper transition. In general, the $\chi_2^{\mathrm{B}}$ 
behave very similar to $s/s_{\mathrm{SB}}$ shown in Fig.~\ref{fig:ps}. 

So far, we have compared the pressure and its low-order derivatives
with respect to the temperature ($s$: first order) and chemical
potential ($\chi_2^{\mathrm{B}}$: second order) within different
truncations. All the comparisons indicate that the nontrivial momentum
dependence of propagators enhances the transition strength of the
chiral crossover. This effect should be even more pronounced in
higher-order thermodynamical quantities, such as
$\chi_4^{\mathrm{B}}$. This expectation is confirmed in
Fig.~\ref{fig:chi2B}, right panel and Fig.~\ref{fig:R42QM}, where
$\chi_4^{\mathrm{B}}$ and the kurtosis $\kappa\sigma^2$ are
shown. Both $\chi_4^{\mathrm{B}}$ and $\kappa\sigma^2$, calculated in
the full and $\mathrm{LPA}^{\prime}$ approximations, vary rapidly
during the crossover. In contradistinction, those obtained from LPA
and $\mathrm{LPA}+h_{k}$ as functions of $T$ are much smoother. Note
that the difference is quite large in the quartic fluctuations, in
contrast to that observed in low-order thermodynamical quantities.

In light of an experimental observation of (critical) fluctuations
this is good news. It has been shown previously, that effects of
mesonic multi-scattering in the medium, as included in the FRG
approach in a LPA truncation, soften the chiral crossover and suppress
thermal fluctuations compared with the mean-field approximation, see
e.g.\ \cite{Skokov:2010wb} for the kurtosis. This limits the chances
of an experimental observation of critical fluctuations and hence the
critical point. However, our present calculations reveal that the
nontrivial quantum and thermal dispersions of quarks and mesons have
the opposite effect: they increase the transition strength of the
crossover and enhances thermal fluctuations. While this effect was
already visible in \cite{Pawlowski:2014zaa,Helmboldt:2014iya} for the
order parameter, it is particularly obvious for high-order thermal
fluctuations, such as non-Gaussian fluctuations of the baryon number,
see Fig.~\ref{fig:R42QM}.

\section{QCD-enhanced Polyakov--quark-meson model}
\label{sec:PQM}

The quark-meson model studied so far is a good and clean testing
ground for studying the relative strength of flcutuation effects such
as multi-meson scattering and non-trivial dispersions. However, it
does not include the confinement dynamics that is necessary to obtain
the correct physics of baryon number fluctuations in the hadronic regime 
at temperatures below $T_c$. 

As already argued in Section~\ref{sec:QCDlow} the quark-meson model can
be embedded in low energy/low temperature QCD by adding a background
gluonic potential, see \eq{eq:Gasplit}. This directly leads to the
usual Polyakov-loop--extended chiral models, such as the
Polyakov--Nambu--Jona-Lasinio
model~\cite{Fukushima:2003fw,Ratti:2005jh,Fu:2007xc} and the
Polyakov--quark-meson (PQM) model~\cite{Schaefer:2007pw}. As explained
in Section~\ref{sec:QCDlow}, at sufficiently low scales the gluon
fluctuations decouple due to the QCD mass gap, and the gluon field can
be treated as a temporal background field. Its contribution to the
thermodynamics, however, is prominent, via the expectation value of
the traced Polyakov loop, viz.
\begin{align}\label{eq:Lloop}
  L(\vec{x})=\0{1}{N_c} \left\langle \Tr\, {\cal P}(\vec
    x)\right\rangle \,,\quad {\rm and }\quad  \bar L=\0{1}{N_c} \langle
  \Tr\,{\cal P}^{\dagger}(\vec x)\rangle \,,
\end{align}
with 
\begin{align}\label{eq:Ploop}
  {\cal P}(\vec x)= \mathcal{P}\exp\Big(ig\int_0^{\beta}d\tau
    A_0(\vec{x},\tau)\Big)\,.
\end{align}
The expectation value of the traced Polyakov loop, $L, \bar L$ serve as order
parameters for the deconfinement phase transition. 

In the presence of a gluonic background field the quark contribution
to the flow is changed. This is accounted for by simply replacing
the fermion distribution functions $n_F(x,T)=1/(\exp( x/T)+1)$ in
Appendix~\ref{app:threshold} with the Polyakov-loop modified ones,
i.e.\
\begin{align}\label{eq:nFL}
  n_F(x,T,L,\bar L)=\frac{1+2\bar L\,e^{x/T}+L\,
    e^{2x/T}}{1+3\bar L\,e^{x/T}+3L\, e^{2x/T}+e^{3x/T}}\,. 
\end{align}
In \eq{eq:nFL} we have introduced the notation  
\begin{align}\nonumber 
x&=\frac{k}{z_{q}}(1+\bar{m}_{q,k}^{2})^{1/2}-\mu\,,\\[2ex] 
\bar x&=\frac{k}{z_{q}}(1+\bar{m}_{q,k}^{2})^{1/2}+\mu\,,
\label{eq:xbarx}
\end{align}
with $z_{q}=1$, see Appendix~\ref{app:threshold}. The distribution
function of the anti-quark, $n_F(\bar x ,T)$ changes to
\begin{align}\label{eq:barnFL}
n_F(\bar x ,T)\to n_F(\bar x,T,\bar L,L)\,. 
\end{align}
In this work we present results in two different
approximations. In the first approximation we aim at exploiting the
existing results of the Polyakov loop expectation value, e.g.\ from
the lattice or continuum QCD computations. Then the Polyakov loop
expectation value serve as a background in which the flow of the matter
fluctuations is computed. Furthermore, $\bar L=L$ is assumed and their
dependence on the small chemical potential is neglected. Here we simply test
the consistency of such an approach as well as evaluating the
sensitive input parameters such as the field derivatives of the
Polyakov loop potential. The respective results in this approach are
collected in Appendix~\ref{app:sPQM}.

The second approximation utilises a given Polyakov loop potential, and
the equations of motion for $L$ and $\bar L$ are solved together with
that of the $\sigma$-meson. The matter fluctuations are computed on
this self-consistent background, leading to a fully coupled
system. This approach is numerically more challenging, in particular
if it comes to extensions of the present approacimation. However, it
has the advantage of self-consistency which turns out to be of eminent
importance for the computation of baryonic fluctuations, or higher
moments in general.

\subsection{Glue potential and higher moments}
\label{sec:PQMglue}

Now we introduce the simple polynomial Polyakov loop model potential
$\mathcal{U}(L,\bar{L})$, see \cite{Ratti:2005jh}. We shall also
discuss the model parameters relevant for a reliable extraction of
higher moments in QCD. A more thorough investigation of the model
dependence of the results will be considered elsewhere. The
dimensionless Polyakov-loop potential $ V_{\text{\tiny{YM}}}[L,\bar
L;t]$ is obtained by a rescaling of $\mathcal{U}(L,\bar{L})$ with
$1/T^4$. It can be expressed in terms of the dimensionless reduced
temperature $t=(T-T_{\mathrm{cr}})/T_{\mathrm{cr}}$ and reads
\begin{align}
  V_{\text{\tiny{YM}}}(L,\bar L;t)=& 
  \frac{\mathcal{U}(L,\bar{L})}{T^4}\nonumber\\[2ex] 
=& -\frac{b_2(t)}{2}L\bar{L}
  -\frac{b_3}{6}(L^3+\bar{L}^3)
  +\frac{b_4}{4}(L\bar{L})^2\,.\label{eq:YMpot}
\end{align}
Only the coefficient $b_2(t)$ of the
quadratic term depends on temperature. For large temperatures we have
$b_2>0$, while for small ones we have $b_2<0$. Its temperature
dependence above $T_c$ is fixed by the Yang-Mills expectation value of
the Polyakov loop computed on the lattice and the Yang-Mills
pressure. It has a divergence for $t\to -1$, that is for vanishing
temperature $T\to0$.
\begin{equation}\label{eq:b2}
b_2(t)=a_0+\frac{a_1}{1+t}+\frac{a_2}{(1+t)^2}+\frac{a_3}{(1+t)^3}\,,
\end{equation}
with the parameters $a_0=6.75$, $a_1=-1.95$, $a_2=2.625$, $a_3=-7.44$,
$b_3=0.75$, $b_4=7.5$. Our computations here are based on a
QCD-enhanced glue potential, that has the correct thermal dependence
on the reduces temperature of QCD instead that of Yang-Mills theory.
It has been shown in
\cite{Pawlowski:2010ht,Haas:2013qwp,Herbst:2013ufa} that in the
vicinity of the phase transition the thermal QCD scaling is obtained
from that in Yang-Mills theory in (\ref{eq:YMpot}) through a simple
linear rescaling of $t$, 
\begin{equation}
t_{\mathrm{YM}}(t_{\mathrm{glue}})\approx 0.57\, t_{\mathrm{glue}}\,.
\end{equation}
Thus, the QCD-enhanced glue potential reads
\begin{equation}\label{eq:gluepot}
  V_{\mathrm{glue}}(L,\bar{L};t_{\mathrm{glue}})=
V_{\text{\tiny{YM}}}(L,\bar L;0.57\, t_{\mathrm{glue}})\,,
\end{equation}
with
$t_{\mathrm{glue}}=(T-T_{\mathrm{cr}}^{\mathrm{glue}})/T_{\mathrm{cr}}^{\mathrm{glue}}$
and $T_{\mathrm{cr}}^{\mathrm{glue}}\approx 200 -210\,\mathrm{MeV}$
for two flavours~\cite{Herbst:2013ufa}. Here we choose
$208\,\mathrm{MeV}$ as already used in \cite{Schaefer:2007pw}.

The QCD enhancement of the potential provides the correct temperature
scaling of the glue potential in QCD. QCD-enhanced computations of
thermodynamical quantities, i.e. pressure and trace anomaly, agree
remarkably well with recent results from lattice QCD for the $N_f=2+1$
flavour case already in LPA, see \cite{Herbst:2013ufa}. Moreover,
despite the absence of the strange quark fluctuations, already the
$N_f=2$ flavour thermodynamical observables are close to that in the
$N_f=2+1$ case, after the overall temperature scale is normalised, for
a comparison of $N_f=2+1$ to $N_f=2+1+1$ see
\cite{Fischer:2014ata}. This supports the interpretation that the
strange quark (and even more so the charm quark for that matter)
contribute to the UV fluctuations above their mass scale and hence
predominantly only change the overall dynamical scale
$\Lambda_{\text{\tiny{QCD}}}$. In turn, at low momentum and
temperature scales the strange quark decouples and its fluctuations
only leads to quantitative modifications of the thermodynamical
observables. In the present work we shall test this interpretation
also for the baryonic fluctuations.

\begin{figure}[t]
\includegraphics[scale=0.6]{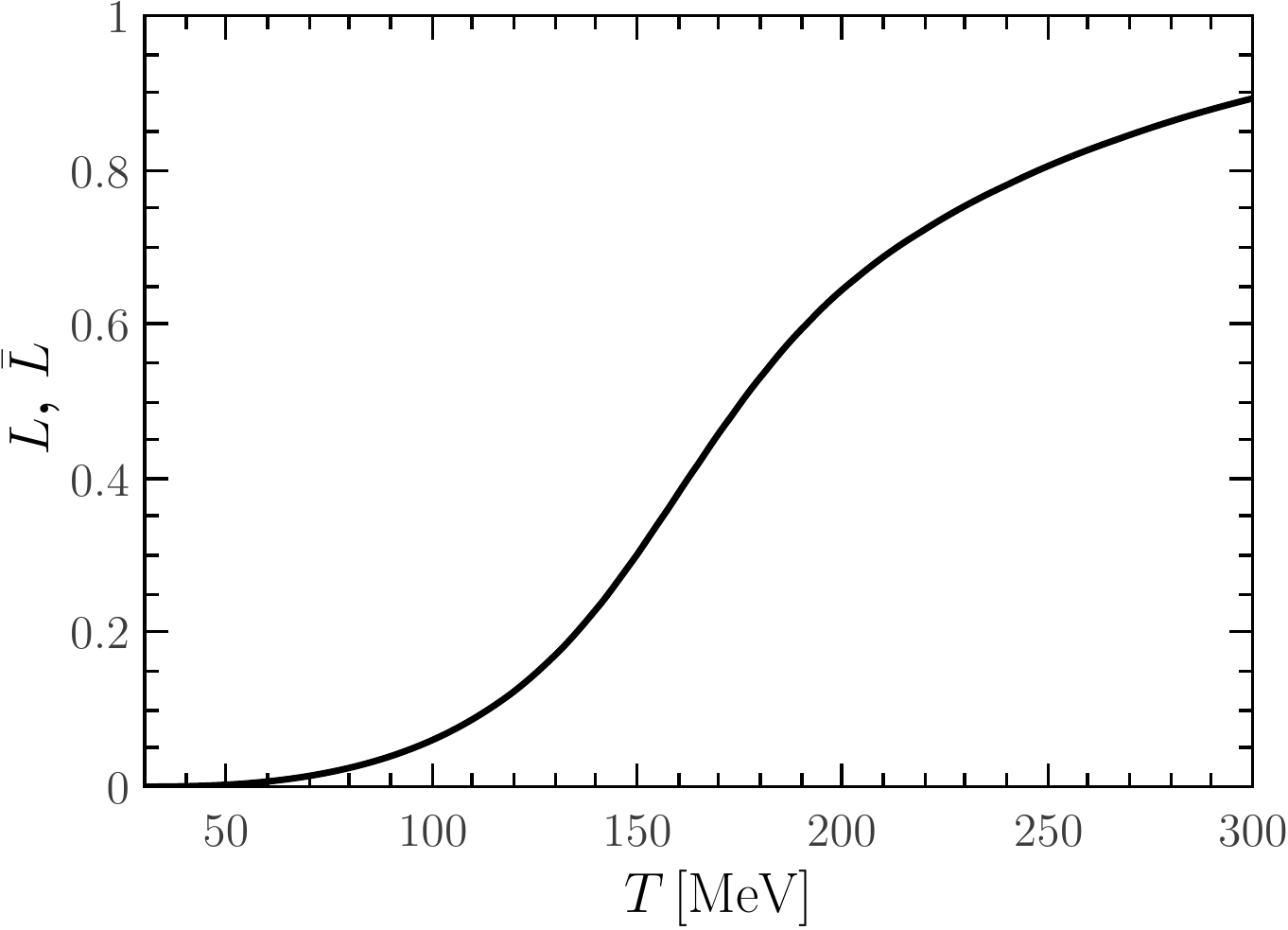}
\caption{Polyakov loop as a function of $T$. The result is obtained by employing the
  QCD-enhanced glue potential
  in~\cite{Herbst:2013ufa}.}\label{fig:PolyaLoop}
\end{figure}
The respective Polyakov loop expectation values within the present
self-consistent two flavour computation are depicted in
Fig.~\ref{fig:PolyaLoop} as a function of $T$. We emphasise again that
the potential itself is a model potential that reproduces the pure glue
pressure and Polyakov loop expectation values. It may fall short of
reproducing higher Polyakov loop correlation functions. The latter
information can e.g.\ be introduced to the models by computing higher
order correlation functions or Polyakov loop fluctuations, see
\cite{Lo:2013etb,Lo:2013hla}.  A full resolution of this task is given
by either computing the full glue potential from functional methods,
\cite{Braun:2007bx,Braun:2009gm,Fister:2013bh,Fischer:2013eca}, or by
computing the full potential on the lattice
\cite{Greensite:2012dy,Langfeld:2013xbf,Smith:2013msa,%
Diakonov:2013lja,Greensite:2014isa}.
The full QCD-embedding with the computation of the glue potential with
the FRG will be considered elsewhere.

In any case it is of chief interest to evaluate the sensitivity of the
results on the higher moments of the baryon multiplicity distribution
on the higher order $L,\bar L$-derivatives of the glue potential. We
will see in the next subsection that the absolute values of the higher
moments $\chi_n^{\mathrm{B}}$, in asymptotic $T/T_c$-regime, crucially
depend on parameters of the glue-potential
$\Omega_{\text{\tiny{glue}}}$. Moreover, the transition regime depends
on the relative strength of the couplings in
$\Omega_{\text{\tiny{glue}}}$ and $\Omega_{\text{\tiny{matt}}}$. This
emphasises the necessity to compute and utilise quantitative versions
of the glue potential, as well as a detailed study of the
input-dependence. This is beyond the scope of the present work and
shall be discussed elsewhere, \cite{FHPR}. 

Here we put forward an analytical method for the computation of higher
moments that clearly demonstrate the relation between
$\chi_n^{\mathrm{B}}$ and the $L,\bar L$ expansion coefficients
(higher moments) of the glue potential. The higher moments
$\chi_n^{\mathrm{B}}$ are proportional to total $\mu$-derivatives of
$\Omega$. Hence, it requires the computation of $\partial_\mu L$ and
$\partial_\mu \bar L$ from the correlation functions, that are the
field derivatives of $\Omega$ at vanishing $\mu$. This rather
technical computation is deferred to Appendix~\ref{app:flucanalytic},
here we only briefly describe the main results. 

Total $\mu$-derivatives can be written as partial $\mu$-derivatives at
fixed fields and field-derivatives multiplied with $\partial_\mu
\Phi_i$ with $\Phi=(L,\bar L, q,\bar q,\sigma,\vec \pi)$, see
\eq{eq:dmushort} in Appendix~\ref{app:flucanalytic},
\begin{align}
  \0{d}{d\mu} =\left.\partial_\mu\right|_{\Phi} - \Omega_{\mu i}
  G_{ij}\0{\delta}{\delta \Phi_j} \,.
\label{eq:dmushortmain}
\end{align}
Here $G_{ij}$ stands for the propagator, see \eq{eq:shortnot} and
$\Omega_{\mu i}$ stands for a mixed partial $\mu$ and
$\Phi_i$-derivative, see \eq{eq:shortnot}. The indices $i,j$ sum over
species of fields, momentum or space-time indices as well as internal
indices. Within this condensed notation, the second moment takes the
form \eq{eq:dmu2Omegafin},
\begin{align}
  \0{d^2\Omega[\Phi_{\text{\tiny EoM}}]}{d\mu^2} = & 
  \Omega_{\mu^2}- \Omega_{\mu i} G_{ij} \Omega_{\mu j} \,.
\label{eq:dmu2Omegafinmain}
\end{align}
In the low temperature limit both terms, $\Omega_{\mu^2}$ as well as
that proportional to $\Omega_{\mu i}$ contribute. We re-order
\eq{eq:dmu2Omegafinmain} and concentrate on
$\chi_2^{\mathrm{B}}$. Note also that in the present approximation to
the full theory we have $\Omega_{\mu\pi}=\Omega_{\mu q} =0$. This
leads us to
\begin{align}\label{eq:Lsigmasplit2}
  \chi_2^{\mathrm{B}}\simeq & \019 \left( \Omega_{\mu^2} - \Omega_{\mu
      \sigma}^2 G_{\sigma\sigma} \right)-\019 \Delta\Omega_2 \,,
\end{align}
with 
\begin{align}\nonumber 
  \Delta\Omega_2 =& 2 \left( \Omega_{\mu \sigma} \Omega_{\mu
      L}G_{\sigma L} +\Omega_{\mu \sigma}\Omega_{\mu \bar L} G_{\sigma
      \bar L}\right) \\[2ex]  & +\Omega_{\mu L}^2 G_{L L} +
  \Omega_{\mu \bar L}^2 G_{\bar L \bar L} +2 \Omega_{\mu
    L}\Omega_{\mu\bar L}G_{L \bar L}\,.
\label{eq:DeltaOmega}\end{align}
The term $\Delta\Omega_2$ carries the direct sensitivity to the model
input, the choice of the Polyakov loop potential
$\Omega_{\text{\tiny{glue}}}=T^4V_{\text{\tiny{glue}}}$. Most
importantly this is reflected in the propagators $G_{LL}, G_{\bar
  L\bar L}, G_{L\bar L}$ that depend in leading order on
$\Omega_{\text{\tiny{glue}}}$. Note that also the other propagators
depend on $\Omega_{\text{\tiny{glue}}}$ as $\Omega^{(2)}_{ij}$ is not
diagonal, but these dependencies are sub-leading. In summary,
\eq{eq:DeltaOmega} entails that for an accurate determination of the
second moment $\chi_2^{\mathrm{B}}$ the glue potential used in model
computations should reproduce the (connected) two-point correlation
functions $G_{LL}=\langle L L\rangle_c $, $G_{L\bar L}=\langle L\bar
L\rangle _c$, and $G_{\bar L\bar L}=\langle \bar L\bar L\rangle_c$ in
QCD. Alternatively the subdominance of these contributions has to be
shown.

The equation for $\chi_4^{\mathrm{B}}$ or rather $d^4/d\mu^4 \Omega$
in terms of partial $\mu$- and $\Phi_i$-derivatives is given in
Appendix~\ref{app:flucanalytic} in \eq{eq:dmu4Omegashort}. Similarily
to \eq{eq:Lsigmasplit2} it can be split in all terms without
$L$-derivatives of $\Omega$ and the rest, $\Delta \Omega_4$.  Apart
from the dependence on the two-point correlation functions $\Delta
\Omega_4$ also carries a dependence of the three- and four point
correlation function of the glue potential.

\subsection{Confinement and the hadronic phase}
\label{sec:confhad}
The analysis of the last section concerning the sensitivity of the
fluctuations observables on the details of the Polyakov loop potential
necessitates a detailed discussions of the mechanisms and scales
behind the asymptotic behaviour of higher moments in the asymptotic
regimes $T/T_c\to 0,\infty$. A particularly interesting regime,
however, is the crossover regime at about $T_c$, as the fluctuations
there may give indirect access to the question of the existence of a
critical endpoint in the phase diagram. It is therefore very important
to understand, the quantitative knowledge of which couplings is
required to make predictions there. It is also of
chief interest to pin down the relevant mechanisms in view of the
model input, that is not fully controlled. This gives us both, access
to the systematic error of the computation as well as the physics in
the transition regime. 

A particularly helpful observables within this discussion is the
kurtosis $\chi_4^{\mathrm{B}}/\chi_2^{\mathrm{B}}$. For $T/T_c\to 0$
it simply counts the hadronic degrees of freedom.  Hence we expect
$\chi_4^{\mathrm{B}}/\chi_2^{\mathrm{B}} \to 1$. In turn, for
$T/T_c\to \infty$ the kurtosis tends towards its perturbative
behaviour, $\chi_4^{\mathrm{B}}/\chi_2^{\mathrm{B}}
\to 2/3 \pi^2$.

While the high temperature dependence is easily understood in
terms of perturbation theory, for $T/T_c\to 0$ the common picture is the
following: for these temperatures the Polyakov loop expectation values
tend towards zero, $L,\bar L \to 0$.  Accordingly, the quark thermal
distribution $n_F(x,T,L,\bar L)$ in the presence of the confining
backround signaled by $L,\bar L\approx 0$ supposedly tends towards a
baryonic one. Indeed, taking $L,\bar L=0$ in \eq{eq:nFL} leads us to
\begin{align}\label{eq:qtoB}
\frac{1+2\bar L\,e^{x/T}+L\,
    e^{2x/T}}{1+3\bar L\,e^{x/T}+3L\, e^{2x/T}+e^{3x/T}} 
\stackrel{L,\,\bar L\to 0}{\longrightarrow} \frac{1}{1+e^{3x/T}}\,.
\end{align}
The right hand side of \eq{eq:qtoB} simply is a baryonic distribution
function. Hence, a vanishing Polyakov loop expectation value would
effectively lead to baryonic properties. At $\mu=0$ we deduce from
\eq{eq:xbarx},
\begin{align}\label{eq:xm}
x= \frac{k}{z_{q}}\sqrt{1+\bar{m}_{q,k}^{2}}\gtrsim m_{q,\text{\tiny con}} 
\quad {\rm with}\quad
m_{q,\text{\tiny con}}= \left. \0{\bar h \bar \sigma}{2z_q} \right|_{k=0}\,, 
\end{align}
is bounded from below by the sum of the temperature-dependent
constituent quark mass $m_{q,\text{\tiny con}}$at $k=0$. The $T\to 0$
limit in \eq{eq:qtoB} then entails, that $e^{3 x/T}\to\infty$ and $n_F
\to e^{-3 x/T}$. The quark part of the grand potential $\Omega$ is
proportional to a sum of $n_F(x,T,0,0)+n_F(\bar x,T,0,0)$ and the
$2n$th-order $\mu$-derivatives simply pull down $3^{2n}$, leading to the 
desired result of counting baryonic degrees of freedom.  Indeed this is
seen when directly using $L=\bar L=0$ in the model, see \Fig{fig:R42PQM} 
in Appendix \ref{app:flucanalytic}. 

%
\begin{figure}[t]
\centering
\includegraphics[width=8.5cm]{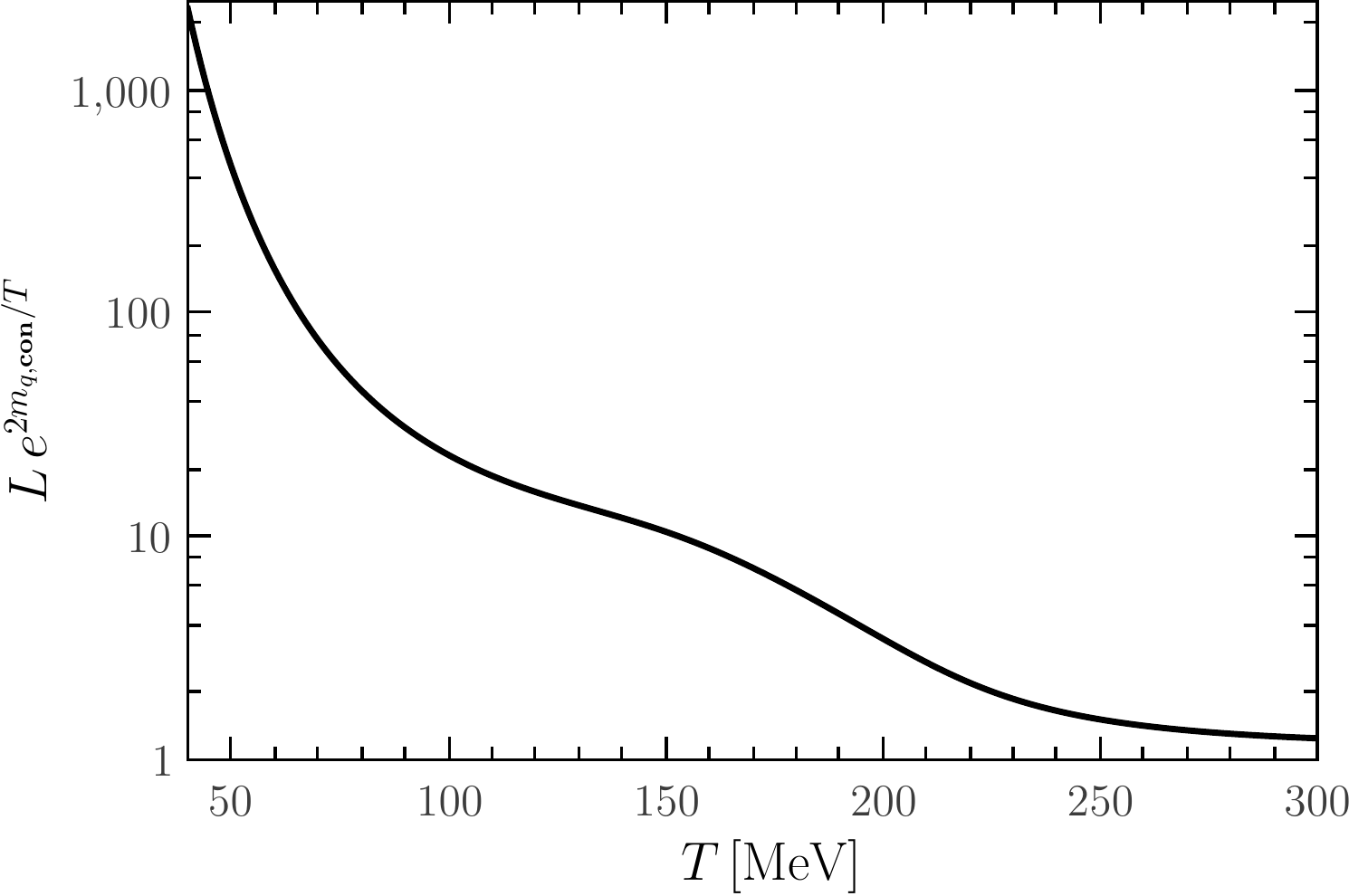}
\caption{$L\,e^{2 m_{q,\text{\tiny con}} /T}$ as a function of
  temperature, obtained in the full computation.}
\label{fig:failureL}
\end{figure} 
%
Note however, that the above picture involves taking $L,\bar L\to 0$
for $T\to 0$ and then taking $e^{x/T}$ to $\infty$. A close
inspection reveals that this explanation falls short: both limits
relate to $T\to 0$ and the thermal distribution in \eq{eq:qtoB}
depends on products of $L,\bar L$ and powers of $e^{x/T}$. Indeed, a
necessary ingredient for the validity of \eq{eq:qtoB}  is
\begin{align}\label{eq:Llim}
L  \,e^{2 m_{q,\text{\tiny con}} /T}\ll 1\,. 
\end{align}
We shall argue now, that \eq{eq:Llim} does not hold anywhere, see
\Fig{fig:failureL} for the PQM model under investigation here. We
emphasise that the following argument is done without explicit
reference to the specific model and also holds in QCD. To that end we
discuss the equations of motion,
\begin{align}\label{eq:EoML}
  \partial_L \Omega = \partial_L
  \Omega_{\text{\tiny{glue}}}+\partial_L \Omega_{\text{\tiny matt}}
  =0\,.
\end{align}
In a flow equation approach to QCD, $\Omega_{\text{\tiny{glue}}}$
comprises the contributions from the gluonic and ghost loop in
\Fig{fig:FRG-QCD_rebosonised}, while $\Omega_{\text{\tiny{matt}}}$
comprises that from the quark and mesonic loop in
\Fig{fig:FRG-QCD_rebosonised}. The representation in terms of a flow
equation facilitates the discussion of the respective scales. Dropping
$\Omega_{\text{\tiny matt}} $ the grand potential reduces to the pure
glue potential with a transition temperature $T_{\text{\tiny glue}}
\approx 200$\,MeV, see \cite{Pawlowski:2010ht,Haas:2013qwp,Herbst:2013ufa}. The glue
potential (in the current approximation) features a first order phase
transition, and below $T_{\text{\tiny glue}}$ the Polyakov loop
expectation vanishes. The possible minor center-breaking corrections
of the glue propagation due to the vacuum polarisation are neglected.
We shall see, that for the present argument we can safely drop these
contributions as they only increase the expectation value of the
Polyakov loop.

We conclude that the non-vanishing expectation values of $L$ and $\bar
L$ in QCD are dominantly triggered by the quark loop in
$\Omega_{\text{\tiny matt}}$. Starting at $T=0$, where the Polyakov
loop vanishes, we can estimate its $T$-dependence by computing the
thermal derivative. To that end we use the dimensionless potential 
\begin{align}\label{eq:VT} 
  V[\Phi_{\text{\tiny{EoM}}}] = \0{\Omega}{T^4} =
  V_{\text{\tiny{glue}}} + V_{\text{\tiny{matt}}}\,,
\end{align}
where we have extracted the canonical $T^4$ scaling. With \eq{eq:VT}
the $T$-derivative of \eq{eq:EoML} normalised with $1/T^4$ reads for
small $T/T_c\to 0$
\begin{align}
  \partial_T \partial_L V = V_{L L} \partial_T L +
  V_{L\bar L} \partial_T \bar L + V_{L\Phi_i}\partial_T \Phi_i+V_{TL} =0\,,
\label{eq:dTEoML}\end{align}
where $V_{TL}=\left.\partial_T\right|_{\Phi} V_L$ and index $i$ runs
over all fields without $L,\,\bar L$. At $\mu=0$ we have
$L=\bar L$, $\partial_T L=\partial_T\bar L$ and we are led to
\begin{align} 
\partial_T L =&-\0{ V_{L\Phi_i}\partial_T
    \Phi_i+V_{TL}}{V_{LL}+ V_{L\bar L}} 
\label{eq:dTL}
\end{align}
Now we explore the existing knowledge in QCD about the curvature
coefficients that contribute to \eq{eq:dTL} as well as the mixed
derivative $V_{TL}$: 

First of all, in the limit $T/T_c\to 0$ the glue potential
$V_{\text{\tiny{glue}},T}$ settles to a finite potential
$V_{\text{\tiny{glue}},T=0}$ with non-vanishing curvature coefficients
$V_{LL}$, $V_{L\bar L}$, $V_{\bar L\bar L}$, see e.g.\
\cite{Fister:2013bh}. Hence the denominator in \eq{eq:dTL} settles at
a finite temperature-independent value. It is left to provide a lower
bound for the temperature-dependence of the numerator. The
temperature-dependence of the expectation values of $\sigma$ decays at
least with the smallest mass scale in QCD, the pion mass. For example,
in the PQM model considered in the present work, the only
non-vanishing expectation value apart from $L,\bar L$ is that of the
sigma meson, $\sigma_0$. In turn, the mixed field derivatives, e.g.\
$V_{L\sigma}$ can only couple via the quark sector and hence decay
exponentially at least with $e^{- m_{q,\text{\tiny con}}/T}$.

It is left to give an estimate for the last term in the numerator,
$V_{TL}$. To that end we split the potential $V$ in a center-
symmetric part and a center-breaking one. The lowest term in $L,\bar
L$ in the center-symmetric part is proportional to $L \bar L$ and its
$L$-derivative is proportional to $\bar L$. In turn, the
center-breaking terms originate in the quark dynamics and their
contributions decay exponentially at least with $e^{- m_{q,\text{\tiny
      con}}/T}$, as is the case in the present model. In QCD the
fermionic loop in the model is substituted by a fully interacting one
and the constituent quark mass in the decay goes to half of the
free energy energy of an interacting quark--anti-quark pair. This leads
to the final estimate for the temperature-dependence of $L$ at
$\mu=0$,
\begin{align} 
 \partial_T L  
\gtrsim   & \max\left( c_{\text{\tiny
        q}} \,e^{- m_{q,\text{\tiny
        con}}/T}\,,\, c_{\text{\tiny
        L}}\, L\right) \,, 
\label{eq:dTLfin}
\end{align}
with prefactors $c_{\text{\tiny 1}}$ and $c_{\text{\tiny L}}$ that
carry a potential polynomial dependence on temperature. As the
Polyakov loop expectation value is vanishing at $T=0$ we deduce from
\eq{eq:dTLfin} that
\begin{align} 
  \lim_{T/T_c\to 0} L \to c(T)\, e^{-
    m_{q,\text{\tiny con}}/T}\,, 
\label{eq:TL}\end{align}
where the prefactor $c(T) $ carries a polynomial temperature
dependence. Consequently we do not reach \eq{eq:Llim} but rather have
\begin{align}\label{eq:Llimfin}
L  \,e^{2x/T}\gg 1\,, 
\end{align}
which concludes our argument. We emphasise again that we have not
resorted to any model properties but only have used the scaling
properties of the glue and matter parts, $V_{\text{\tiny
    glue}}=\Omega_{\text{\tiny glue}}/T^4$ and $V_{\text{\tiny
    matt}}=\Omega_{\text{\tiny matt}}/T^4$ respectively, of the grand
potential.

\subsection{Confinement and baryonic fluctuations}
\label{sec:confkurt}

\begin{figure}[t]
\centering
\includegraphics[width=8.5cm]{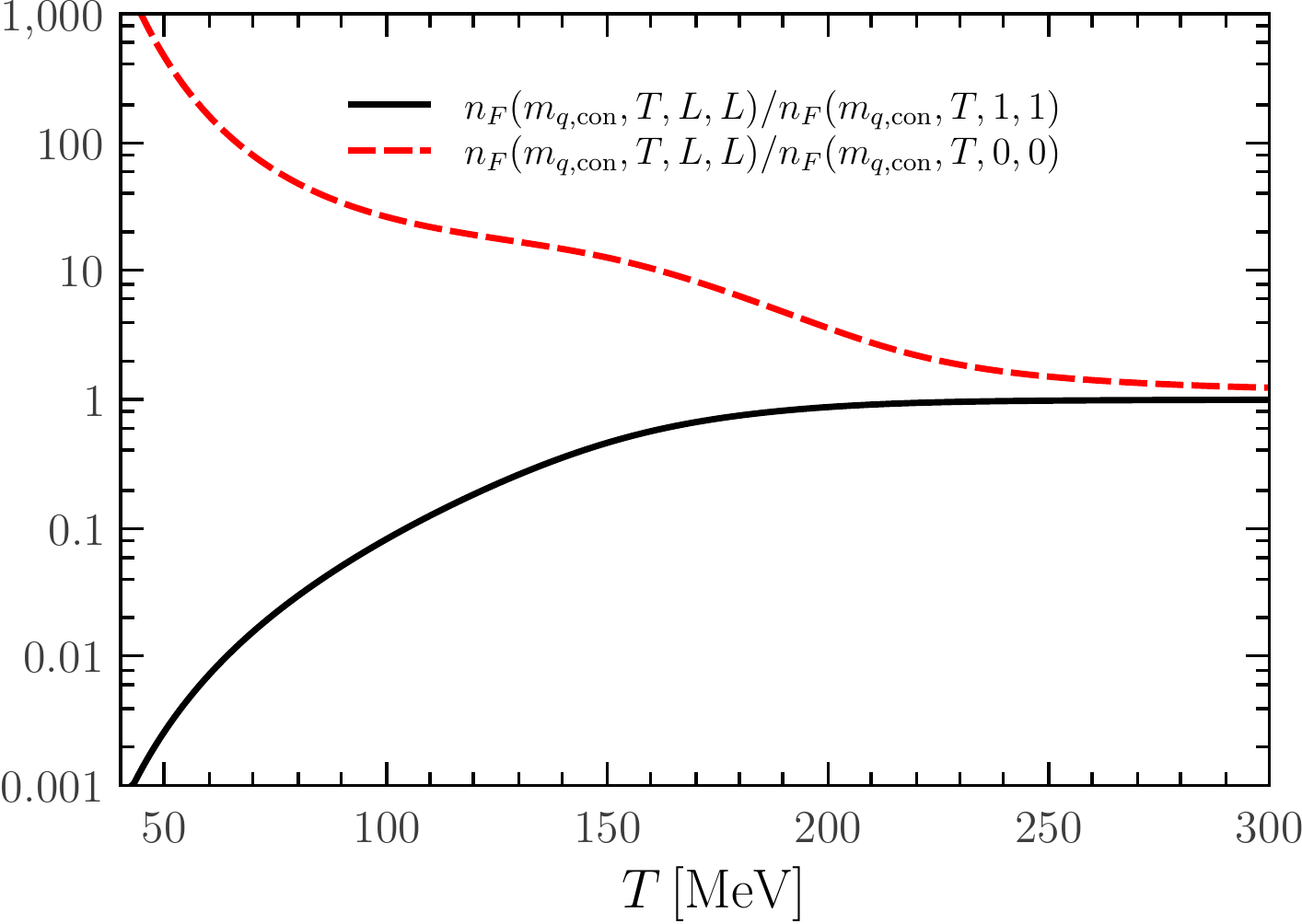}
\caption{Ratios of $n_F(m_{q,\text{\tiny
      con}},T,L,L)/n_F(m_{q,\text{\tiny con}},T,0,0)$ (full over
  baryonic distribution) and $n_F(m_{q,\text{\tiny
      con}},T,L,L)/n_F(m_{q,\text{\tiny con}},T,1,1)$ (full over quark
  distribution) as a function of temperature, obtained in the full
  computation.}
\label{fig:ratiosn}
\end{figure}

The analysis in the last section entails the failure of the standard
statistical confinement picture based on \eq{eq:qtoB}. Indeed, the
latter limit does not hold for any temperature and the Polyakov loop
augmented thermal quark distribution never comes even close the
baryonic one, see \Fig{fig:ratiosn}. 

Instead one can deduce, that the
hadronic limit is reached if \eq{eq:Llimfin} is satisfied. In the PQM
model used here this happens for $T/T_c\approx 0.8$, below which we
only expect to see the trivial thermal scaling. It is left to show
that in this limit we see the hadronic nature of the low temperature
phase, that is in particular
$\chi_4^{\mathrm{B}}/\chi_2^{\mathrm{B}}\to 1$, despite the the
thermal distribution for $T/T_c$ settling well in between the quark
and baryon distribution, see \Fig{fig:ratiosn}.
%
%
Indeed, at $\mu=0$ and $T/T_c\to 0$, and with \eq{eq:TL}, the thermal
distribution at $k=0$ is well approximated by
\begin{align}\label{eq:nFlim}
n_F(x,T,L,\bar L)\to L\, e^{-x/T}\approx e^{- 2
    m_{q,\text{\tiny con}}/T}\,, 
\end{align}
where $x$ in \eq{eq:xbarx}, \eq{eq:xm} is bounded by $m_{q,\text{\tiny
    con}}/T$ for all $k$. Hence the thermal distribution decays with a thermal
correlation length related to the inverse of twice the constituent
quark mass.  Note also that the quark contribution of (the flow of)
the grand potential $\Omega$ is directly proportional to $n_F$, see
\eq{eq:qOmega} in Appendix~\ref{app:threshold}.

With the scaling properties for $T/T_c\to 0$ deduced above we can
expand the grand potential in orders of $\exp\{ -m_{q,\text{\tiny
    con}}/T\}$ on the equations of motion. This leaves us with
\begin{align}\nonumber 
  \Omega =& - \0{b_2}{2}L\bar L -\0{b_3}{6}\left( L^3 +\bar L^3
  \right) +\0{c_1}{2}
  \Bigl[ L \left( e^{- x}+2 e^{-2\bar x}\right)\\[2ex]
  &+\bar L \left( e^{-\bar x}+2 e^{-2 x}\right) \Bigr]+
  \Omega_{\text{\tiny{mes}}}+O\left( e^{-4\0{m_{q,\text{\tiny
            con}}}{T}}\right)\,.
\label{eq:expandOmega}\end{align}
where the $b_2,b_3$ terms stem from the expansion of the glue
potential $\Omega_{\rm glue}$ in powers of $L,\bar L$, see
e.g.~\cite{Schaefer:2007pw}, and we have dropped the subleading term
$b_4/4(L\bar L)^4$. The $c_1$-term stems from the quark contribution
to the grand potential and introduces an explicit breaking of center
symmetry. The exponential decay factors such as $e^{-x}$ would carry
free energies of half of an interacting quark--anti-quark pair at
infinite distance. Here we simply neglect the binding energy and take
that of a free quark--anti-quark pair with constituent quark masses as
a lower bound. This is also the approximation underlying the current
model approximation. As we shall see, the precise values of the
parameters $b_2,b_3,c_1$ are irrelevant for the $T\to 0$ asymptotics
of the kurtosis. We also have dropped the mesonic terms as
subleading. In \eq{eq:expandOmega}, $T$ is absorbed into $x$ and $\bar
x$ for brevity, and $L,\bar L$ on the EoM read
\begin{align}\nonumber 
  L = \0{1}{ b_2}\left[ c_1 \left( e^{-\bar x} +2 e^{-2 x}\right) -  b_3
    \bar L^2\right]+ O(
  e^{-3\0{m_{q,\text{\tiny con}}}{T}})\,,  \\[2ex]
  \bar L = \0{1}{b_2}\left[ c_1 \left( e^{- x} +2 e^{-2 \bar
        x}\right) - b_3 L^2\right]+ O( e^{-3\0{m_{q,\text{\tiny
          con}}}{T}})\,.
\label{eq:LEoM}\end{align}
\Eq{eq:LEoM} makes explicit the expansion in powers of
$\exp\{-m_{q,\text{\tiny con}}/T\}$. All terms dropped in
\eq{eq:expandOmega} are at least of the order
$\exp\{-4m_{q,\text{\tiny con}}/T\}$. It also supports the nice
heuristic interpretation of the Polyakov loop expectation values
$L,\bar L$ being related to the free energy of a test 
quark/anti-quark: the leading order of $L,\bar L$ decay exponentially
with $\exp\{-(m_{q,\text{\tiny con}}\pm \mu)/T\}$ respectively. The
implicit $\bar L^2$ and $L^2$ terms encode terms of at least quadratic
order in $\exp\{-(m_{q,\text{\tiny con}}\pm \mu)/T\}$. \Eq{eq:LEoM}
also entails that at finite chemical potential $\bar L$ is enhanced
relative to $L$. Inserting the respective equations for $L,\bar L$ on
the right hand sides of \eq{eq:LEoM} leads us to
\begin{align}\nonumber 
  L = \0{c_1}{ b_2}   e^{-\bar x} + \left( 2 \0{c_1}{ b_2}  -b_3  \0{c_1^2}{ b_2^3} 
\right)  e^{-2 x}+ O\left(
  e^{-3\0{m_{q,\text{\tiny con}}}{T}}\right)\,,  \\[2ex]
  \bar L = \0{c_1}{ b_2}   e^{-x} + \left( 2 \0{c_1}{ b_2}  -b_3  \0{c_1^2}{ b_2^3} 
\right)  e^{-2 \bar x}+ O\left(
  e^{-3\0{m_{q,\text{\tiny con}}}{T}}\right)\,.
\label{eq:LEoMfin}\end{align}
\Eq{eq:LEoMfin} reveals the composite nature of the Polyakov
loop expectation values. For example, $L$ has an expansion in quarks
and anti-diquarks all of which carry the same $SU(N)$ representation.
Hence we expect $L\bar L$ to carry both mesonic ($q\bar q$) and
baryonic properties ($qqq$, $\bar q\bar q\bar q$). Indeed, it follows from 
\eq{eq:LEoMfin} that
\begin{align}\nonumber 
  L\bar L=& \0{c_1^2}{ b_2^2} e^{-2\0{m_{q,\text{\tiny con}}}{T}} \\[2ex]
  &+ \left( 2 \0{c^2_1}{ b^2_2} -b_3 \0{c_1^3}{ b_2^4} \right)
  \left[e^{-3 x}+e^{-3 \bar x}\right]+ O\left( e^{-4\0{m_{q,\text{\tiny
            con}}}{T}}\right)\,.
  \label{eq:LbarLbaryons}\end{align}
The first term in \eq{eq:LbarLbaryons} carries no baryon number and
heuristically relates to mesons, while the second one carries baryon
number, there are no diquark contributions. The occurance of the decay
with multiples of the constituent quark mass $e^{-2 x}$ for $q\bar q$
and $e^{-3 x}$ for $qqq$ relates to the fact that the Polyakov loop
terms describe pairs and triplets of quarks rather than the bound states. 

In case of the baryons this is related to the missing higher order
quark interactions in the current model that carry the interaction
energy of the hadronic states. This can be amended by taking into
account multi-quark interactions with or without dynamical
hadronisation. 

In case of the mesons the missing contributions are hidden in the
mesonic part of the grand potential. To see this more clearly we
proceed with the expansion of the full grand potential. Note first
that $\Omega$ in \eq{eq:expandOmega} only has linear and bilinear
terms in $L$ and $\bar L$ evaluated on their EoMs, \eq{eq:LEoM}, as
well as cubic terms. The latter cubic terms carry baryon number in
leading order. Using $(L\partial_L + \bar L\partial_{\bar L})
\Omega=0$ to express the linear terms with the bilinear and cubic ones
we arrive at
\begin{align}\nonumber 
  \Omega =& \0{b_2}{2} \bar L L +\0{b_3}{3}\left( L^3 +\bar L^3
  \right) + \Omega_{\text{\tiny{mes}}}+O\left(
    e^{-4\0{m_{q,\text{\tiny con}}}{T}}\right)\\[2ex]=&
  \Omega_{\tiny{qqq}}+\Omega_{\tiny{q\bar
        q}}+\Omega_{\text{\tiny{mes}}} \,, 
\label{eq:expandOmegafin}\end{align}
with 
\begin{align}\nonumber 
\Omega_{\tiny{qqq}}= & 
\left( \0{c^2_1}{ b_2}
    -\0{b_3}{6} \0{c_1^3}{ b_2^3} 
\right) \left(e^{-3 x}+e^{-3 \bar x}\right)+O\left(
    e^{-4\0{m_{q,\text{\tiny con}}}{T}}\right)\\[2ex]  
\Omega_{\tiny{q\bar
        q}} = & \0{c_1^2}{2 b_2}
  e^{-2\0{m_{q,\text{\tiny con}}}{T}} + O(
  e^{-4\0{m_{q,\text{\tiny con}}}{T}})\,.
\label{eq:expandOmegabarmes}\end{align}
Note that the EoM also imply that the linear terms have the form in
\eq{eq:LbarLbaryons} with additional contributions in the first line
in \eq{eq:expandOmegabarmes} steming from the $L^3+\bar
L^3$-term. With \eq{eq:expandOmegafin} the grand potential takes its
final baryonic-type form. It has a $\mu$-independent offset and
subleading contributions. The $\Omega_{\tiny{qqq}}$ term carries
baryon number with contributions from all three terms in the grand
potential. In terms of the Polyakov loop it has the form $L^3+\bar
L^3$.  The mechanism unraveled above only holds for $L\,\bar L\to 0$
which signals the hadronic phase.

Note also that the total mesonic contribution is given by
$\Omega_{q\bar q} +\Omega_{\text{\tiny{mes}}}$.  The latter part
dominates in the hadronic phase and shows the correct mesonic mass
scales. We expect a similar phenomenon for the baryonic part once the
baryonic states are included in the effective action. Note however,
that the baryons do not affect the dynamics of the theory due to their 
heavy masses which suppressed any baryonic off-shell fluctuations at all 
momentum scales. 

Now we are in the position to discuss the fate of the kurtosis deep in
the hadronic regime, that is for low temperatures $T/T_c \ll 1$. From
\eq{eq:expandOmegafin} we deduce, that even powers of total
$\mu$-derivatives of the grand potential are simply proportional to
the potential in the $T\to 0$ limit,
\begin{align}\label{eq:muOmegaT0}
  \0{d^{2n}\Omega}{d (\mu/T)^{2n}}= 3^{2n} \Omega_{\tiny{qqq}} +O(
  e^{-4\0{m_{q,\text{\tiny con}}}{T}})\,,
\end{align}
with $\Omega_{\tiny{qqq}}$ counts three-quark
states, the effective baryons. \Eq{eq:muOmegaT0} implies in particular
\begin{align}\label{eq:KurtosisT0}
  \0{\chi_4^{\mathrm{B}}}{\chi_2^{\mathrm{B}}}= 1 +O(
  e^{-\0{m_{q,\text{\tiny con}}}{T}})\,,
\end{align}
In summary, despite the final outcome being the same, the mechanism
behind the asymptotic $T$-scaling of the kurtosis and also the higher
moments is different to that suggested in the literature: in
contradistinction to former claims the asymptotic low energy scaling
is obtained in a regime where the thermal distribution are far from
being baryonic, see
Figs.~\ref{fig:failureL},\ref{fig:ratiosn}. Indeed, the Polyakov loop
enhanced thermal distribution is far from the baryonic one for any
temperature, as is clear from the above analysis.

\begin{figure}[t]
\includegraphics[scale=0.6]{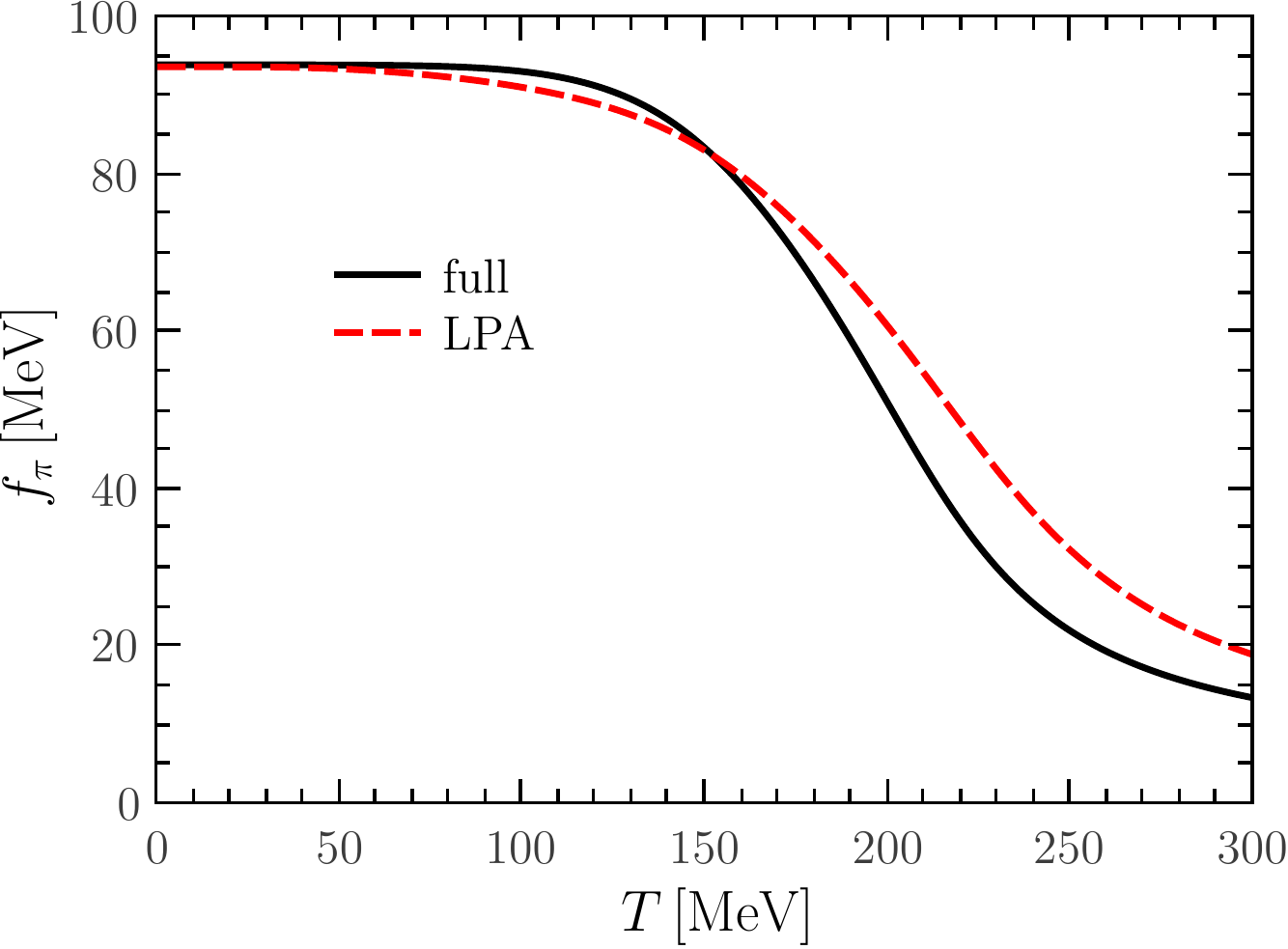}
\caption{Pion decay constant as a function of $T$,
  calculated in the PQM with LPA and full
  approximation.}\label{fig:fpiPQM}
\end{figure}
\subsection{Results}
\label{sec:Results}
\begin{figure*}[t]
\includegraphics[scale=0.6]{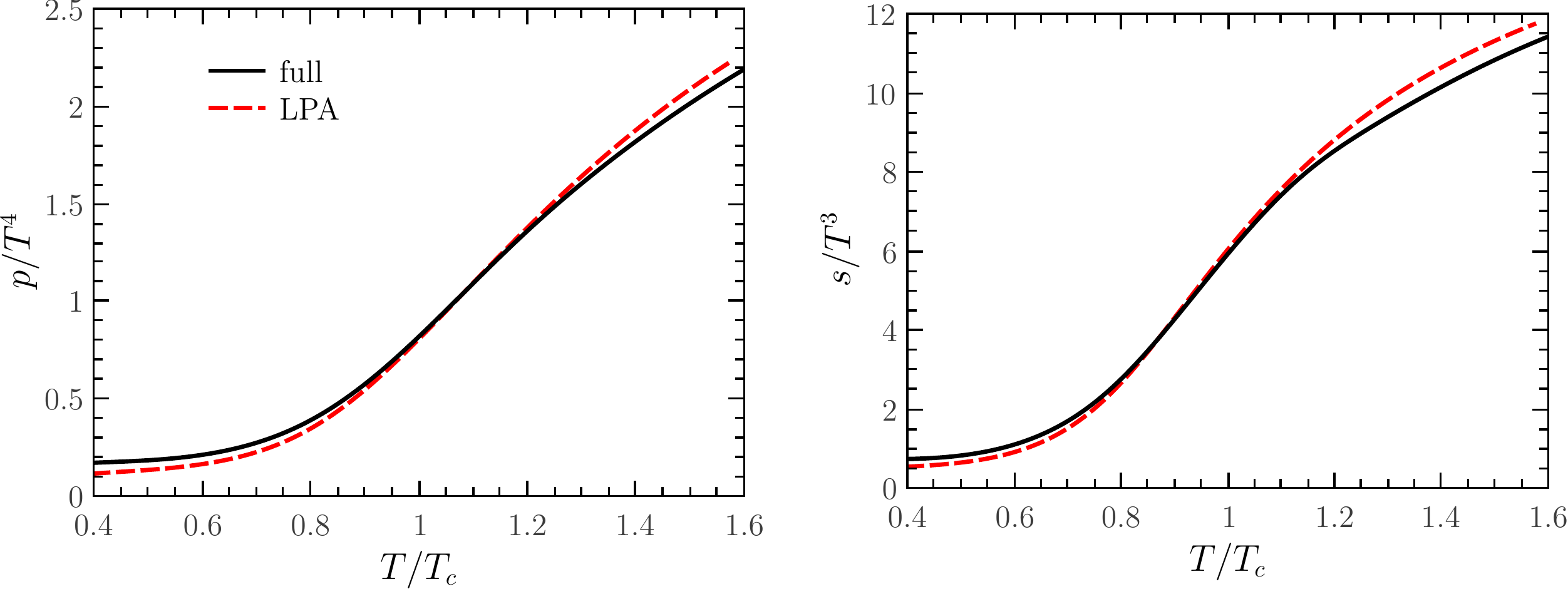}
\caption{Pressure (left panel) and entropy density
  (right panel) as functions of $T/T_c$. The psudo-critical
  temperature is determined by the peak of $|\partial
  \bar{\rho}_{0,k=0}/\partial T|$, $T_c= 181\,\mathrm{MeV}$,
  $190\,\mathrm{MeV}$ for full and LPA,
  respectively.}\label{fig:PQMthermo}
\end{figure*}
\begin{figure*}[t]
\includegraphics[scale=0.6]{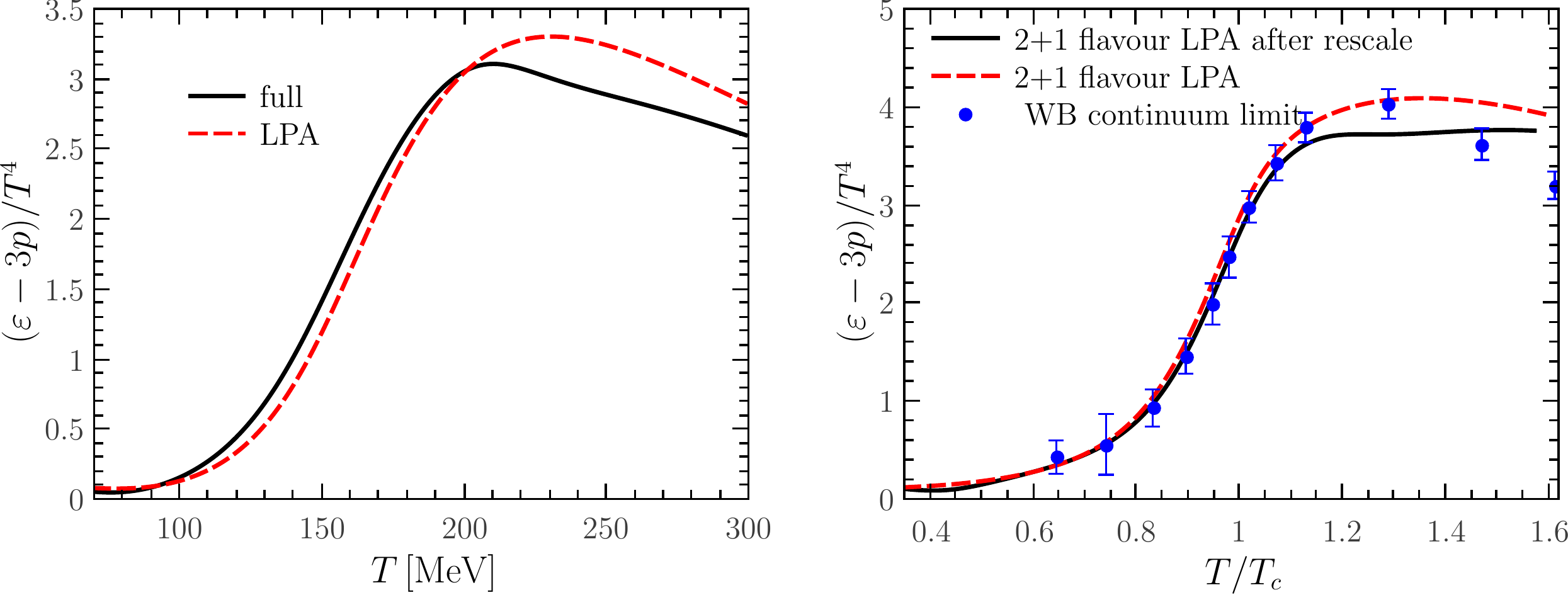}
\caption{Left panel: trace anomaly as a function of
  $T$. Right panel: trace anomaly for $N_f=2+1$ in LPA from
  \cite{Herbst:2013ufa} (red dashed line), multiplied by the ratio of
  the full $N_f=2$ data over the LPA data (black solid line), in
  comparison to lattice result from Wuppertal-Budapest collaboration
  \cite{Borsanyi:2010cj}.}\label{fig:traceano}
\end{figure*}
In the last section we have reached a better understanding of the
mechnisms taking place in the hadronic low energy/low temperature
regime of QCD. Most of the properties discussed do indeed not depend
on the model under investigation but are properties of
QCD. With this in mind we finally come to the numerical computations
of this paper for the QCD-enhanced $N_f=2$ PQM model. Before we
present the results, let us briefly summarise the approximations used
here: We build upon the approximation tested in Section
\ref{sec:QMmodel} for the quark-meson model. We use a full effective
meson potential $V_k(\phi;,L,\bar L)$ that includes multi-meson
scatterings. The dependence on the Polyakov loop variables $L,\bar L$
is implicit as the flow is evaluated on given backgrounds $L,\bar
L$. We also introduce scale-dependent wave function renormalisations
$Z_\phi(L,\bar L), Z_q(L,\bar L)$ with the same implicit dependence on
$L,\bar L$. The $Z$'s account for the change of the dispersions for
both mesons and quarks. Additionally we allow for scale-dependent
quark-meson interaction $h_k(L,\bar L)$. The quark-meson part in the
$L,\bar L$ background is amended with a QCD-enhanced Polyakov loop
potential $V_{\text{\tiny{glue}}}(L,\bar L)$ in
Eq.~(\ref{eq:gluepot}). The potential is a standard model potential
with the correct temperature scaling of QCD, see
\cite{Pawlowski:2010ht,Haas:2013qwp,Herbst:2013ufa}. Our full results
are compared with results in the standard LPA approximation as has
been used in \cite{Herbst:2013ufa} for $N_f=2$ and $N_f=2+1$ flavour
QCD.
\begin{figure*}[t]
\includegraphics[scale=0.6]{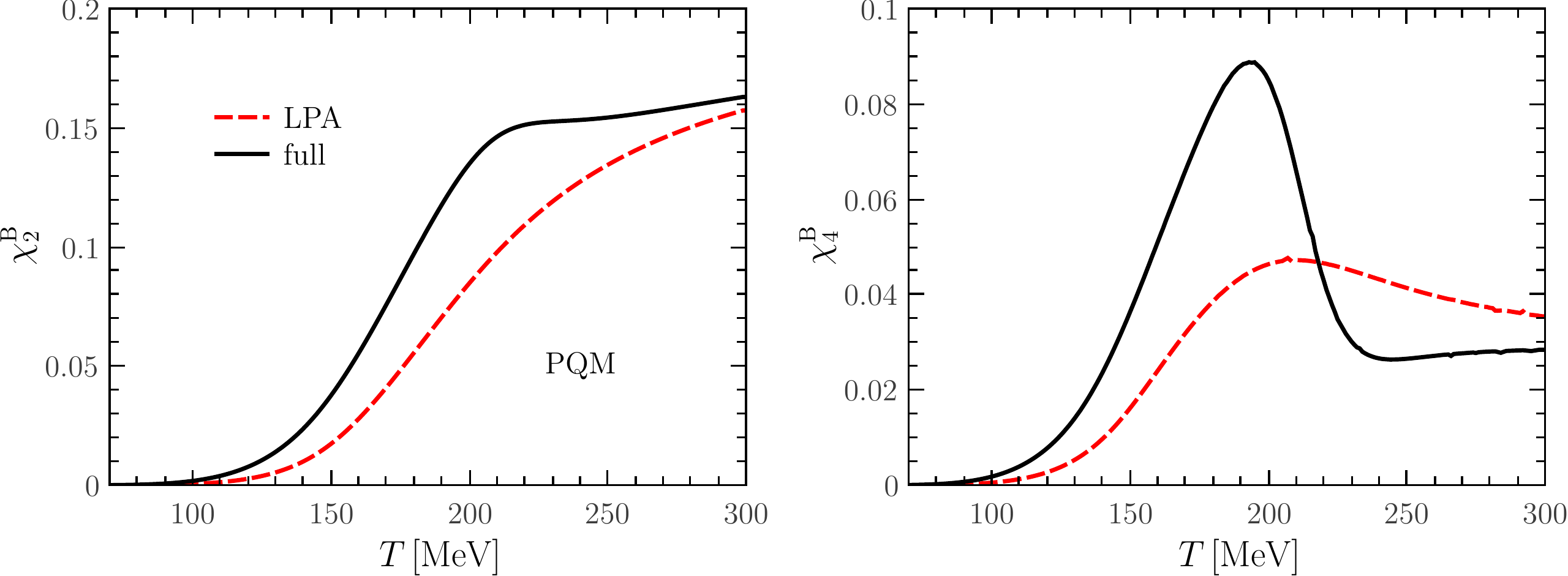}
\caption{Quadratic (left panel) and quartic (right
  panel) baryon number fluctuations as functions of $T$ in the PQM
  model in LPA and full truncation.}\label{fig:chi2PQMconsi}
\end{figure*}

In \Fig{fig:fpiPQM} we show the pion decay constant as a function of
temperature. As in the QM model the additional fluctuations from the
wave function renormalisation and the Yukawa coupling lead to a
sharper chiral crossover. In \cite{Herbst:2013ufa} it has been shown
that in the $2+1$ flavour case there is a very good agreement of both
pressure and trace anomaly as functions of reduced temperature for
$T/T_c \lesssim 1.2$ with the corresponding lattice results of the
Wuppertal Budapest group, see Figure 5 there. Note that the pressure
even agrees well above this scale. However, the trace anomaly, which
is more sensitive to the correct counting of fluctuations, shows a
widening gap. The LPA results overshoot the lattice results which
hints a both, an incorrect counting of fluctuations at large scale as
well as the impact of the ultraviolet cutoff.

As already mentioned before, the agreement between the $2+1$ and $2$
flavour results as functions of reduced temperature is still
remarkable for $T/T_c \lesssim 1$. This supports the point of view
that the dominant effect of the strange quark on the thermodynamics at
temperatures $T\lesssim T_c$ is to change the overall scale. For
larger momentum or temperature scales, however, its dynamics
contributes inevitably. Keeping this in mind we compare the LPA and
full pressure and entropy in \Fig{fig:PQMthermo} as functions of
reduced temperature. The additional fluctuations introduces via the
wave function renormalisations and the Yukawa couplings have little
impact on the thermodynamics. We expect this property to hold also for
2+1 flavours. Hence the already very good agreement of the pressure and
entropy in LPA is not spoiled. 

The trace anomaly, left panel of \Fig{fig:traceano}, also does not
change significantly for temperatures below $T_c$. Above $T_c$, the
full results undershoots the LPA result. Again we expect the $2+1$
result to show the same change. However, in this regime we are
sensitive to cutoff effects and missing fluctuations in the first place. 

We conclude that the additional quantum and thermal fluctuations
considered in the present work have very little impact on the
thermodynamics below $T_c$, despite their significant change of the
matter fluctuations as shown in the QM model. Above $T_c$ they have a
larger impact as can be seen from the trace anomaly, left panel of
\Fig{fig:traceano}. In a bold attempt to estimate the results of a
$2+1$-flavour computation we also show the $N_f=2+1$ data in LPA for
the trace anomaly from \cite{Herbst:2013ufa}, multiplied by the ratio
of the full $N_f=2$ data over the LPA data, see right panel of
\Fig{fig:traceano}. As expected, the quantitative agreement with the
lattice data is not spoiled for $T\lesssim T_c$. For temperatures
$T\gtrsim T_c$ one clearly sees the effects of the ultraviolet cutoff
and the missing fluctuations.

Finally we come to the baryonic fluctuations.The numerical results for
the second and fourth moment are shown in
Figure~\ref{fig:chi2PQMconsi}. We emphasise that the computation of
the $\mu$-derivatives of the grand potential at small temperatures
requires the careful evaluation of the genuine thermal and chemical
potential scaling of quark anomalous dimension and Yukawa coupling, for more details see
Appendix~\ref{app:Yukawa}. Interestingly, the
difference of the full results to that in LPA is significant in clear
contradistinction to the situation for the thermodynamics. This is
expected as the higher moments by definition carry the details of the
fluctuation physics in the model or theory at hand. From
Fig.~\ref{fig:chi2PQMconsi} one observes once more that the quadratic
and quartic baryon number fluctuations obtained with the full
approximation, are larger and change more rapidly during the chiral
crossover. Note that larger fluctuations are important in view of
experimental measurements, whose statistical error increase 
significantly for higher fluctuations.

\begin{figure*}[t]
\includegraphics[scale=0.6]{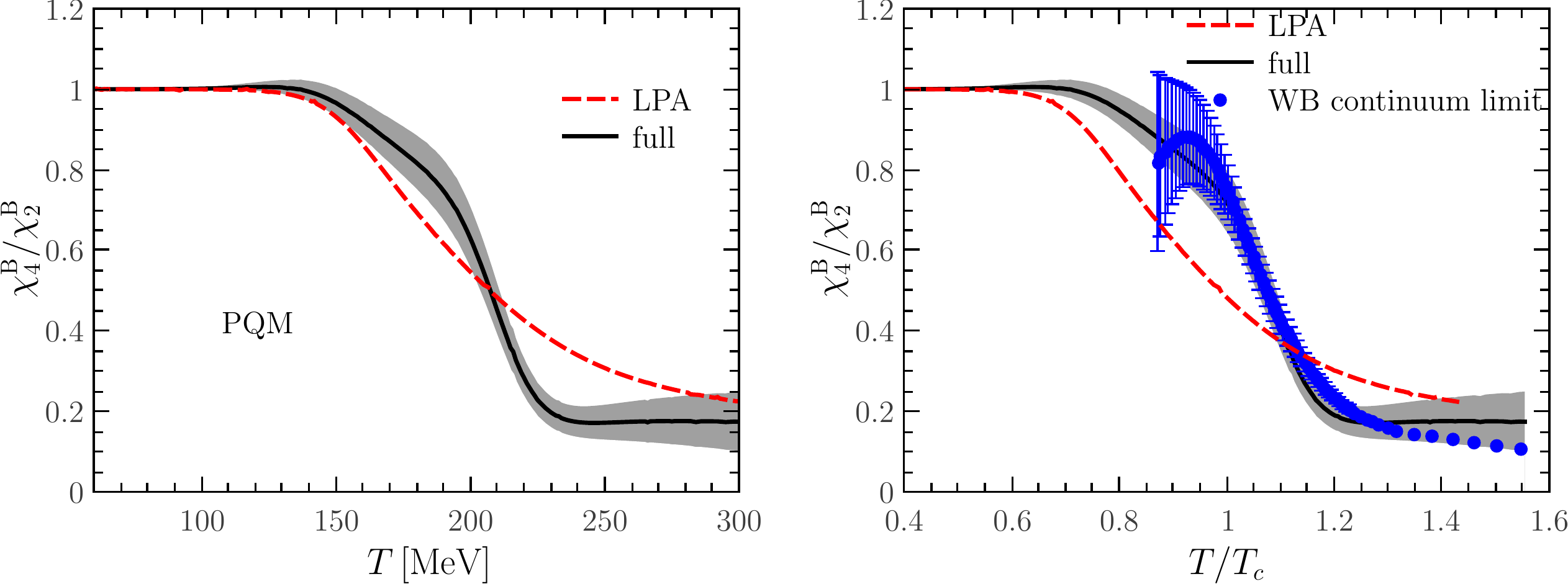}
\caption{Left panel: Kurtosis of baryon number distribution in the PQM
  model. The grey band shows an error estimate for the full
  computation, resulting from the UV-cutoff scale. Right panel:
  Comparison with the continuum-extrapolated lattice results from
  Wuppertal-Budapest collaboration \cite{Borsanyi:2013hza}.  The
  temperature is rescaled by their corresponding pseudo-critical
  temperature $T_c$.}\label{fig:R42PQMconsi}
\end{figure*}
In Fig.~\ref{fig:R42PQMconsi} we present the final results on the
kurtosis. As expected, the significant change from the LPA results to
the result of the full approximation propagates from that in the
second and fourth moments $\chi_{2}^{\rm B}, \chi_{4}^{\rm B}$. This is 
good news in terms of experimental measurements, and
provides a very good test for the convergence of the present
approximation scheme towards the full result. Such an analysis also
requires a discussion of the systematic error. The grey band in
Fig.~\ref{fig:R42PQMconsi} provides a rough estimate of the systematic
error of the full computation arising from the UV-cutoff scale
$\Lambda$, through the following simple formula \cite{Herbst:2013ufa}:
\begin{align}\label{eq:kurtosiserror}
  \0{\chi_4^{\mathrm{B}}}{\chi_2^{\mathrm{B}}}\pm \Delta
  \0{\chi_4^{\mathrm{B}}}{\chi_2^{\mathrm{B}}}
  =\0{\chi_4^{\mathrm{B}}}{\chi_2^{\mathrm{B}}}\Big(1\pm
  \0{4}{e^{\Lambda/T}-1}\Big)\,,
\end{align}
with $\Lambda=700\,\mathrm{MeV}$. Note that this error estimate does
not include the effects of the correct hadronic decoupling above $T_c$
which in the present framework is most easily implemented with
dynamical hadronisation. Moreover, at even higher scales gluonic
fluctuations have to be added. Finally, for temperatures $T\lesssim
0.8\, T_c$ we expect frequency-dependences in particular of the
fermionic couplings to further improve the reliability of the
results. This estimate is based on a computation of fermionic
couplings one the first fermionic Matsubara frequency $\pi T$ and the
improved one used in the present work, that takes into account the
correct thermal and $\mu$-scaling. The latter computation implements
effectively the effects of full frequency dependences, for details see
the Appendix~\ref{app:Yukawa}. The results start to disagree for
$T\lesssim 0.8\, T_c$ and we expect a compution with full
frequency-dependences to decrease the systematic error further in this
regime. A more detailed analysis in an upgraded approximation will be
discussed elsewhere.

The expected asymptotic value 1 for the kurtosis is obtained at low
temperature in both LPA and the full approximation. We also compare
the kurtosis of baryon multiplicity distribution with that of the
continuum-extrapolated lattice result from Wuppertal-Budapest
collaboration \cite{Borsanyi:2013hza}. The latter is a 2+1 quark
flavour result with physical quark masses. In the spirit of the
previous discussion about the dominant strange quark effect as a
different total scale setting in 2+1 flavour QCD in comparison to 2
flavour QCD we rescale the temperature in units of the pseudo-critical
temperature $T_c$. For the definition of the absolut temperature scale we
choose the peak position in the fourth moment $\chi_4^{\mathrm{B}}$,
as this is easily accessible in both approaches. From the right panel
of Fig.~\ref{fig:chi2PQMconsi}, one can easily read
$T_c=193\,\mathrm{MeV}$, $210\,\mathrm{MeV}$ for full and LPA,
respectively. For the 2+1 flavor lattice simulation,
$T_c=155\,\mathrm{MeV}$ is obtained from the results of
$\chi_4^{\mathrm{B}}$ \cite{Bellwied:2015lba}, which is also
consistent with the calculations in \cite{Borsanyi:2010bp}. With such
a normalisation one finds that the full calculation is in very good
agreement with the lattice result for temperature $T\lesssim T_c$,
while the LPA is not, see Fig.~\ref{fig:R42PQMconsi}. The most
significant difference is the rapid convergence of the kurtosis in the
full computation and the lattice result towards $1$. This entails that
the dominance of the hadronic nature for $T\leq T_c$ is approached
very quickly, which also has an impact on the rapid chemical
freeze-out that is observed experimentally. The present approach
allows to study the relative importance of glue, quark, mesonic and
baryonic flcutuations which will provide further insights in this
intriguing question.

\section{Summary and outlook}
\label{sec:summa}

In this work we have studied the phase structure,
thermodynamics, and in particular higher moments of baryon
multiplicity distribution, in low energy effective models with matter
and glue fluctuations. The fluctuations have been taken into account
with the functional renormalisation group. This has allowed us to
investigate in detail the impact of the matter and glue dynamics on baryon
number fluctuations and the kurtosis of baryon number distribution.

Furthermore, we have provided new insights on the mechanism of quark
confinement in terms of effective degrees of freedom: the common
picture in low energy effective models is that effective quark
confinement is realized through an effective baryonic thermal
distributions, called statistical confinement. A detailed study
unravels that the effective thermal distribution is quite far from the
baryonic one, and the deviation grows exponentially when $T/T_c$
approaches zero. Instead of an effective baryonic thermal distribution
we find that the center symmetry of the glue potential plays a key
role, which guarantees that the degree of freedom is baryonic at low
temperature. 

For a detailed study of the r$\hat{\rm o}$le of fluctuations several
different approximations to the full low energy effective theory of 2
flavour QCD have been employed. These extended approximations have
been compared with the standard LPA truncation, that allows for a very
efficient inclusion of meson fluctuations and has been considered in the
literature. We find that the non-trivial frequency and momentum
dependence of the propagators strenghens the temperature-driven chiral
crossover. In another words, the non-trivial dispersion introduced by
quantum and thermal fluctuations increases the transition
strength. This effect is observed in all observables considered in
this work: the order parameter of phase transition, the pressure, the
entropy, trace anomaly and baryonic fluctuation observables
$\chi_2^{\rm B}$ and $\chi_4^{\rm B}$. Interestingly, while the
effects on order parameter and thermodynamical observables is very
small, it is significant for the higher moments. This is very well
compatible with the fact, that the results for 2+1 QCD for
order parameter and thermodynamics including the trace anamaly were
already in quantitative agreement with the lattice results for
$T\lesssim T_c$ for the LPA approximation.

We have also provided estimates for the trace anomaly, see
\Fig{fig:traceano} and the kurtosis, see Fig.~\ref{fig:R42PQMconsi}
in 2+1 flavour QCD. This is based on the comparison of 2 flavour with
2+1 flavour computations in \cite{Herbst:2013ufa} as well as 2+1 with
2+1+1 flavour computations in \cite{Fischer:2014ata}. This suggests
that the predominant effect of strange and charm quark is the simple
rescaling of the total momentum scale which is done by presenting the
results as functions of the reduced temperature. In the validity
regime of the present truncation to low energy QCD, for temperatures
$T\lesssim T_c$, the present results agree very well with the
corresponding lattice results. This provides further support for the
reliability and the convergence of the present approximation scheme of the
QCD-enhanced model for these temperature regime. 

Finally we discuss our findings in view of the experiments: the
present analysis predicts larger baryonic fluctuations in QCD at
finite density in comparison to model results in the literature
obtained in LPA. This effect is quite prominent and, thus, it may play
an important role in the observed kurtosis of the net-proton
multiplicity distributions at RHIC. 

Technically, it has been shown in the literature that meson
fluctuations beyond the mean-field approximation smoothen the chiral
crossover. This also leads to a significantly reduced critical region
in comparison to mean field results as well as small baryonic
fluctuations. In summary the mesonic fluctuation analysis made the
search for a critical end point far more difficult. The present
analysis emphasises the relevance of fermionic fluctuations. The
latter naturally strengthen the crossover and in particular enhance
baryonic fluctuations observables. This increases both, the likelihood
of a critical endpoint as well as its observability via fluctuation
measurements. This is in line with the observation that the
measurements of higher moments of the net-proton distributions in the
first phase of BES program at RHIC show that their energy dependence
can neither be reproduced by transport models without critical point
contributions, nor by a hadron resonance gas
model~\cite{Adamczyk:2013dal}. Hence, although more analyses and
explanations are needed to clarify the measurements, the QCD critical
point is still a fascinating candidate.

The calculations in this work have been performed in a QCD-enhanced
low energy effective model. Its limitations at $T\gtrsim T_c$ as well
as the missing frequency-dependence of the couplings have been
discussed in detail. Ongoing extensions of the current work concern
2+1 flavour QCD, the frequency-dependence relevant in particular at
small temperatures $T\lesssim 0.8\, T_c$, the missing glue fluctuations
and hadronic decoupling at $T\gtrsim T_c$. The latter is considered
within dynamical hadronisation. These improvements entail a full
embedding of the present low energy effective theory in continuum QCD
in the FRG framework as done in the fQCD collaboration \cite{fQCD}. We
hope to report on related results in near future.
 
\begin{acknowledgments}
  We thank T.~K.~Herbst, M.~Mitter, F.~Rennecke, B.-J. Schaefer and
  N.~Strodthoff for valuable discussions and work on related
  subjects. W.-J. Fu thanks also
  the support of Alexander von Humboldt Foundation. This work is
  supported by the Helmholtz Alliance HA216/EMMI, and by ERC-AdG-290623.
\end{acknowledgments}
\appendix

\section{Regulators and threshold functions}
\label{app:threshold}
In the present work we use $3d$ flat regulators,
\cite{Litim:2000ci,Litim:2001up}, for quarks and mesons. They read
\begin{equation}
  \label{eq:regu}
  \begin{split}
    R^{\phi}_{k}(q_{0},\vec{q})&
    =Z_{\phi,k}\vec{q}^{2}r_{B}(\vec{q}^{2}/k^2)\,,\\[2ex]
    R^{q}_{k}(q_{0},\vec{q})&
    =Z_{q,k}i\vec{\gamma}\cdot\vec{q}r_{F}(\vec{q}^{2}/k^2)\,,
  \end{split}
\end{equation}
with 
\begin{equation}
  \label{eq:rBF}
  \begin{split}
    r_{B}(x)&=\Big(\frac{1}{x}-1\Big)\Theta(1-x)\,,\\[2ex] 
    r_{F}(x)&=\Big(\frac{1}{\sqrt{x}}-1\Big)\Theta(1-x)\,.
  \end{split}
\end{equation}
In order to simplify our expressions for the threshold functions, we
define
  \begin{align}\nonumber 
    G_{\phi}(q,\bar{m}_{\phi,k}^{2})&=
    \frac{1}{z_{\phi}\tilde{q}_0^2+1+\bar{m}_{\phi,k}^{2}},\\[2ex]
    G_{q}(q,\bar{m}_{q,k}^{2})&=
    \frac{1}{z_{q}^2(\tilde{q}_0+i\tilde{\mu})^2+1+\bar{m}_{q,k}^{2}}\,,
  \end{align}
with $\tilde{q}_0=q_0/k$, $\tilde{\mu}=\mu/k$, and $q_0=(2n_q+1)\pi T$
($n_q\in \mathbb{Z}$) for fermions and $2n_q\pi T$ for bosons. Note
that $z=Z^{\parallel}/Z^{\perp}$, which are the ratios between the
longitudinal and transverse wave function renormalisation factors, are
chosen to be 1 throughout this paper.  First of all, we define
\begin{align}\nonumber 
  \mathcal{F}_{(1)}(\bar{m}_{q,k}^{2},z_{q};T,\mu)&=
  \frac{T}{k}\sum_{n_q}G_{q}(q,\bar{m}_{q,k}^{2}),\\[2ex]
  \mathcal{B}_{(1)}(\bar{m}_{\phi,k}^{2},z_{\phi};T,)&=
  \frac{T}{k}\sum_{n_q}G_{\phi}(q,\bar{m}_{\phi,k}^{2})\,.
\end{align}
Summing up the Matsubara frequencies yields 
\begin{align}\nonumber 
  &\mathcal{F}_{(1)}(\bar{m}_{q,k}^{2},z_{q};T,\mu)=
  \frac{1}{2z_{q}\sqrt{1+\bar{m}_{q,k}^{2}}}\\[2ex]
  &\hspace{.3cm}\times\Big(1-n_{F}(\bar{m}_{q,k}^{2},z_{q};
  T,\mu)-n_{F}(\bar{m}_{q,k}^{2},z_{q};T,-\mu)\Big)\,,
\label{eq:qOmega}\end{align}
and
\begin{align}\nonumber 
    &\hspace{-.7cm}\mathcal{B}_{(1)}(\bar{m}_{\phi,k}^{2},z_{\phi};T)\\[2ex]
    =&\frac{1}{z_{\phi}^{1/2}\sqrt{1+\bar{m}_{\phi,k}^{2}}}
    \Big(\frac{1}{2}+n_{B}(\bar{m}_{\phi,k}^{2},z_{\phi};T)\Big)\,,
\end{align}
with the distribution functions
\begin{align}\nonumber 
  &n_{B}(\bar{m}_{\phi,k}^{2},z_{\phi};T)\\[2ex]
  =&\frac{1}{\exp\bigg\{\frac{1}{T}\frac{k}{z_{\phi}^{1/2}}
    \Big(1+\bar{m}_{\phi,k}^{2}\Big)^{1/2}\bigg\}-1}\,,
\end{align}
and 
\begin{equation}
\begin{split}
  &n_{F}(\bar{m}_{q,k}^{2},z_{q};T,\mu)\\[2ex]
  =&\frac{1}{\exp\bigg\{\frac{1}{T}
    \Big[\frac{k}{z_{q}}(1+\bar{m}_{q,k}^{2})^{1/2}
    -\mu\Big]\bigg\}+1}\,.
\end{split}
\end{equation}
The threshold functions appearing in the flow equation for the
effective potential, i.e. Eq.~(\ref{eq:Vflow}), are given by
\begin{align}\nonumber 
      &l_{0}^{(B,d)}(\bar{m}_{\phi,k}^{2},\eta_{\phi,k};T)\\[2ex]
=&\frac{2}{d-1}\Big(1-\frac{\eta_{\phi,k}}{d+1}\Big)
\mathcal{B}_{(1)}(\bar{m}_{\phi,k}^{2},z_{\phi}=1;T)\,,
\end{align}
and
\begin{align}\nonumber 
    &l_{0}^{(F,d)}(\bar{m}_{q,k}^{2},\eta_{q,k};T,\mu)\\[2ex]
    =&\frac{2}{d-1}\Big(1-\frac{\eta_{q,k}}{d}\Big)
\mathcal{F}_{(1)}(\bar{m}_{q,k}^{2},z_{q}=1;T,\mu)\,.
\end{align}
Furthermore, we also need
\begin{align}
  \mathcal{F}_{(n)}(\bar{m}_{q,k}^{2},z_{q};T,\mu)
  =\frac{T}{k}\sum_{n_q}\big(G_{q}(q,
\bar{m}_{q,k}^{2})\big)^n\,,
\end{align}
which are easily obtained from $\mathcal{F}_{(1)}$ through 
\begin{equation}
  \mathcal{F}_{(n+1)}(\bar{m}_{q,k}^{2},z_{q};T,\mu)=
-\frac{1}{n}\frac{\partial}{
    \partial \bar{m}_{q,k}^{2}}\mathcal{F}_{(n)}(
\bar{m}_{q,k}^{2},z_{q};T,\mu)\,.
\end{equation}
The threshold function $\mathcal{BB}_{(2,2)}$ is derived from
\begin{align}\nonumber 
  &\mathcal{BB}_{(2,2)}(\bar{m}_{\phi_a,k}^{2},
  \bar{m}_{\phi_b,k}^{2},z_{\phi};T)\\[2ex]
  =&\frac{\partial^2}{\partial \bar{m}_{\phi_a,k}^{2}\partial
    \bar{m}_{\phi_b,k}^{2}}\mathcal{BB}_{(1,1)}(
  \bar{m}_{\phi_a,k}^{2},\bar{m}_{\phi_b,k}^{2},z_{\phi};T)\,,
\end{align}
with
\begin{align}\nonumber 
&\mathcal{BB}_{(1,1)}(\bar{m}_{\phi_a,k}^{2},
\bar{m}_{\phi_b,k}^{2},z_{\phi};T)\\[2ex]
=&\frac{T}{k}\sum_{n_q}G_{\phi}(q,
\bar{m}_{\phi_a,k}^{2})G_{\phi}(q,\bar{m}_{\phi_b,k}^{2})\,.
\end{align}
Its analytic expression, after the Matsubara frequencies are summed
up, is given by
\begin{align}\nonumber 
  &\mathcal{BB}_{(1,1)}(\bar{m}_{\phi_a,k}^{2},\bar{m}_{\phi_b,k}^{2},z_{\phi};T)\\[2ex]
\nonumber 
  =&-\frac{1}{z_{\phi}^{1/2}}\bigg\{\Big(\frac{1}{2}+n_{B}(\bar{m}_{\phi_a,k}^{2},z_{\phi};T)
  \Big)\frac{1}{\big(1+\bar{m}_{\phi_a,k}^{2}\big)^{1/2}}\\[2ex]
\nonumber 
  &\times\frac{1}{\big(\bar{m}_{\phi_a,k}^{2}-\bar{m}_{\phi_b,k}^{2}\big)}+
  \Big(\frac{1}{2}+n_{B}(\bar{m}_{\phi_b,k}^{2},z_{\phi};T)\Big)\\[2ex]
  &\times\frac{1}{\big(1+\bar{m}_{\phi_b,k}^{2}\big)^{1/2}}\frac{1}{\big(
    \bar{m}_{\phi_b,k}^{2}-\bar{m}_{\phi_a,k}^{2}\big)}\bigg\}\,,
\end{align}
In the same way, by means of derivatives with respect to appropriate
masses, threshold functions $\mathcal{FB}$'s in
Eq.~(\ref{eq:etapsiexp}) are easily obtained with 
\begin{widetext}
\begin{align}\nonumber 
  &FB_{(1,1)}(\bar{m}_{q,k}^{2},\bar{m}_{\phi,k}^{2},z_{q},z_{\phi};T,\mu,p_0)
  =\frac{T}{k}\sum_{n_q}
  G_{\phi}(p-q,\bar{m}_{\phi,k}^{2})G_{q}(q,\bar{m}_{q,k}^{2})\\
\nonumber 
  =&\frac{1}{2}\frac{k^2}{z_{\phi}z_{q}^2}\bigg\{-
  n_{B}(\bar{m}_{\phi,k}^{2},z_{\phi};T)
  \frac{z_{\phi}^{1/2}}{\big(1+\bar{m}_{\phi,k}^{2}
    \big)^{1/2}}\frac{1}{\Big(ip_0-\mu+\frac{k}{z_{\phi}^{1/2}}
    \big(1+\bar{m}_{\phi,k}^{2}\big)^{1/2}\Big)^2-\big(1
    +\bar{m}_{q,k}^{2}\big)\Big(\frac{k}{z_{q}}\Big)^2}\\
\nonumber 
  &-\big(n_{B}(\bar{m}_{\phi,k}^{2},z_{\phi};T)
  +1\big)\frac{z_{\phi}^{1/2}}{\big(1+\bar{m}_{\phi,k}^{2}
    \big)^{1/2}}\frac{1}{\Big(ip_0-\mu
    -\frac{k}{z_{\phi}^{1/2}}\big(1+\bar{m}_{\phi,k}^{2}\big)^{1/2}
    \Big)^2-\big(1+\bar{m}_{q,k}^{2}\big)\Big(\frac{k}{z_{q}}\Big)^2}\\
\nonumber 
  &+n_{F}(\bar{m}_{q,k}^{2},z_{q};T,
  -\mu)\frac{z_{q}}{\big(1+\bar{m}_{q,k}^{2}\big)^{1/2}}\frac{1}{
    \Big(ip_0-\mu-\frac{k}{z_{q}}\big(1
    +\bar{m}_{q,k}^{2}\big)^{1/2}\Big)^2-\big(1+
    \bar{m}_{\phi,k}^{2}\big)\frac{k^2}{z_{\phi}}}\\
  &+\big(n_{F}(\bar{m}_{q,k}^{2},z_{q};T,\mu)-1\big)\frac{z_{q}}{\big(1+
    \bar{m}_{q,k}^{2}\big)^{1/2}}
  \frac{1}{\Big(ip_0-\mu+\frac{k}{z_{q}}\big(1+
    \bar{m}_{q,k}^{2}\big)^{1/2}\Big)^2-\big(1
    +\bar{m}_{\phi,k}^{2}\big)\frac{k^2}{z_{\phi}}}\bigg\}\,.
\label{eq:f1b1}
\end{align}
\end{widetext}
The $T\to 0$ limit of \eq{eq:f1b1} is achieved by simply dropping the
bosonic thermal distribution functions, $n_B\to 0$, as well as
reducing the fermionic ones to Heaviside theta functions,
$n_F(\bar{m}_{q,k}^{2},z_{q};T,\pm \mu) \to
\theta(k/z_{q}(1+\bar{m}_{q,k}^{2})^{1/2} \mp\mu)$. The latter limit
signals the silver blaze property: below the critical chemical
potential $\mu=\mu_*$ no observable shows a dependence on the chemical
potential. Note that for $T\to 0$ there is still a $\mu$-dependence on
$p_0- i\,\mu$. Again this relates to the silver blaze property as
observables relate to $p_0$-integrals, and the $\mu$-dependence can be
removed by a simple shift of the integration contour below the first
pole in the complex plane at $\mu=\mu_*$. Consequently, on the level
of the threshold functi$FB_{(1,1)}$ in \eq{eq:f1b1} there is a
qualitative difference between the $\mu$-dependence in the quark
distribution functions $n_F$ and that on the frequency $p_0 +i\,\mu$:
only upon frequency integration both $\mu$-dependences show the
characteristic exponential dependence on $\exp \{\pm \mu/T\}$ relevant
for the silver blaze property. This is investigated in detail in
Appendix~\ref{app:Yukawa}. 

Here we proceed with the relevant threshold functions 
\begin{align}\nonumber 
  &FB_{(2,1)}(\bar{m}_{q,k}^{2},\bar{m}_{\phi,k}^{2},z_{q},z_{\phi};T,
  \mu,p_0)\\[2ex]\nonumber
  =&\frac{T}{k}\sum_{n_q}G_{\phi}(p-q,\bar{m}_{\phi,k}^{2})
  \big(G_{q}(q,\bar{m}_{q,k}^{2})\big)^2\\[2ex]
  =&-\frac{\partial}{\partial
    \bar{m}_{q,k}^{2}}FB_{(1,1)}(\bar{m}_{q,k}^{2},
  \bar{m}_{\phi,k}^{2},z_{q},z_{\phi};T,\mu,p_0)\,,
\end{align}
and 
\begin{align} \nonumber 
 &FB_{(1,2)}(\bar{m}_{q,k}^{2},
  \bar{m}_{\phi,k}^{2},z_{q},z_{\phi};T,\mu,p_0)\\[2ex]\nonumber 
  =&\frac{T}{k}\sum_{n_q}
  \big(G_{\phi}(p-q,\bar{m}_{\phi,k}^{2})\big)^2G_{q}(q,\bar{m}_{q,k}^{2})\\[2ex]
  =&-\frac{\partial}{\partial \bar{m}_{\phi,k}^{2}}
  FB_{(1,1)}(\bar{m}_{q,k}^{2},\bar{m}_{\phi,k}^{2},z_{q},z_{\phi};T,\mu,p_0)\,.
\end{align}
Note that the functions $FB_{(1,1)}$, $FB_{(2,1)}$, and $FB_{(1,2)}$ are complex valued at
non-zero chemical potential and, thus, we have to project them onto the
real axis, as explained in the text, viz.
\begin{align}
\mathcal{FB}_{(n,m)}&=\mathrm{Re}(FB_{(n,m)})\,,
\end{align}
with $z_{\phi}=z_{q}=1$. The threshold function in the
flow of the Yukawa coupling~(\ref{eq:hexp}) can be expressed as
\begin{align}\nonumber
  &L_{(1,1)}^{(d)}(\bar{m}_{q,k}^{2},\bar{m}_{\phi,k}^{2},\eta_{q,k},\eta_{\phi,k};T,\mu)\\[2ex]
  \nonumber =&\frac{2}{d-1}\bigg[\Big(1-\frac{\eta_{\phi,k}}{d+1}\Big)
  \mathcal{FB}_{(1,2)}(\bar{m}_{q,k}^{2},\bar{m}_{\phi,k}^{2};T,\mu,p_{0,\text{\tiny{ex}}})\\[2ex]
  &+\Big(1-\frac{\eta_{q,k}}{d}\Big)\mathcal{FB}_{(2,1)}(
  \bar{m}_{q,k}^{2},\bar{m}_{\phi,k}^{2};T,\mu)\bigg]\,,
\end{align}
with the fixed frequency $p_{0,\text{\tiny{ex}}}$ to be specified in
Appendix~\ref{app:Yukawa}. 

\section{Anomalous dimensions of the mesons} 
\label{app:mesons} 

We have discussed the flow equation for the zero-point correlation
function, i.e. the effective potential, in Section~\ref{sec:FRG}. In
order to investigate more interesting properties embedded in the meson
and quark propagators, one has to go beyond this
order. Differentiating both sides of Eq.~(\ref{eq:QMflow}) with
respect to fields twice, one arrives at the flow equations for
two-point functions. After performing appropriate projects, one can
obtain the desired flow equations. For example, the meson anomalous
dimension can be obtained with
\begin{equation}
  \label{eq:etaphi}
  \eta_{\phi,k}(p_0,\vec p)=
  -\frac{1}{Z_{\phi,k}}\frac{1}{N_f^2-1}\frac{\partial^2}{\partial |\vec{p}|^2}
  \frac{\delta^2 \partial_t \Gamma_k}{\delta \pi_i(-p) \delta \pi_i(p)}
  \bigg|_{\rho=\kappa}\,, 
\end{equation}
where a sum over the $N_f^2-1$ pions is implied. The flow is evaluated
at the expansion point $\kappa$ of the Taylor expansion in the mesonic
field, as the right hand sides of the flows are evaluated at this
point. Note that another projection procedure for $\eta$ relates to
finite differences, e.g.\ $\partial_t
\Gamma_k^{(2)}(p_0=0,\vec p^2=k^2) - \partial_t
\Gamma^{(2)}_k(0)$, as has been used in
\cite{Christiansen:2014raa,Christiansen:2015rva}. This gives better
access to the global change of the wave function in the relevant
momentum regime $\vec p^2\leq k^2$. However, it has been shown in
\cite{Helmboldt:2014iya} that a Taylor expansion about $ q=0$ works
quantitatively (for a mesonic $O(N)$-model) for momenta $q^2\lesssim
k^2$. This origintaes in the infrared regularisation of these
momenta. As $\eta_{\phi,k}$ is only inserted in the right hand side of
flows with loop momenta $\vec q^2\leq k^2$ we use $\vec q^2=0$ as the
evaluation momentum. We also choose vanishing frequency $q_0=0$ in
order to keep an expansion point that has the underlying Euclidean
$O(4)$-symmetry. Note, however, that the regulator used breaks the
$O(4)$-symmetry and we will re-evaluate this choice in the case of the
fermionic anomalous dimension.

In summary this leads us to 
\begin{align}\nonumber 
  \eta_{\phi,k}=&\eta_{\phi,k}(0,0) \\[2ex]\nonumber 
  =&\frac{1}{6\pi^2}\Big\{\frac{4}{k^{2}}\bar{
    \kappa}_{k}(\bar{V}^{''}_{k}(\bar{\kappa}_{k}))^{2}
  \mathcal{BB}_{(2,2)}(\bar{m}_{\pi,k}^{2},\bar{m}_{\sigma,k}^{2};T)\\[2ex]
\nonumber 
  &+N_{c}\bar{h}_k^{2}
  \big[(2\eta_{q,k}-3)\mathcal{F}_{(2)}(\bar{m}_{q,k}^{2};T,\mu)\\[2ex] 
  &-4(\eta_{q,k}-2)\mathcal{F}_{(3)}(\bar{m}_{q,k}^{2};T,\mu)\big]\Big\}\,,
\label{eq:etaphiexp}
\end{align}
where the threshold functions $\mathcal{BB}_{(2,2)}$ and
$\mathcal{F}_{(n)}$ are defined in \cite{Pawlowski:2014zaa}, and are
presented in Appendix~\ref{app:threshold} as well.

\section{Flows of the fermionic couplings}
\label{app:Yukawa}

In the spirit of the derivative expansion used for the present class of models 
the flow of the fermionic couplings is obtained by evaluating the corresponding 
derivatives w.r.t.\ the fields at low frequencies and momenta. These flows are given by
\begin{align}\nonumber 
  \eta_{q,k}(p_0,\vec p)=&\frac{1}{Z_{q,k}(p_0,\vec p)}\frac{1}{4 N_c N_f}\\[2ex] 
  \times\mathrm{Re}&\Bigg[\frac{\partial^2}{\partial
    |\vec{p}|^2}\mathrm{Tr}\bigg(i \vec{\gamma}\cdot\vec{p}\Big(
-\frac{\delta^2 \partial_t
    \Gamma_k}{\delta \bar{q}(-p) \delta
    q(p)}\Big)\bigg)\bigg|_{\rho=\kappa}\Bigg]\,,
\label{eq:etapsi}\end{align}
for the anomalous dimension and 
\begin{align}\nonumber 
  \partial_t h_k(p_0,\vec p)=&\frac{\sqrt{2N_f}}{\sigma}\frac{1}{4 N_c N_f}\\[2ex] 
  \times\mathrm{Re}&\Bigg[\mathrm{Tr}\Big(-\frac{\delta^2 \partial_t
    \Gamma_k}{\delta \bar{q}(-p) \delta
    q(p)}\Big)\bigg|_{\rho=\kappa}\Bigg]\,,
  \label{eq:hflow}\end{align}
for the Yukawa coupling. The results in \cite{Helmboldt:2014iya} for
the simple spatial momentum dependence for $\vec q^2\lesssim k^2$ of
the mesons carries over to the quarks as it originates in the infrared
regularisation of these momenta. Hence we use $\vec q=0$. 

It is left to specify the frequency at which \eq{eq:etapsi} and
\eq{eq:hflow} are evaluated. We have already discussed briefly below
\eq{eq:f1b1} in Appendix~\ref{app:threshold} that the $\mu$-dependence
in the argument $p_0 + i\,\mu$ only reflects the correct physical
behavior if keeping the full frequency dependence. Hence, within an
evaluation of the fermionic flows at a fixed frequency we face
subtleties that relate to the silver blaze problem: all couplings with
fermionic legs have an explicit $T$ and $\mu$-dependence as well as
one in the frequency argument. At finite temperature and chemical
potential the latter is a sum of the fermionic matsubara frequency and
the chemical potential, $p_0 + i\,\mu = 2 \pi T (n+1/2 ) + i \mu$ with
$n\in\mathbb{Z}$. The former explicit dependence shows the standard
thermal suppression as well as that of the parameters such as the
masses and couplings. This is clearly seen in \eq{eq:f1b1} where the
explicit $\mu$ and $T$-dependence is that of the fermionic
distribution functions $n_F$ as well as that of $\bar m^2_{\phi/q}$,
$z_{\phi/q}$.

We conclude that inserting $p_0+i\,\mu$-dependent anomalous dimensions
in the loops leads to standard thermal sums, and finally to additional
thermal factors $n_F$ and $1+ n_F$ related to thermal distribution
functions as in \eq{eq:f1b1}. Related results will be considered
elsewhere. 

In turn, frequency-independent wave function renormalisations are obtained by
evaluating $\eta_{q,k}(p_0+i\, \mu,\vec p)$ at some fixed small fermionic 
Matsubara frequency $p_0=2 \pi T (n+1/2 )$. This introduces an
artificial temperature-dependence. Note also that this subtlety is not
related to the specific regulator used but rather applies to all
regulators. For small cutoff scales $k\lesssim 
T$ this is a subleading effects, for $k\gg T$, however, thermal
effects should be suppressed exponentially with $\exp\left\{- c(R_k)
  k/T\right\}$ with a cutoff dependent coefficient $c(R_k)$, see
\cite{Fister:2015eca}. For the present 3d flat cutoff we have
$c(R_k)=1/2$ leading to the standard thermal distribution functions
already for the anomalous dimensions.

\begin{figure}[t]
\includegraphics[scale=0.6]{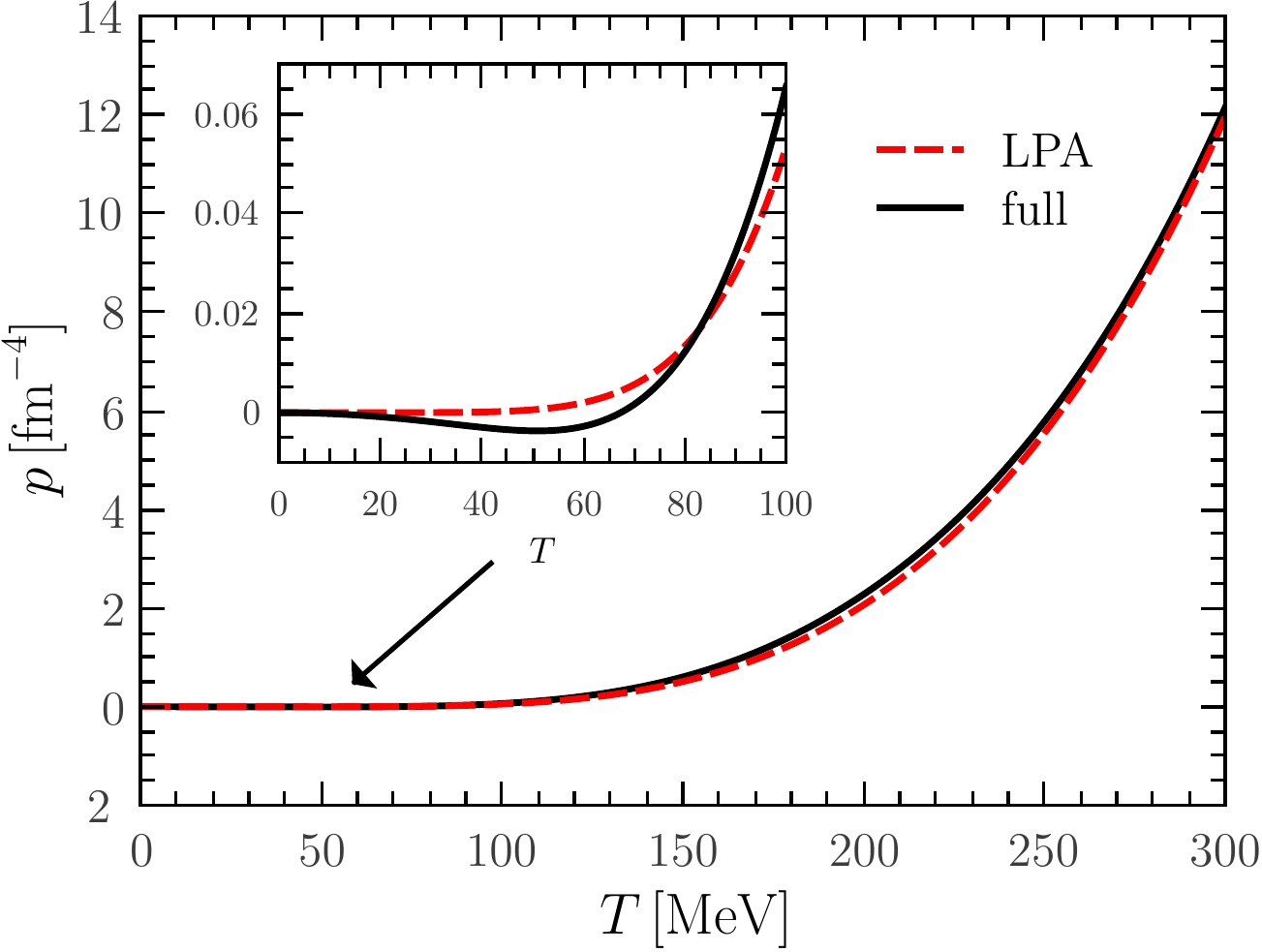}
\caption{Pressure in unit of $\mathrm{fm}^{-4}$ as a
  function of $T$ with truncations full and LPA. The inset plot
  zooms in to the region $T=0\sim 100\,\mathrm{MeV}$.}\label{fig:pre1}
\end{figure}
The articficial temperature-dependence is sub-leading as it is
polynomially suppressed with powers of $T/k$ but it may play a
qualitative r$\hat{\rm o}$le in temperature regimes where the
observables at hand is small. Moreover, the larger the canonical
momentum dimension is, the bigger is the effect. In \Fig{fig:pre1} we
show the pressure of the QM model in the low temperature
regime. There, it is exponentially suppressed with
$\exp\left\{-m_\pi/T\right\}$ and hence the above cutoff effects play
a r$\hat{\rm o}$le. Here, $\eta_{q,k}$ and $h_k$ are obtained by
evaluating the corresponding flows, \eq{eq:etapsi} and \eq{eq:hflow}
simply on the first Matsubara frequency $\pi T$. The discussion above
explains the negative pressure seen in \Fig{fig:pre1} for the QM-model
in the full approximation. 

\begin{figure*}[t]
\includegraphics[scale=0.6]{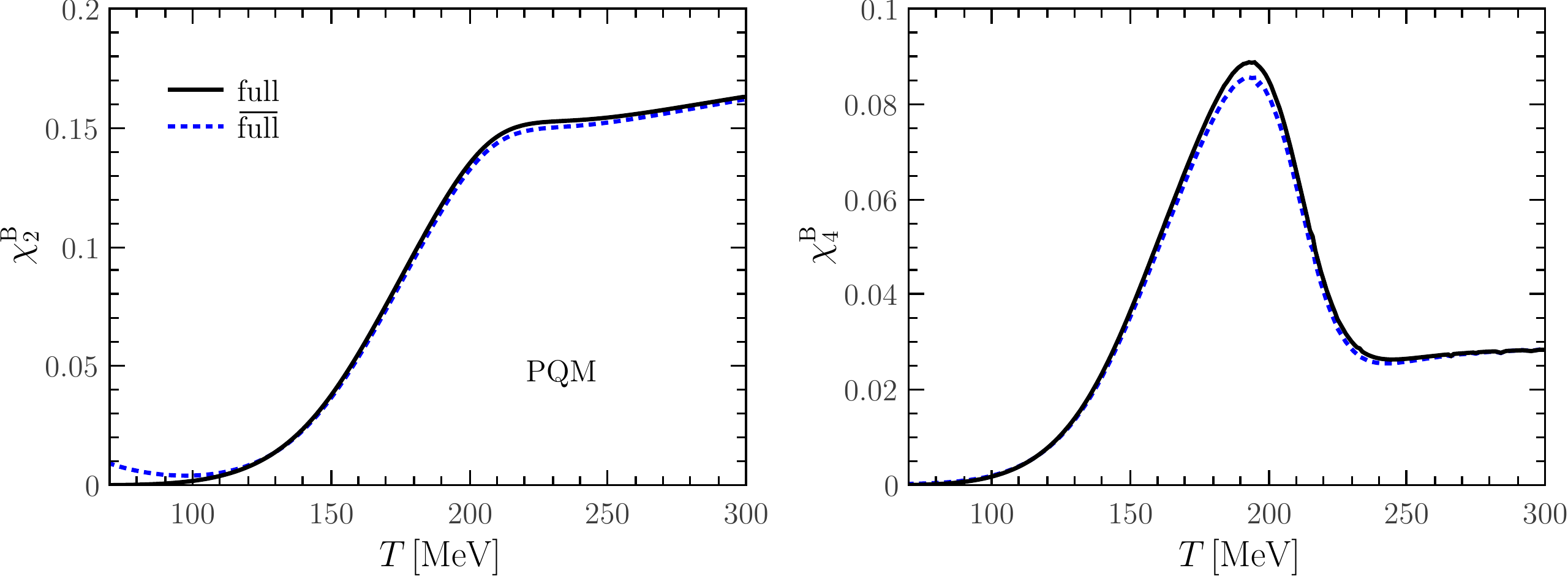}
\caption{Quadratic (left panel) and quartic (right
  panel) baryon number fluctuations in the full approximation with
  external frequency $p_0 + i\, \mu = p_{0,\text{\tiny{ex}}}$, denoted
  by $\mathrm{full}$, and that with external
  frequencies $p_0+i\, \mu = \pi T +i\, \mu$, 
  denoted by
  $\overbar {\rm full} $.}
\label{fig:chi2PQMmu}
\end{figure*}
\begin{figure}[t]
\includegraphics[scale=0.6]{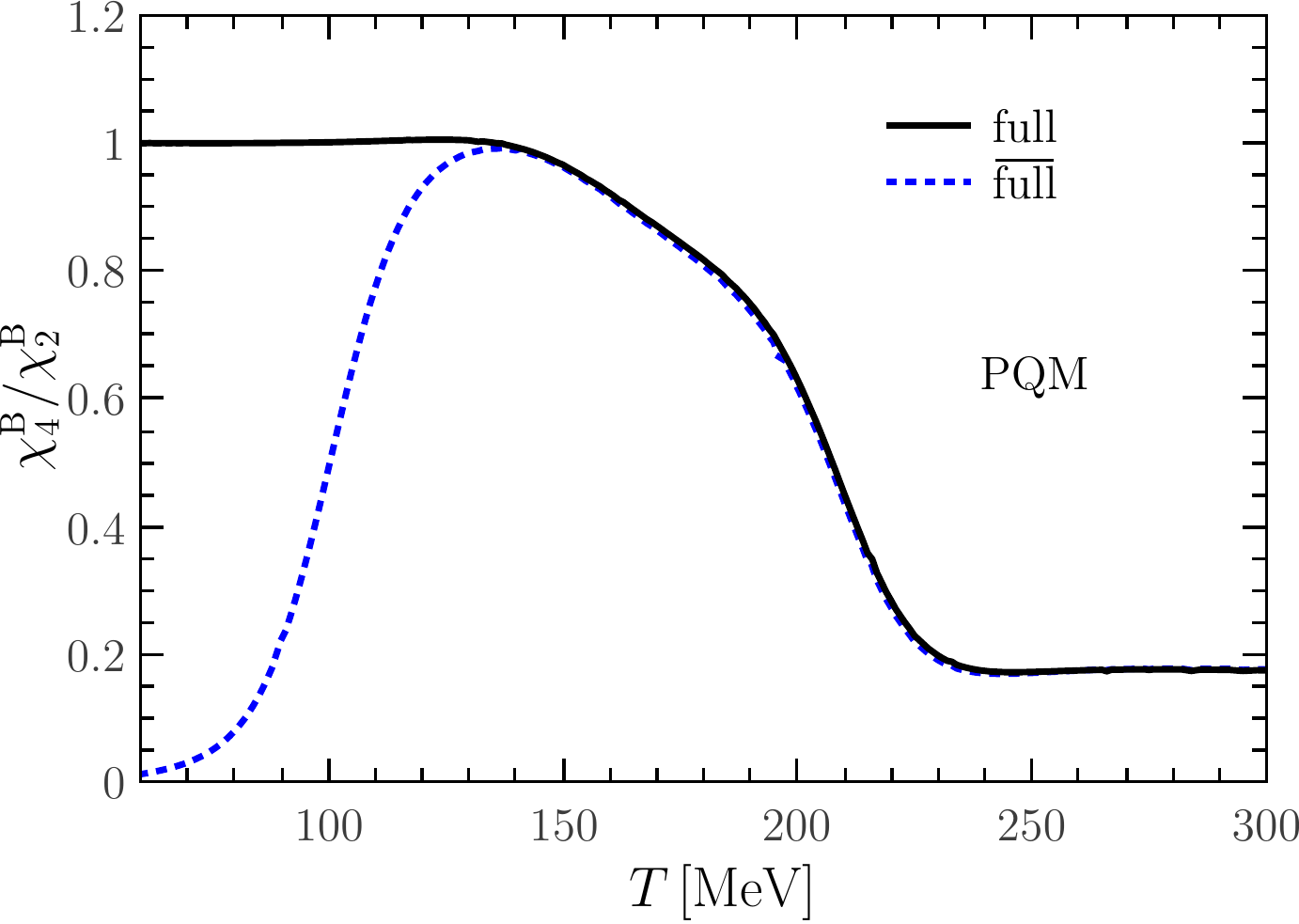}
\caption{Same as Fig.~\ref{fig:chi2PQMmu}, but for the kurtosis of
  baryon number distribution.}\label{fig:R42PQMmu}
\end{figure}
Evidently, this problem is resolved by improving the approximation to
frequency-dependent quark anomalous dimensions $\eta_q(p)$ and Yukawa
couplings $h(p)$. Here we resort to an effective resolution that does
not force us to crank up the approximation: on the right hand side of
the flows both $\eta_q$ and $h$ are evaluated for momenta $\vec q^2
\leq k^2$.  The integrands decay proportional to $\vec q^2$ for small
momenta $\vec q^2/k^2\to 0$ and are maximal for momenta $\vec q^2/k^2
\approx 1$. This anyway suggests an evaluation of the flows for
$\eta_q$ and $h$ at $\vec q^2 \approx k^2$. The present approximation
assumes $O(4)$-symmetry of the effective action also at finite $k$,
and hence the couplings are assumed to be functions of $q_0^2 +\vec
q^2$. Thus, we will evaluate the flows \eq{eq:etapsi} and
\eq{eq:hflow} at $p_0 =k$ for $k\gtrsim \pi T$. This resolves the
problem of the artificial temperature dependence.

For $k\lesssim \pi T$
the minimal momentum squared is bounded by the lowest Matsubara
frequency and hence $p^2 \geq (\pi T)^2$. This suggests an evaluation at 
\begin{align}\label{eq:p0ferm}
p_{0,\text{\tiny{ex}}}^2 = k^2 + (\pi T)^2 \theta_T (k/T)\,,
\end{align}
with some smoothened version $\theta_T$ of the heavyside function with
$\theta_T(k/T \ll 1) \to 1$ and $\theta( k/T \gg 1) \to 0$ reflecting
the known thermal decay, e.g. $\theta_T(x) = \exp \{ -2 x \}$. Such a
$\theta_T (2 k/T)$ overestimates the effect of the Matsubara sum and
we take the conservative choice $\exp \{ -2/5 x \}$ which guarantees
the decoupling for large cutoff scales $k/T \gg 1$ and does not
influence the thermal behaviour for $k/T \lesssim 1$.

The above discussion of the temperature-dependence applies as well to
the dependence on the chemical potential. An evaluation of $\eta_q$ at
a fixed Matsubara-frequency destroys the relation between $\mu$ and
$p_0$-derivatives and induces an artificial $\mu$-dependencies.  Again
this may play a r$\hat {\rm o}$le for small temperatures where
$\mu$-derivaties tend towards zero. Hence, in the present work we
simply drop the sub-leading $\mu$-dependencies and set $p_0 +i\,\mu\to
p_0$ in \eq{eq:f1b1} and the following equations. In summary we arrive
at the anomalous dimension
\begin{align}\nonumber 
  \eta_{q,k}=&\frac{1}{24\pi^2N_{f}}(4-\eta_{\phi,k})\bar{h}_{k}^{2}\\[2ex] \nonumber 
  &\times\Big\{(N_{f}^{2}-1)
  \mathcal{FB}_{(1,2)}(\bar{m}_{q,k}^{2},\bar{m}_{\pi,k}^{2};T,\mu,p_{0,\text{\tiny{ex}}})\\[2ex] 
  &+\mathcal{FB}_{(1,2)}(\bar{m}_{q,k}^{2},\bar{m}_{\sigma,k}^{2};T,\mu,,p_{0,\text{\tiny{ex}}})
\Big\}\,.
\label{eq:etapsiexp}
\end{align}
and the flow of the Yukawa coupling 
\begin{align}\nonumber 
  \partial_t \bar{h}_k=&(\frac{\eta_{\phi,k}}{2}+\eta_{q,k})\bar{h}_k
  +\frac{1}{4\pi^2 N_f}\bar{h}_k^3\\[2ex] \nonumber 
  \times&\Big\{-(N_{f}^{2}-1)
  L_{(1,1)}^{(4)}(\bar{m}_{q,k}^{2},\bar{m}_{\pi,k}^{2},\eta_{q,k},\eta_{\phi,k};T,\mu,
p_{0,\text{\tiny{ex}}})\\[2ex]
  &+L_{(1,1)}^{(4)}(\bar{m}_{q,k}^{2},\bar{m}_{\sigma,k}^{2},
  \eta_{q,k},\eta_{\phi,k};T,\mu,p_{0,\text{\tiny{ex}}})\Big\}\,.
\label{eq:hexp}
\end{align}
Let us now compare the results for $\chi_2^{\rm B}$, $\chi_4^{\rm B}$
and the kurtosis $\chi_2^{\rm B}/\chi_2^{\rm B}$ for the full
aproximation with consistent $T$ and $\mu$-dependence leading to
\eq{eq:etapsiexp}, \eq{eq:hexp} with the computation with fixed
frequency $p_0+i\, \mu = \pi T+ i\,\mu$, denoted by
$\overbar {\rm full}$. The results
for the second and fourth moment $\chi_2^{\rm B}$, $\chi_4^{\rm B}$
and for the kurtosis $\chi_2^{\rm B}$, $\chi_4^{\rm B}$ are presented
in Figs.~\ref{fig:chi2PQMmu},\ref{fig:R42PQMmu}. Evidently, both
approximations are consistent with each other except for low
temperatures, as can be seen in particular for the second moment in
Fig.~\ref{fig:chi2PQMmu}. These deviations are even more obvious in
the kurtosis, Fig.~\ref{fig:R42PQMmu}, where the deviation below
$T\approx 150$\,MeV is significant, that is for $T\lesssim 0.8\,T_c$.
The full apropximation, expanded at $p_0+ i\, \mu \to
p_{0,\text{\tiny{ex}}}$ has a $T,\mu$-consistent scaling and exhibits
the correct asymptotic behavior of the kurtosis for low temperature in
Eq.~(\ref{eq:KurtosisT0}). In turn, an expansion at $p_0+i \,\mu
\to p_{0,\text{\tiny{ex}}}+i \,\mu $ fails for temperatures $T
\lesssim 0.8\, T_c$. 

\section{Baryon number fluctuations and vertex expansions}
\label{app:flucanalytic}
The generalised suceptibilities in \eq{eq:suscept} involve total n$th$
$\mu_B$-derivatives of the pressure $p =- \Omega[\Phi_{\text{\tiny
    EoM}}]$, where $\Phi_{\text{\tiny EoM}}$ are the solutions of the
equations of motion of all fields in the model, $\Phi=(L,\bar L,
q,\bar q,\sigma,\vec \pi)$. These derivatives hit the explicit
$\mu_B$-dependence of the grand potential, and its implicit ones in
the couplings as well as that of $\Phi_{\text{\tiny EoM}}$. 

In the following we discuss $\mu=\mu_B/3$-derivatives of the grand
potential $\Omega$. For example, its first $\mu$-derivative reads
\begin{align}
  \0{d\Omega[\Phi_{\text{\tiny EoM}}]}{d\mu} =
  &\left. \partial_\mu\right|_{\Phi}\Omega+ \partial_\mu\Phi_{i,\text{\tiny
      EoM}}\0{\partial\Omega}{\partial {\Phi_i}}
  =\partial_\mu\Omega\,,
\label{eq:dmuOmega}
\end{align}
where the evaluation of the right hand side at $\Phi=\Phi_{\text{\tiny
    EoM}}$ is understood.  In \eq{eq:dmuOmega} we
have used the equations of motion, and the partial derivatives are at
fixed $\Phi$. In \eq{eq:dmuOmega} and the following relation we shall
use, that the total $\mu$-derivative can be split in a partial
$\mu$-derivative at fixed field and a part that hits the implicit
$\mu$-dependence of $\Phi_{\text{\tiny EoM}}$, to wit
\begin{align}
  \0{d}{d\mu} =\partial_\mu + \partial_\mu\Phi_{i,\text{\tiny
      EoM}} \0{\partial }{\partial {\Phi_i} }\,,
\label{eq:dmu}
\end{align}
With \eq{eq:dmu} the second
$\mu$-derivative follows as
\begin{align}
  \0{d^2\Omega[\Phi_{\text{\tiny EoM}}]}{d\mu^2} = &\0{d}{d\mu} \partial_\mu
    \Omega= \partial^2_\mu\Omega+ \partial_\mu\Phi_{i,\text{\tiny
      EoM}}
  \0{\partial^2 \Omega}{\partial\Phi_i \partial \mu}\,.
\label{eq:dmu2Omega}
\end{align}
\Eq{eq:dmu2Omega} depends on the $\partial_\mu\Phi_{i,\text{\tiny
    EoM}}$. This can be computed from the $\mu$-derivatives of the
respective EoM, to wit
\begin{align} 
  \0{d}{d\mu} \0{\partial\Omega}{\partial\Phi_i}
  =\0{\partial^2\Omega}{\partial\mu \partial \Phi_i}
  +\partial_\mu\Phi_{j,\text{\tiny EoM}} \Omega^{(2)}_{ij}=0\,,
\end{align}
with 
\begin{align}
 \Omega^{(2)}_{ij}= \0{\partial^2\Omega}{\partial
    \Phi_i \partial \Phi_j}\,.
\end{align} 
This can be resolved for $\partial_\mu\Phi_{\text{\tiny EoM}} $ leading to 
\begin{align}\label{eq:dmuPhi}
\partial_\mu\Phi_{i,\text{\tiny EoM}}  = - G_{ij} 
\Omega_{\mu j}\,, 
\end{align}
where, for the sake of brevity, we have introduced the notation 
\begin{align}\label{eq:shortnot} 
  \Omega_{\mu^n i_1...i_m}=\0{\partial^{n+m}\Omega}{\partial
    \mu^n\partial\phi_{i_1}\cdots \partial\phi_{i_m}}\quad {\rm and} \quad 
G_{ij}=\left[\0{1}{
      \Omega^{(2)}}\right]_{ij}\,, 
\end{align} 
which also keeps the higher $\mu$-derivatives simple. Inserting
\eq{eq:dmuPhi} in \eq{eq:dmu2Omega} and using the notation
\eq{eq:shortnot} finally leads to
\begin{align}
  \0{d^2\Omega[\Phi_{\text{\tiny EoM}}]}{d\mu^2} = & 
  \Omega_{\mu^2}- \Omega_{\mu i} G_{ij} \Omega_{\mu j} \,.
\label{eq:dmu2Omegafin}
\end{align}
Now we proceed to the third $\mu$-derivative of $\Omega$. Within our
short hand notation \eq{eq:shortnot}, and using \eq{eq:dmuPhi}, the
total $\mu$-derivative, \eq{eq:dmu}, takes the form
\begin{align}
  \0{d}{d\mu} =\partial_\mu - \Omega_{\mu i} G_{ij}\0{\delta}{\delta \Phi_j} \,.
\label{eq:dmushort}
\end{align}
Applying \eq{eq:dmushort} to \eq{eq:dmu2Omegafin} leads us to 
\begin{align}
  \0{d^3\Omega[\Phi_{\text{\tiny EoM}}]}{d\mu^3}  = & 
 \Omega_{\mu^3}  - \Omega_{\mu i} G_{ij}\, \Omega_{\mu^2 j}  
- \0{d}{d\mu} \left[ \Omega_{\mu i} G_{ij} \Omega_{\mu j}\right] \,.
\label{eq:dmu3Omega}
\end{align}
The total $\mu$-derivative in the second term on the right hand side
can also be performed with
\begin{align}\nonumber
  \0{d}{d\mu} \Omega_{\mu^n i_1...i_m} =&\ \Omega_{\mu^{n+1} i_1...i_m}-
  \Omega_{\mu^{n} i_1...i_m k}
  G_{ks} \Omega_{\mu s}\,,\\[2ex]
  \0{d}{d\mu} G_{ij} =&\ - G_{ik} \left(\Omega_{\mu kl} - \Omega_{ klr}
    G_{rs} \Omega_{\mu s}\right) G_{lj}\,,
 \label{eq:dmurel}\end{align}
following from \eq{eq:dmushort}. Then we finally arrive at 
\begin{align}\nonumber 
  \0{d^3\Omega[\Phi_{\text{\tiny EoM}}]}{d\mu^3} =& 
 \Omega_{\mu^3} - 3 \Omega_{\mu j} G_{ji} \, \Omega_{\mu^2 i}
+3\Omega_{\mu l}G_{lk}\, \Omega_{\mu j}G_{ji}\, \Omega_{\mu ki}  
\\[2ex]
&  -\Omega_{\mu i} G_{ir}\, 
\Omega_{\mu l}  G_{lk}\,   \Omega_{\mu j}  G_{js}\,  \Omega_{krs} \,.
\label{eq:dmu3Omegashort}
\end{align}
For the kurtosis we need the fourth $\mu$-derivative. It can be
straightforwardly derived from \eq{eq:dmu3Omegashort} with
\eq{eq:dmushort} and \eq{eq:dmurel}, and leads us to
\begin{align}\nonumber 
  & \0{d^4\Omega[\Phi_{\text{\tiny EoM}}]}{d\mu^4} \\[2ex]\nonumber =&
  \Omega_{\mu^4} - 4\Omega_{\mu j} G_{ji} \Omega_{\mu^3 i}- 3
  \Omega_{\mu^2 i}G_{ij} \Omega_{\mu^2 j} \\[2ex]\nonumber
  &+12\Omega_{\mu^2 l}G_{lk}\,\Omega_{\mu j} G_{ji} \,\Omega_{\mu ki}
  +6\Omega_{\mu l}G_{lk} \,\Omega_{\mu j}G_{ji} \,\Omega_{\mu^2 ki}
  \\[2ex]\nonumber & -6 \Omega_{\mu l} G_{lk} \, \Omega_{\mu j}
  G_{jr}\, \Omega_{\mu^2 i} G_{is}\, \Omega_{krs} \\[2ex]\nonumber &
  -4 \Omega_{\mu i} G_{ik} \,\Omega_{\mu l} G_{lr} \Omega_{\mu
    j}G_{js} \, \Omega_{\mu krs}\\[2ex]\nonumber &-12\Omega_{\mu i_3}
G_{i_3j_3}\Omega_{\mu i_2}G_{i_2j_2}\Omega_{\mu j_3 i_1}G_{i_1j_1}
\Omega_{\mu j_1 j_2}\\[2ex]\nonumber & +12\Omega_{\mu
    i_1}G_{i_1 i_2} \Omega_{\mu i_2 l}G_{lk}\, \Omega_{\mu j}G_{ji}\,
  \Omega_{\mu j_1}G_{j_1j_2 }\, \Omega_{ ki j_2} \\[2ex]\nonumber &
  -3\Omega_{\mu i_1}G_{i_1 i_2}\,\Omega_{\mu i_3}G_{i_3 i_4}
  \Omega_{i_2 i_4 l}G_{lk}\, \Omega_{\mu j}G_{ji}\, \Omega_{\mu
    j_1}G_{j_1j_2 }\, \Omega_{ k i j_2} \\[2ex] &+\Omega_{\mu
    u} G_{uk}\, \Omega_{\mu i} G_{ir} \, \Omega_{\mu l} G_{ls}
  \,\Omega_{\mu j} G_{jt} \Omega_{ krst}\,.
\label{eq:dmu4Omegashort}
\end{align}
We have also used (\ref{eq:dmu2Omegafin}) and (\ref{eq:dmu4Omegashort}) to 
compute higher moments. A potential problem arises when the determinant of 
$\Omega^{(2)}$ in (\ref{eq:shortnot}) is vanishing, which hampers the inverse 
of $\Omega^{(2)}$ and results in singularity for the propagator $G_{ij}$. Therefore, 
computations based on this analytical method need further investigations.

\begin{figure*}[!htb]
\includegraphics[scale=0.6]{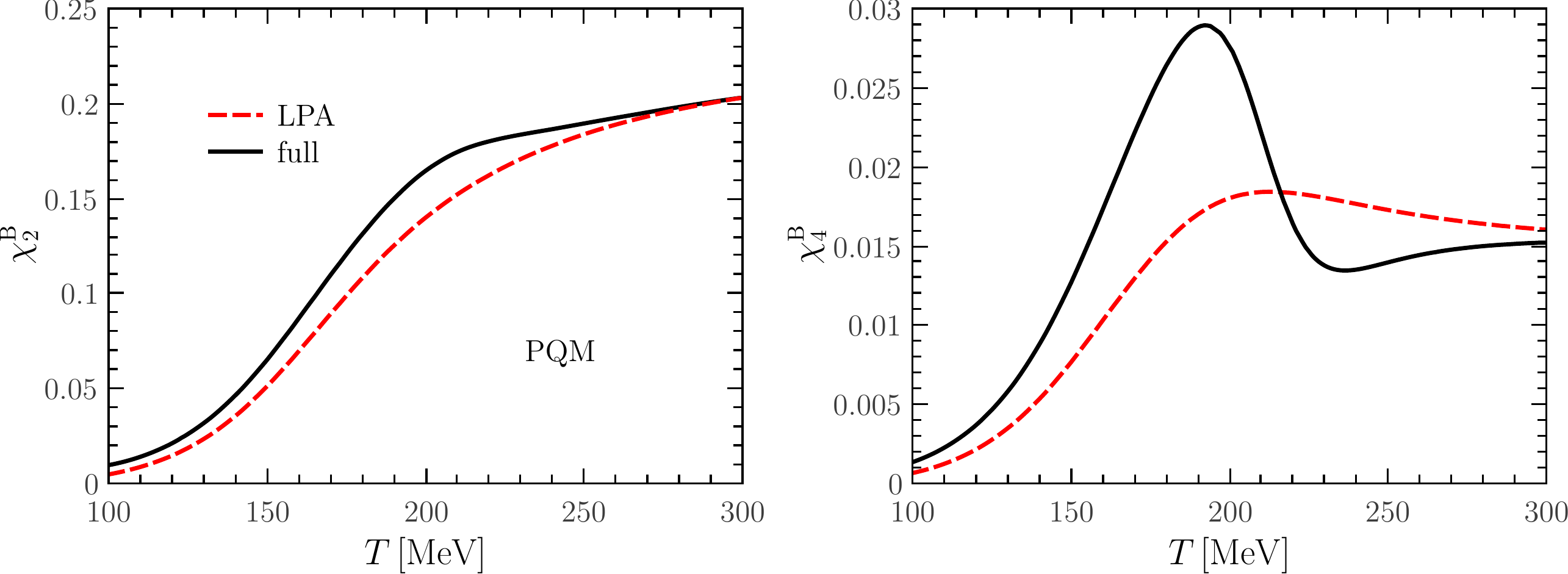}
\caption{Quadratic (left panel) and quartic (right
  panel) baryon number fluctuations as functions of $T$ calculated in
  the simplified PQM model, with the dependence of Polyakov loop on
  $T$ as input. Here we compare two different truncations: LPA and
  full.}\label{fig:chi2BPQM}
\end{figure*}
\section{Kurtosis in a simplified PQM model}
\label{app:sPQM}
\begin{figure*}[!htb]
\includegraphics[scale=0.6]{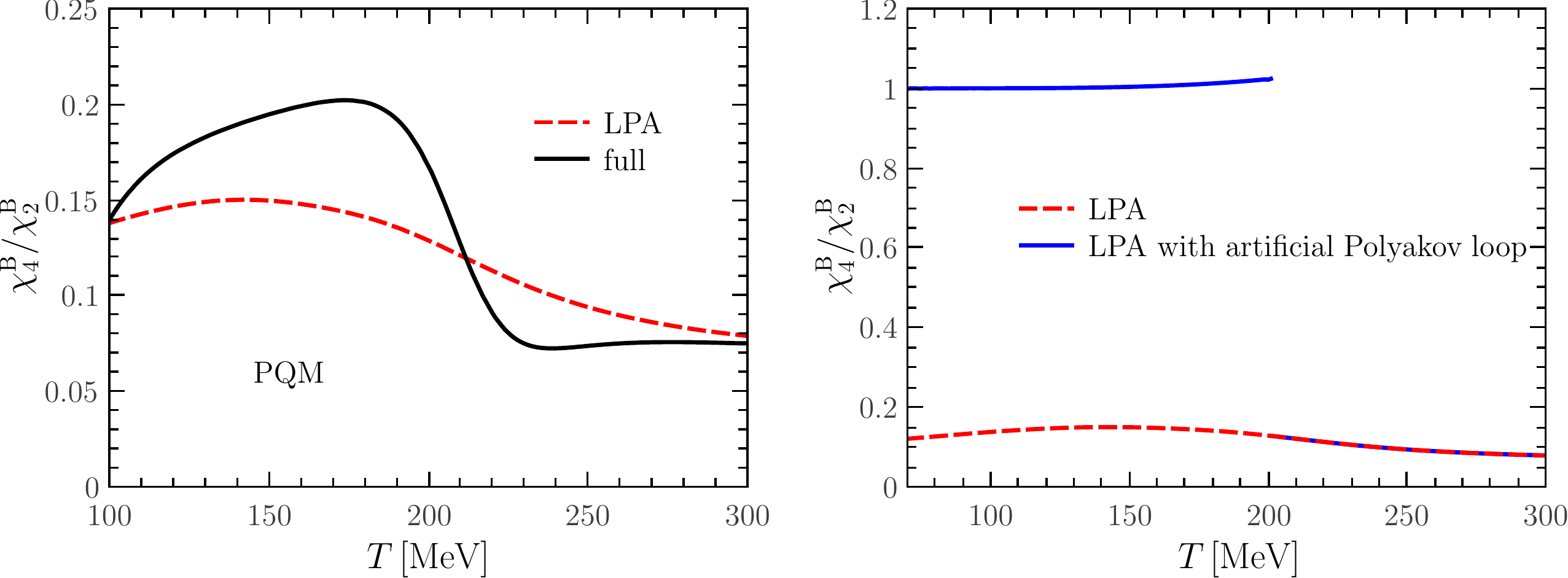}
\caption{Left panel: kurtosis of the baryon number
  distribution in the simplified PQM model. Right panel: the
  red-dashed curve is just the same one in the left panel, in
  comparison to the blue-solid line, which is obtained with the
  Polyakov loop set to be zero artificially below
  $T=200\,\mathrm{MeV}$, while unchanged above this temperature. See
  text for more details.}\label{fig:R42PQM}
\end{figure*}
The representation of the higher moments in
Appendix~\ref{app:flucanalytic} suggests the potential importance of a
self-consistent computation of all quantities, in particular that of
the higher $L,\bar L$-moments of the glue potential. Here we check
this hypothesis in the simplified PQM model: the $T$-dependent
Polyakov loop expectation value is used as an input, and its
dependence on the chemical potential is neglected. As external input
we use the Polyakov loop expectation value as computed
in~\cite{Herbst:2013ufa}. Note that the LPA computation with a full effective
meson potential there is very close to the present LPA computation
within a Taylor expansion. 

\Fig{fig:chi2BPQM} shows the quadratic and quartic baryon number
fluctuations in the simplified PQM model within LPA and full
approximation. One observes that the fluctuations in the improved
approximation are larger, and vary more rapidly with the change of
temperature during the chiral crossover, similarily to the results in
the self-consistent computation, see \Fig{fig:chi2PQMconsi}. 

  The kurtosis in the simplified PQM model, however, looks similar to
  that in the QM model: the missing self-consistency of the glue
  background leads to a qualitative failure in the hadronic phase, see
  Fig.~\ref{fig:R42PQM} in comparison to the self-consistent results
  in \Fig{fig:R42PQMconsi}. Instead, the kurtosis in
    Fig.~\ref{fig:R42PQM} is similar to that in the QM model, in
    agreement to an earlier mean field analysis in \cite{Fu:2009wy},
    see Fig.~5 and Fig.~7 there. If we enforce the hadronic nature with
    $L\bar L=0$ for $T\leq T_c$ and $L\bar L=1$ for $T\geq T_c$ the
    asymptotic low and high temperature behaviour is restored, see the
    right panel of Fig.~\ref{fig:R42PQM}. Note however, that this
    limit does not hold in the hadronic phase, instead we have
    \eq{eq:Llimfin} and \eq{eq:nFlim}.


\bibliography{QMfluc}

\begin{thebibliography}{72}%
\makeatletter
\providecommand \@ifxundefined [1]{%
 \@ifx{#1\undefined}
}%
\providecommand \@ifnum [1]{%
 \ifnum #1\expandafter \@firstoftwo
 \else \expandafter \@secondoftwo
 \fi
}%
\providecommand \@ifx [1]{%
 \ifx #1\expandafter \@firstoftwo
 \else \expandafter \@secondoftwo
 \fi
}%
\providecommand \natexlab [1]{#1}%
\providecommand \enquote  [1]{``#1''}%
\providecommand \bibnamefont  [1]{#1}%
\providecommand \bibfnamefont [1]{#1}%
\providecommand \citenamefont [1]{#1}%
\providecommand \href@noop [0]{\@secondoftwo}%
\providecommand \href [0]{\begingroup \@sanitize@url \@href}%
\providecommand \@href[1]{\@@startlink{#1}\@@href}%
\providecommand \@@href[1]{\endgroup#1\@@endlink}%
\providecommand \@sanitize@url [0]{\catcode `\\12\catcode `\$12\catcode
  `\&12\catcode `\#12\catcode `\^12\catcode `\_12\catcode `\%12\relax}%
\providecommand \@@startlink[1]{}%
\providecommand \@@endlink[0]{}%
\providecommand \url  [0]{\begingroup\@sanitize@url \@url }%
\providecommand \@url [1]{\endgroup\@href {#1}{\urlprefix }}%
\providecommand \urlprefix  [0]{URL }%
\providecommand \Eprint [0]{\href }%
\providecommand \doibase [0]{http://dx.doi.org/}%
\providecommand \selectlanguage [0]{\@gobble}%
\providecommand \bibinfo  [0]{\@secondoftwo}%
\providecommand \bibfield  [0]{\@secondoftwo}%
\providecommand \translation [1]{[#1]}%
\providecommand \BibitemOpen [0]{}%
\providecommand \bibitemStop [0]{}%
\providecommand \bibitemNoStop [0]{.\EOS\space}%
\providecommand \EOS [0]{\spacefactor3000\relax}%
\providecommand \BibitemShut  [1]{\csname bibitem#1\endcsname}%
\let\auto@bib@innerbib\@empty
\bibitem [{\citenamefont {Schwarz}(2003)}]{Schwarz:2003du}%
  \BibitemOpen
  \bibfield  {author} {\bibinfo {author} {\bibfnamefont {D.~J.}\ \bibnamefont
  {Schwarz}},\ }\href {\doibase 10.1002/andp.200310010} {\bibfield  {journal}
  {\bibinfo  {journal} {Annalen Phys.}\ }\textbf {\bibinfo {volume} {12}},\
  \bibinfo {pages} {220} (\bibinfo {year} {2003})},\ \Eprint
  {http://arxiv.org/abs/astro-ph/0303574} {arXiv:astro-ph/0303574 [astro-ph]}
  \BibitemShut {NoStop}%
\bibitem [{\citenamefont {Adams}\ \emph {et~al.}(2005)\citenamefont {Adams}
  \emph {et~al.}}]{Adams:2005dq}%
  \BibitemOpen
  \bibfield  {author} {\bibinfo {author} {\bibfnamefont {J.}~\bibnamefont
  {Adams}} \emph {et~al.} (\bibinfo {collaboration} {STAR Collaboration}),\
  }\href {\doibase 10.1016/j.nuclphysa.2005.03.085} {\bibfield  {journal}
  {\bibinfo  {journal} {Nucl.Phys.}\ }\textbf {\bibinfo {volume} {A757}},\
  \bibinfo {pages} {102} (\bibinfo {year} {2005})},\ \Eprint
  {http://arxiv.org/abs/nucl-ex/0501009} {arXiv:nucl-ex/0501009 [nucl-ex]}
  \BibitemShut {NoStop}%
\bibitem [{\citenamefont {Adcox}\ \emph {et~al.}(2005)\citenamefont {Adcox}
  \emph {et~al.}}]{Adcox:2004mh}%
  \BibitemOpen
  \bibfield  {author} {\bibinfo {author} {\bibfnamefont {K.}~\bibnamefont
  {Adcox}} \emph {et~al.} (\bibinfo {collaboration} {PHENIX Collaboration}),\
  }\href {\doibase 10.1016/j.nuclphysa.2005.03.086} {\bibfield  {journal}
  {\bibinfo  {journal} {Nucl.Phys.}\ }\textbf {\bibinfo {volume} {A757}},\
  \bibinfo {pages} {184} (\bibinfo {year} {2005})},\ \Eprint
  {http://arxiv.org/abs/nucl-ex/0410003} {arXiv:nucl-ex/0410003 [nucl-ex]}
  \BibitemShut {NoStop}%
\bibitem [{\citenamefont {Aamodt}\ \emph {et~al.}(2010)\citenamefont {Aamodt}
  \emph {et~al.}}]{Aamodt:2010pa}%
  \BibitemOpen
  \bibfield  {author} {\bibinfo {author} {\bibfnamefont {K.}~\bibnamefont
  {Aamodt}} \emph {et~al.} (\bibinfo {collaboration} {ALICE Collaboration}),\
  }\href {\doibase 10.1103/PhysRevLett.105.252302} {\bibfield  {journal}
  {\bibinfo  {journal} {Phys.Rev.Lett.}\ }\textbf {\bibinfo {volume} {105}},\
  \bibinfo {pages} {252302} (\bibinfo {year} {2010})},\ \Eprint
  {http://arxiv.org/abs/1011.3914} {arXiv:1011.3914 [nucl-ex]} \BibitemShut
  {NoStop}%
\bibitem [{\citenamefont {Stephanov}(2006)}]{Stephanov:2007fk}%
  \BibitemOpen
  \bibfield  {author} {\bibinfo {author} {\bibfnamefont {M.}~\bibnamefont
  {Stephanov}},\ }\href@noop {} {\bibfield  {journal} {\bibinfo  {journal}
  {PoS}\ }\textbf {\bibinfo {volume} {LAT2006}},\ \bibinfo {pages} {024}
  (\bibinfo {year} {2006})},\ \Eprint {http://arxiv.org/abs/hep-lat/0701002}
  {arXiv:hep-lat/0701002 [hep-lat]} \BibitemShut {NoStop}%
\bibitem [{\citenamefont {Borsanyi}\ \emph {et~al.}(2013)\citenamefont
  {Borsanyi}, \citenamefont {Fodor}, \citenamefont {Katz}, \citenamefont
  {Krieg}, \citenamefont {Ratti} \emph {et~al.}}]{Borsanyi:2013hza}%
  \BibitemOpen
  \bibfield  {author} {\bibinfo {author} {\bibfnamefont {S.}~\bibnamefont
  {Borsanyi}}, \bibinfo {author} {\bibfnamefont {Z.}~\bibnamefont {Fodor}},
  \bibinfo {author} {\bibfnamefont {S.}~\bibnamefont {Katz}}, \bibinfo {author}
  {\bibfnamefont {S.}~\bibnamefont {Krieg}}, \bibinfo {author} {\bibfnamefont
  {C.}~\bibnamefont {Ratti}},  \emph {et~al.},\ }\href {\doibase
  10.1103/PhysRevLett.111.062005} {\bibfield  {journal} {\bibinfo  {journal}
  {Phys.Rev.Lett.}\ }\textbf {\bibinfo {volume} {111}},\ \bibinfo {pages}
  {062005} (\bibinfo {year} {2013})},\ \Eprint {http://arxiv.org/abs/1305.5161}
  {arXiv:1305.5161 [hep-lat]} \BibitemShut {NoStop}%
\bibitem [{\citenamefont {Ding}(2014)}]{Ding:2014kva}%
  \BibitemOpen
  \bibfield  {author} {\bibinfo {author} {\bibfnamefont {H.-T.}\ \bibnamefont
  {Ding}},\ }\bibfield  {booktitle} {\emph {\bibinfo {booktitle} {{Proceedings,
  24th International Conference on Ultra-Relativistic Nucleus-Nucleus
  Collisions (Quark Matter 2014)}}},\ }\href {\doibase
  10.1016/j.nuclphysa.2014.09.053} {\bibfield  {journal} {\bibinfo  {journal}
  {Nucl. Phys.}\ }\textbf {\bibinfo {volume} {A931}},\ \bibinfo {pages} {52}
  (\bibinfo {year} {2014})},\ \Eprint {http://arxiv.org/abs/1408.5236}
  {arXiv:1408.5236 [hep-lat]} \BibitemShut {NoStop}%
\bibitem [{\citenamefont {Bellwied}\ \emph {et~al.}(2015)\citenamefont
  {Bellwied}, \citenamefont {Borsanyi}, \citenamefont {Fodor}, \citenamefont
  {Katz}, \citenamefont {Pasztor}, \citenamefont {Ratti},\ and\ \citenamefont
  {Szabo}}]{Bellwied:2015lba}%
  \BibitemOpen
  \bibfield  {author} {\bibinfo {author} {\bibfnamefont {R.}~\bibnamefont
  {Bellwied}}, \bibinfo {author} {\bibfnamefont {S.}~\bibnamefont {Borsanyi}},
  \bibinfo {author} {\bibfnamefont {Z.}~\bibnamefont {Fodor}}, \bibinfo
  {author} {\bibfnamefont {S.~D.}\ \bibnamefont {Katz}}, \bibinfo {author}
  {\bibfnamefont {A.}~\bibnamefont {Pasztor}}, \bibinfo {author} {\bibfnamefont
  {C.}~\bibnamefont {Ratti}}, \ and\ \bibinfo {author} {\bibfnamefont {K.~K.}\
  \bibnamefont {Szabo}},\ }\href@noop {} {\  (\bibinfo {year} {2015})},\
  \Eprint {http://arxiv.org/abs/1507.04627} {arXiv:1507.04627 [hep-lat]}
  \BibitemShut {NoStop}%
\bibitem [{\citenamefont {Ding}\ \emph {et~al.}(2015)\citenamefont {Ding},
  \citenamefont {Mukherjee}, \citenamefont {Ohno}, \citenamefont {Petreczky},\
  and\ \citenamefont {Schadler}}]{Ding:2015fca}%
  \BibitemOpen
  \bibfield  {author} {\bibinfo {author} {\bibfnamefont {H.~T.}\ \bibnamefont
  {Ding}}, \bibinfo {author} {\bibfnamefont {S.}~\bibnamefont {Mukherjee}},
  \bibinfo {author} {\bibfnamefont {H.}~\bibnamefont {Ohno}}, \bibinfo {author}
  {\bibfnamefont {P.}~\bibnamefont {Petreczky}}, \ and\ \bibinfo {author}
  {\bibfnamefont {H.~P.}\ \bibnamefont {Schadler}},\ }\href@noop {} {\
  (\bibinfo {year} {2015})},\ \Eprint {http://arxiv.org/abs/1507.06637}
  {arXiv:1507.06637 [hep-lat]} \BibitemShut {NoStop}%
\bibitem [{\citenamefont {Karsch}\ and\ \citenamefont
  {Redlich}(2011)}]{Karsch:2010ck}%
  \BibitemOpen
  \bibfield  {author} {\bibinfo {author} {\bibfnamefont {F.}~\bibnamefont
  {Karsch}}\ and\ \bibinfo {author} {\bibfnamefont {K.}~\bibnamefont
  {Redlich}},\ }\href {\doibase 10.1016/j.physletb.2010.10.046} {\bibfield
  {journal} {\bibinfo  {journal} {Phys. Lett.}\ }\textbf {\bibinfo {volume}
  {B695}},\ \bibinfo {pages} {136} (\bibinfo {year} {2011})},\ \Eprint
  {http://arxiv.org/abs/1007.2581} {arXiv:1007.2581 [hep-ph]} \BibitemShut
  {NoStop}%
\bibitem [{\citenamefont {Litim}\ and\ \citenamefont
  {Pawlowski}(1998)}]{Litim:1998nf}%
  \BibitemOpen
  \bibfield  {author} {\bibinfo {author} {\bibfnamefont {D.~F.}\ \bibnamefont
  {Litim}}\ and\ \bibinfo {author} {\bibfnamefont {J.~M.}\ \bibnamefont
  {Pawlowski}},\ }\href@noop {} {\ ,\ \bibinfo {pages} {168} (\bibinfo {year}
  {1998})},\ \Eprint {http://arxiv.org/abs/hep-th/9901063}
  {arXiv:hep-th/9901063 [hep-th]} \BibitemShut {NoStop}%
\bibitem [{\citenamefont {Berges}\ \emph {et~al.}(2002)\citenamefont {Berges},
  \citenamefont {Tetradis},\ and\ \citenamefont {Wetterich}}]{Berges:2000ew}%
  \BibitemOpen
  \bibfield  {author} {\bibinfo {author} {\bibfnamefont {J.}~\bibnamefont
  {Berges}}, \bibinfo {author} {\bibfnamefont {N.}~\bibnamefont {Tetradis}}, \
  and\ \bibinfo {author} {\bibfnamefont {C.}~\bibnamefont {Wetterich}},\ }\href
  {\doibase 10.1016/S0370-1573(01)00098-9} {\bibfield  {journal} {\bibinfo
  {journal} {Phys. Rept.}\ }\textbf {\bibinfo {volume} {363}},\ \bibinfo
  {pages} {223} (\bibinfo {year} {2002})},\ \Eprint
  {http://arxiv.org/abs/hep-ph/0005122} {arXiv:hep-ph/0005122} \BibitemShut
  {NoStop}%
\bibitem [{\citenamefont {Pawlowski}(2007)}]{Pawlowski:2005xe}%
  \BibitemOpen
  \bibfield  {author} {\bibinfo {author} {\bibfnamefont {J.~M.}\ \bibnamefont
  {Pawlowski}},\ }\href {\doibase 10.1016/j.aop.2007.01.007} {\bibfield
  {journal} {\bibinfo  {journal} {Annals Phys.}\ }\textbf {\bibinfo {volume}
  {322}},\ \bibinfo {pages} {2831} (\bibinfo {year} {2007})},\ \Eprint
  {http://arxiv.org/abs/hep-th/0512261} {arXiv:hep-th/0512261 [hep-th]}
  \BibitemShut {NoStop}%
\bibitem [{\citenamefont {Gies}(2012)}]{Gies:2006wv}%
  \BibitemOpen
  \bibfield  {author} {\bibinfo {author} {\bibfnamefont {H.}~\bibnamefont
  {Gies}},\ }\href {\doibase 10.1007/978-3-642-27320-9_6} {\bibfield  {journal}
  {\bibinfo  {journal} {Lect.Notes Phys.}\ }\textbf {\bibinfo {volume} {852}},\
  \bibinfo {pages} {287} (\bibinfo {year} {2012})},\ \Eprint
  {http://arxiv.org/abs/hep-ph/0611146} {arXiv:hep-ph/0611146 [hep-ph]}
  \BibitemShut {NoStop}%
\bibitem [{\citenamefont {Schaefer}\ and\ \citenamefont
  {Wambach}(2008)}]{Schaefer:2006sr}%
  \BibitemOpen
  \bibfield  {author} {\bibinfo {author} {\bibfnamefont {B.-J.}\ \bibnamefont
  {Schaefer}}\ and\ \bibinfo {author} {\bibfnamefont {J.}~\bibnamefont
  {Wambach}},\ }\href {\doibase 10.1134/S1063779608070083} {\bibfield
  {journal} {\bibinfo  {journal} {Phys.Part.Nucl.}\ }\textbf {\bibinfo {volume}
  {39}},\ \bibinfo {pages} {1025} (\bibinfo {year} {2008})},\ \Eprint
  {http://arxiv.org/abs/hep-ph/0611191} {arXiv:hep-ph/0611191 [hep-ph]}
  \BibitemShut {NoStop}%
\bibitem [{\citenamefont {Pawlowski}(2011)}]{Pawlowski:2010ht}%
  \BibitemOpen
  \bibfield  {author} {\bibinfo {author} {\bibfnamefont {J.~M.}\ \bibnamefont
  {Pawlowski}},\ }\href {\doibase 10.1063/1.3574945} {\bibfield  {journal}
  {\bibinfo  {journal} {AIP Conf.Proc.}\ }\textbf {\bibinfo {volume} {1343}},\
  \bibinfo {pages} {75} (\bibinfo {year} {2011})},\ \Eprint
  {http://arxiv.org/abs/1012.5075} {arXiv:1012.5075 [hep-ph]} \BibitemShut
  {NoStop}%
\bibitem [{\citenamefont {Braun}(2012)}]{Braun:2011pp}%
  \BibitemOpen
  \bibfield  {author} {\bibinfo {author} {\bibfnamefont {J.}~\bibnamefont
  {Braun}},\ }\href {\doibase 10.1088/0954-3899/39/3/033001} {\bibfield
  {journal} {\bibinfo  {journal} {J.Phys.}\ }\textbf {\bibinfo {volume}
  {G39}},\ \bibinfo {pages} {033001} (\bibinfo {year} {2012})},\ \Eprint
  {http://arxiv.org/abs/1108.4449} {arXiv:1108.4449 [hep-ph]} \BibitemShut
  {NoStop}%
\bibitem [{\citenamefont {von Smekal}(2012)}]{vonSmekal:2012vx}%
  \BibitemOpen
  \bibfield  {author} {\bibinfo {author} {\bibfnamefont {L.}~\bibnamefont {von
  Smekal}},\ }\href {\doibase 10.1016/j.nuclphysbps.2012.06.006} {\bibfield
  {journal} {\bibinfo  {journal} {Nucl.Phys.Proc.Suppl.}\ }\textbf {\bibinfo
  {volume} {228}},\ \bibinfo {pages} {179} (\bibinfo {year} {2012})},\ \Eprint
  {http://arxiv.org/abs/1205.4205} {arXiv:1205.4205 [hep-ph]} \BibitemShut
  {NoStop}%
\bibitem [{\citenamefont {Braun}\ \emph {et~al.}(2011)\citenamefont {Braun},
  \citenamefont {Haas}, \citenamefont {Marhauser},\ and\ \citenamefont
  {Pawlowski}}]{Braun:2009gm}%
  \BibitemOpen
  \bibfield  {author} {\bibinfo {author} {\bibfnamefont {J.}~\bibnamefont
  {Braun}}, \bibinfo {author} {\bibfnamefont {L.~M.}\ \bibnamefont {Haas}},
  \bibinfo {author} {\bibfnamefont {F.}~\bibnamefont {Marhauser}}, \ and\
  \bibinfo {author} {\bibfnamefont {J.~M.}\ \bibnamefont {Pawlowski}},\ }\href
  {\doibase 10.1103/PhysRevLett.106.022002} {\bibfield  {journal} {\bibinfo
  {journal} {Phys.Rev.Lett.}\ }\textbf {\bibinfo {volume} {106}},\ \bibinfo
  {pages} {022002} (\bibinfo {year} {2011})},\ \Eprint
  {http://arxiv.org/abs/0908.0008} {arXiv:0908.0008 [hep-ph]} \BibitemShut
  {NoStop}%
\bibitem [{\citenamefont {Braun}(2010)}]{Braun:2009si}%
  \BibitemOpen
  \bibfield  {author} {\bibinfo {author} {\bibfnamefont {J.}~\bibnamefont
  {Braun}},\ }\href {\doibase 10.1103/PhysRevD.81.016008} {\bibfield  {journal}
  {\bibinfo  {journal} {Phys.Rev.}\ }\textbf {\bibinfo {volume} {D81}},\
  \bibinfo {pages} {016008} (\bibinfo {year} {2010})},\ \Eprint
  {http://arxiv.org/abs/0908.1543} {arXiv:0908.1543 [hep-ph]} \BibitemShut
  {NoStop}%
\bibitem [{\citenamefont {Haas}\ \emph {et~al.}(2013)\citenamefont {Haas},
  \citenamefont {Stiele}, \citenamefont {Braun}, \citenamefont {Pawlowski},\
  and\ \citenamefont {Schaffner-Bielich}}]{Haas:2013qwp}%
  \BibitemOpen
  \bibfield  {author} {\bibinfo {author} {\bibfnamefont {L.~M.}\ \bibnamefont
  {Haas}}, \bibinfo {author} {\bibfnamefont {R.}~\bibnamefont {Stiele}},
  \bibinfo {author} {\bibfnamefont {J.}~\bibnamefont {Braun}}, \bibinfo
  {author} {\bibfnamefont {J.~M.}\ \bibnamefont {Pawlowski}}, \ and\ \bibinfo
  {author} {\bibfnamefont {J.}~\bibnamefont {Schaffner-Bielich}},\ }\href
  {\doibase 10.1103/PhysRevD.87.076004} {\bibfield  {journal} {\bibinfo
  {journal} {Phys.Rev.}\ }\textbf {\bibinfo {volume} {D87}},\ \bibinfo {pages}
  {076004} (\bibinfo {year} {2013})},\ \Eprint {http://arxiv.org/abs/1302.1993}
  {arXiv:1302.1993 [hep-ph]} \BibitemShut {NoStop}%
\bibitem [{\citenamefont {Herbst}\ \emph {et~al.}(2014)\citenamefont {Herbst},
  \citenamefont {Mitter}, \citenamefont {Pawlowski}, \citenamefont {Schaefer},\
  and\ \citenamefont {Stiele}}]{Herbst:2013ufa}%
  \BibitemOpen
  \bibfield  {author} {\bibinfo {author} {\bibfnamefont {T.~K.}\ \bibnamefont
  {Herbst}}, \bibinfo {author} {\bibfnamefont {M.}~\bibnamefont {Mitter}},
  \bibinfo {author} {\bibfnamefont {J.~M.}\ \bibnamefont {Pawlowski}}, \bibinfo
  {author} {\bibfnamefont {B.-J.}\ \bibnamefont {Schaefer}}, \ and\ \bibinfo
  {author} {\bibfnamefont {R.}~\bibnamefont {Stiele}},\ }\href {\doibase
  10.1016/j.physletb.2014.02.045} {\bibfield  {journal} {\bibinfo  {journal}
  {Phys.Lett.}\ }\textbf {\bibinfo {volume} {B731}},\ \bibinfo {pages} {248}
  (\bibinfo {year} {2014})},\ \Eprint {http://arxiv.org/abs/1308.3621}
  {arXiv:1308.3621 [hep-ph]} \BibitemShut {NoStop}%
\bibitem [{\citenamefont {{Pawlowski}}\ and\ \citenamefont
  {{Rennecke}}(2014)}]{Pawlowski:2014zaa}%
  \BibitemOpen
  \bibfield  {author} {\bibinfo {author} {\bibfnamefont {J.~M.}\ \bibnamefont
  {{Pawlowski}}}\ and\ \bibinfo {author} {\bibfnamefont {F.}~\bibnamefont
  {{Rennecke}}},\ }\href {\doibase 10.1103/PhysRevD.90.076002} {\bibfield
  {journal} {\bibinfo  {journal} {\prd}\ }\textbf {\bibinfo {volume} {90}},\
  \bibinfo {eid} {076002} (\bibinfo {year} {2014})},\ \Eprint
  {http://arxiv.org/abs/1403.1179} {arXiv:1403.1179 [hep-ph]} \BibitemShut
  {NoStop}%
\bibitem [{\citenamefont {Helmboldt}\ \emph {et~al.}(2015)\citenamefont
  {Helmboldt}, \citenamefont {Pawlowski},\ and\ \citenamefont
  {Strodthoff}}]{Helmboldt:2014iya}%
  \BibitemOpen
  \bibfield  {author} {\bibinfo {author} {\bibfnamefont {A.~J.}\ \bibnamefont
  {Helmboldt}}, \bibinfo {author} {\bibfnamefont {J.~M.}\ \bibnamefont
  {Pawlowski}}, \ and\ \bibinfo {author} {\bibfnamefont {N.}~\bibnamefont
  {Strodthoff}},\ }\href {\doibase 10.1103/PhysRevD.91.054010} {\bibfield
  {journal} {\bibinfo  {journal} {Phys.Rev.}\ }\textbf {\bibinfo {volume}
  {D91}},\ \bibinfo {pages} {054010} (\bibinfo {year} {2015})},\ \Eprint
  {http://arxiv.org/abs/1409.8414} {arXiv:1409.8414 [hep-ph]} \BibitemShut
  {NoStop}%
\bibitem [{\citenamefont {Mitter}\ \emph {et~al.}(2015)\citenamefont {Mitter},
  \citenamefont {Pawlowski},\ and\ \citenamefont
  {Strodthoff}}]{Mitter:2014wpa}%
  \BibitemOpen
  \bibfield  {author} {\bibinfo {author} {\bibfnamefont {M.}~\bibnamefont
  {Mitter}}, \bibinfo {author} {\bibfnamefont {J.~M.}\ \bibnamefont
  {Pawlowski}}, \ and\ \bibinfo {author} {\bibfnamefont {N.}~\bibnamefont
  {Strodthoff}},\ }\href {\doibase 10.1103/PhysRevD.91.054035} {\bibfield
  {journal} {\bibinfo  {journal} {Phys.Rev.}\ }\textbf {\bibinfo {volume}
  {D91}},\ \bibinfo {pages} {054035} (\bibinfo {year} {2015})},\ \Eprint
  {http://arxiv.org/abs/1411.7978} {arXiv:1411.7978 [hep-ph]} \BibitemShut
  {NoStop}%
\bibitem [{\citenamefont {Braun}\ \emph {et~al.}(2014)\citenamefont {Braun},
  \citenamefont {Fister}, \citenamefont {Pawlowski},\ and\ \citenamefont
  {Rennecke}}]{Braun:2014ata}%
  \BibitemOpen
  \bibfield  {author} {\bibinfo {author} {\bibfnamefont {J.}~\bibnamefont
  {Braun}}, \bibinfo {author} {\bibfnamefont {L.}~\bibnamefont {Fister}},
  \bibinfo {author} {\bibfnamefont {J.~M.}\ \bibnamefont {Pawlowski}}, \ and\
  \bibinfo {author} {\bibfnamefont {F.}~\bibnamefont {Rennecke}},\ }\href@noop
  {} {\  (\bibinfo {year} {2014})},\ \Eprint {http://arxiv.org/abs/1412.1045}
  {arXiv:1412.1045 [hep-ph]} \BibitemShut {NoStop}%
\bibitem [{\citenamefont {Pawlowski}(2014)}]{Pawlowski:2014aha}%
  \BibitemOpen
  \bibfield  {author} {\bibinfo {author} {\bibfnamefont {J.~M.}\ \bibnamefont
  {Pawlowski}},\ }\href {\doibase 10.1016/j.nuclphysa.2014.09.074} {\bibfield
  {journal} {\bibinfo  {journal} {Nucl.Phys.}\ }\textbf {\bibinfo {volume}
  {A931}},\ \bibinfo {pages} {113} (\bibinfo {year} {2014})}\BibitemShut
  {NoStop}%
\bibitem [{\citenamefont {Skokov}\ \emph
  {et~al.}(2010{\natexlab{a}})\citenamefont {Skokov}, \citenamefont {Stokic},
  \citenamefont {Friman},\ and\ \citenamefont {Redlich}}]{Skokov:2010wb}%
  \BibitemOpen
  \bibfield  {author} {\bibinfo {author} {\bibfnamefont {V.}~\bibnamefont
  {Skokov}}, \bibinfo {author} {\bibfnamefont {B.}~\bibnamefont {Stokic}},
  \bibinfo {author} {\bibfnamefont {B.}~\bibnamefont {Friman}}, \ and\ \bibinfo
  {author} {\bibfnamefont {K.}~\bibnamefont {Redlich}},\ }\href {\doibase
  10.1103/PhysRevC.82.015206} {\bibfield  {journal} {\bibinfo  {journal}
  {Phys.Rev.}\ }\textbf {\bibinfo {volume} {C82}},\ \bibinfo {pages} {015206}
  (\bibinfo {year} {2010}{\natexlab{a}})},\ \Eprint
  {http://arxiv.org/abs/1004.2665} {arXiv:1004.2665 [hep-ph]} \BibitemShut
  {NoStop}%
\bibitem [{\citenamefont {Skokov}\ \emph {et~al.}(2011)\citenamefont {Skokov},
  \citenamefont {Friman},\ and\ \citenamefont {Redlich}}]{Skokov:2010uh}%
  \BibitemOpen
  \bibfield  {author} {\bibinfo {author} {\bibfnamefont {V.}~\bibnamefont
  {Skokov}}, \bibinfo {author} {\bibfnamefont {B.}~\bibnamefont {Friman}}, \
  and\ \bibinfo {author} {\bibfnamefont {K.}~\bibnamefont {Redlich}},\ }\href
  {\doibase 10.1103/PhysRevC.83.054904} {\bibfield  {journal} {\bibinfo
  {journal} {Phys.Rev.}\ }\textbf {\bibinfo {volume} {C83}},\ \bibinfo {pages}
  {054904} (\bibinfo {year} {2011})},\ \Eprint {http://arxiv.org/abs/1008.4570}
  {arXiv:1008.4570 [hep-ph]} \BibitemShut {NoStop}%
\bibitem [{\citenamefont {Friman}\ \emph {et~al.}(2011)\citenamefont {Friman},
  \citenamefont {Karsch}, \citenamefont {Redlich},\ and\ \citenamefont
  {Skokov}}]{Friman:2011pf}%
  \BibitemOpen
  \bibfield  {author} {\bibinfo {author} {\bibfnamefont {B.}~\bibnamefont
  {Friman}}, \bibinfo {author} {\bibfnamefont {F.}~\bibnamefont {Karsch}},
  \bibinfo {author} {\bibfnamefont {K.}~\bibnamefont {Redlich}}, \ and\
  \bibinfo {author} {\bibfnamefont {V.}~\bibnamefont {Skokov}},\ }\href
  {\doibase 10.1140/epjc/s10052-011-1694-2} {\bibfield  {journal} {\bibinfo
  {journal} {Eur. Phys. J.}\ }\textbf {\bibinfo {volume} {C71}},\ \bibinfo
  {pages} {1694} (\bibinfo {year} {2011})},\ \Eprint
  {http://arxiv.org/abs/1103.3511} {arXiv:1103.3511 [hep-ph]} \BibitemShut
  {NoStop}%
\bibitem [{\citenamefont {Skokov}\ \emph {et~al.}(2012)\citenamefont {Skokov},
  \citenamefont {Friman},\ and\ \citenamefont {Redlich}}]{Skokov:2011rq}%
  \BibitemOpen
  \bibfield  {author} {\bibinfo {author} {\bibfnamefont {V.}~\bibnamefont
  {Skokov}}, \bibinfo {author} {\bibfnamefont {B.}~\bibnamefont {Friman}}, \
  and\ \bibinfo {author} {\bibfnamefont {K.}~\bibnamefont {Redlich}},\ }\href
  {\doibase 10.1016/j.physletb.2012.01.022} {\bibfield  {journal} {\bibinfo
  {journal} {Phys. Lett.}\ }\textbf {\bibinfo {volume} {B708}},\ \bibinfo
  {pages} {179} (\bibinfo {year} {2012})},\ \Eprint
  {http://arxiv.org/abs/1108.3231} {arXiv:1108.3231 [hep-ph]} \BibitemShut
  {NoStop}%
\bibitem [{\citenamefont {Skokov}\ \emph {et~al.}(2013)\citenamefont {Skokov},
  \citenamefont {Friman},\ and\ \citenamefont {Redlich}}]{Skokov:2012ds}%
  \BibitemOpen
  \bibfield  {author} {\bibinfo {author} {\bibfnamefont {V.}~\bibnamefont
  {Skokov}}, \bibinfo {author} {\bibfnamefont {B.}~\bibnamefont {Friman}}, \
  and\ \bibinfo {author} {\bibfnamefont {K.}~\bibnamefont {Redlich}},\ }\href
  {\doibase 10.1103/PhysRevC.88.034911} {\bibfield  {journal} {\bibinfo
  {journal} {Phys. Rev.}\ }\textbf {\bibinfo {volume} {C88}},\ \bibinfo {pages}
  {034911} (\bibinfo {year} {2013})},\ \Eprint {http://arxiv.org/abs/1205.4756}
  {arXiv:1205.4756 [hep-ph]} \BibitemShut {NoStop}%
\bibitem [{\citenamefont {Morita}\ \emph {et~al.}(2014)\citenamefont {Morita},
  \citenamefont {Skokov}, \citenamefont {Friman},\ and\ \citenamefont
  {Redlich}}]{Morita:2012kt}%
  \BibitemOpen
  \bibfield  {author} {\bibinfo {author} {\bibfnamefont {K.}~\bibnamefont
  {Morita}}, \bibinfo {author} {\bibfnamefont {V.}~\bibnamefont {Skokov}},
  \bibinfo {author} {\bibfnamefont {B.}~\bibnamefont {Friman}}, \ and\ \bibinfo
  {author} {\bibfnamefont {K.}~\bibnamefont {Redlich}},\ }\href {\doibase
  10.1140/epjc/s10052-013-2706-1} {\bibfield  {journal} {\bibinfo  {journal}
  {Eur.Phys.J.}\ }\textbf {\bibinfo {volume} {C74}},\ \bibinfo {pages} {2706}
  (\bibinfo {year} {2014})},\ \Eprint {http://arxiv.org/abs/1211.4703}
  {arXiv:1211.4703 [hep-ph]} \BibitemShut {NoStop}%
\bibitem [{\citenamefont {Morita}\ \emph {et~al.}(2013)\citenamefont {Morita},
  \citenamefont {Friman}, \citenamefont {Redlich},\ and\ \citenamefont
  {Skokov}}]{Morita:2013tu}%
  \BibitemOpen
  \bibfield  {author} {\bibinfo {author} {\bibfnamefont {K.}~\bibnamefont
  {Morita}}, \bibinfo {author} {\bibfnamefont {B.}~\bibnamefont {Friman}},
  \bibinfo {author} {\bibfnamefont {K.}~\bibnamefont {Redlich}}, \ and\
  \bibinfo {author} {\bibfnamefont {V.}~\bibnamefont {Skokov}},\ }\href
  {\doibase 10.1103/PhysRevC.88.034903} {\bibfield  {journal} {\bibinfo
  {journal} {Phys.Rev.}\ }\textbf {\bibinfo {volume} {C88}},\ \bibinfo {pages}
  {034903} (\bibinfo {year} {2013})},\ \Eprint {http://arxiv.org/abs/1301.2873}
  {arXiv:1301.2873 [hep-ph]} \BibitemShut {NoStop}%
\bibitem [{\citenamefont {Morita}\ \emph {et~al.}(2015)\citenamefont {Morita},
  \citenamefont {Friman},\ and\ \citenamefont {Redlich}}]{Morita:2014fda}%
  \BibitemOpen
  \bibfield  {author} {\bibinfo {author} {\bibfnamefont {K.}~\bibnamefont
  {Morita}}, \bibinfo {author} {\bibfnamefont {B.}~\bibnamefont {Friman}}, \
  and\ \bibinfo {author} {\bibfnamefont {K.}~\bibnamefont {Redlich}},\ }\href
  {\doibase 10.1016/j.physletb.2014.12.037} {\bibfield  {journal} {\bibinfo
  {journal} {Phys. Lett.}\ }\textbf {\bibinfo {volume} {B741}},\ \bibinfo
  {pages} {178} (\bibinfo {year} {2015})},\ \Eprint
  {http://arxiv.org/abs/1402.5982} {arXiv:1402.5982 [hep-ph]} \BibitemShut
  {NoStop}%
\bibitem [{\citenamefont {Morita}\ and\ \citenamefont
  {Redlich}(2015)}]{Morita:2014nra}%
  \BibitemOpen
  \bibfield  {author} {\bibinfo {author} {\bibfnamefont {K.}~\bibnamefont
  {Morita}}\ and\ \bibinfo {author} {\bibfnamefont {K.}~\bibnamefont
  {Redlich}},\ }\href {\doibase 10.1093/ptep/ptv047} {\bibfield  {journal}
  {\bibinfo  {journal} {PTEP}\ }\textbf {\bibinfo {volume} {2015}},\ \bibinfo
  {pages} {043D03} (\bibinfo {year} {2015})},\ \Eprint
  {http://arxiv.org/abs/1409.8001} {arXiv:1409.8001 [hep-ph]} \BibitemShut
  {NoStop}%
\bibitem [{\citenamefont {Fu}\ \emph {et~al.}(2010)\citenamefont {Fu},
  \citenamefont {Liu},\ and\ \citenamefont {Wu}}]{Fu:2009wy}%
  \BibitemOpen
  \bibfield  {author} {\bibinfo {author} {\bibfnamefont {W.-j.}\ \bibnamefont
  {Fu}}, \bibinfo {author} {\bibfnamefont {Y.-x.}\ \bibnamefont {Liu}}, \ and\
  \bibinfo {author} {\bibfnamefont {Y.-L.}\ \bibnamefont {Wu}},\ }\href
  {\doibase 10.1103/PhysRevD.81.014028} {\bibfield  {journal} {\bibinfo
  {journal} {Phys.Rev.}\ }\textbf {\bibinfo {volume} {D81}},\ \bibinfo {pages}
  {014028} (\bibinfo {year} {2010})},\ \Eprint {http://arxiv.org/abs/0910.5783}
  {arXiv:0910.5783 [hep-ph]} \BibitemShut {NoStop}%
\bibitem [{\citenamefont {Fu}\ and\ \citenamefont {Wu}(2010)}]{Fu:2010ay}%
  \BibitemOpen
  \bibfield  {author} {\bibinfo {author} {\bibfnamefont {W.-j.}\ \bibnamefont
  {Fu}}\ and\ \bibinfo {author} {\bibfnamefont {Y.-l.}\ \bibnamefont {Wu}},\
  }\href {\doibase 10.1103/PhysRevD.82.074013} {\bibfield  {journal} {\bibinfo
  {journal} {Phys. Rev.}\ }\textbf {\bibinfo {volume} {D82}},\ \bibinfo {pages}
  {074013} (\bibinfo {year} {2010})},\ \Eprint {http://arxiv.org/abs/1008.3684}
  {arXiv:1008.3684 [hep-ph]} \BibitemShut {NoStop}%
\bibitem [{\citenamefont {Skokov}\ \emph
  {et~al.}(2010{\natexlab{b}})\citenamefont {Skokov}, \citenamefont {Friman},
  \citenamefont {Nakano}, \citenamefont {Redlich},\ and\ \citenamefont
  {Schaefer}}]{Skokov:2010sf}%
  \BibitemOpen
  \bibfield  {author} {\bibinfo {author} {\bibfnamefont {V.}~\bibnamefont
  {Skokov}}, \bibinfo {author} {\bibfnamefont {B.}~\bibnamefont {Friman}},
  \bibinfo {author} {\bibfnamefont {E.}~\bibnamefont {Nakano}}, \bibinfo
  {author} {\bibfnamefont {K.}~\bibnamefont {Redlich}}, \ and\ \bibinfo
  {author} {\bibfnamefont {B.-J.}\ \bibnamefont {Schaefer}},\ }\href {\doibase
  10.1103/PhysRevD.82.034029} {\bibfield  {journal} {\bibinfo  {journal}
  {Phys.Rev.}\ }\textbf {\bibinfo {volume} {D82}},\ \bibinfo {pages} {034029}
  (\bibinfo {year} {2010}{\natexlab{b}})},\ \Eprint
  {http://arxiv.org/abs/1005.3166} {arXiv:1005.3166 [hep-ph]} \BibitemShut
  {NoStop}%
\bibitem [{\citenamefont {Karsch}\ \emph {et~al.}(2011)\citenamefont {Karsch},
  \citenamefont {Schaefer}, \citenamefont {Wagner},\ and\ \citenamefont
  {Wambach}}]{Karsch:2010hm}%
  \BibitemOpen
  \bibfield  {author} {\bibinfo {author} {\bibfnamefont {F.}~\bibnamefont
  {Karsch}}, \bibinfo {author} {\bibfnamefont {B.-J.}\ \bibnamefont
  {Schaefer}}, \bibinfo {author} {\bibfnamefont {M.}~\bibnamefont {Wagner}}, \
  and\ \bibinfo {author} {\bibfnamefont {J.}~\bibnamefont {Wambach}},\ }\href
  {\doibase 10.1016/j.physletb.2011.03.013} {\bibfield  {journal} {\bibinfo
  {journal} {Phys.Lett.}\ }\textbf {\bibinfo {volume} {B698}},\ \bibinfo
  {pages} {256} (\bibinfo {year} {2011})},\ \Eprint
  {http://arxiv.org/abs/1009.5211} {arXiv:1009.5211 [hep-ph]} \BibitemShut
  {NoStop}%
\bibitem [{\citenamefont {Schaefer}\ and\ \citenamefont
  {Wagner}(2012)}]{Schaefer:2011ex}%
  \BibitemOpen
  \bibfield  {author} {\bibinfo {author} {\bibfnamefont {B.~J.}\ \bibnamefont
  {Schaefer}}\ and\ \bibinfo {author} {\bibfnamefont {M.}~\bibnamefont
  {Wagner}},\ }\href {\doibase 10.1103/PhysRevD.85.034027} {\bibfield
  {journal} {\bibinfo  {journal} {Phys. Rev.}\ }\textbf {\bibinfo {volume}
  {D85}},\ \bibinfo {pages} {034027} (\bibinfo {year} {2012})},\ \Eprint
  {http://arxiv.org/abs/1111.6871} {arXiv:1111.6871 [hep-ph]} \BibitemShut
  {NoStop}%
\bibitem [{\citenamefont {Wagner}\ \emph {et~al.}(2010)\citenamefont {Wagner},
  \citenamefont {Walther},\ and\ \citenamefont {Schaefer}}]{Wagner:2009pm}%
  \BibitemOpen
  \bibfield  {author} {\bibinfo {author} {\bibfnamefont {M.}~\bibnamefont
  {Wagner}}, \bibinfo {author} {\bibfnamefont {A.}~\bibnamefont {Walther}}, \
  and\ \bibinfo {author} {\bibfnamefont {B.-J.}\ \bibnamefont {Schaefer}},\
  }\href {\doibase 10.1016/j.cpc.2009.12.008} {\bibfield  {journal} {\bibinfo
  {journal} {Comput.Phys.Commun.}\ }\textbf {\bibinfo {volume} {181}},\
  \bibinfo {pages} {756} (\bibinfo {year} {2010})},\ \Eprint
  {http://arxiv.org/abs/0912.2208} {arXiv:0912.2208 [hep-ph]} \BibitemShut
  {NoStop}%
\bibitem [{\citenamefont {Gies}\ and\ \citenamefont
  {Wetterich}(2002)}]{Gies:2001nw}%
  \BibitemOpen
  \bibfield  {author} {\bibinfo {author} {\bibfnamefont {H.}~\bibnamefont
  {Gies}}\ and\ \bibinfo {author} {\bibfnamefont {C.}~\bibnamefont
  {Wetterich}},\ }\href {\doibase 10.1103/PhysRevD.65.065001} {\bibfield
  {journal} {\bibinfo  {journal} {Phys.Rev.}\ }\textbf {\bibinfo {volume}
  {D65}},\ \bibinfo {pages} {065001} (\bibinfo {year} {2002})},\ \Eprint
  {http://arxiv.org/abs/hep-th/0107221} {arXiv:hep-th/0107221 [hep-th]}
  \BibitemShut {NoStop}%
\bibitem [{\citenamefont {Gies}\ and\ \citenamefont
  {Wetterich}(2004)}]{Gies:2002hq}%
  \BibitemOpen
  \bibfield  {author} {\bibinfo {author} {\bibfnamefont {H.}~\bibnamefont
  {Gies}}\ and\ \bibinfo {author} {\bibfnamefont {C.}~\bibnamefont
  {Wetterich}},\ }\href {\doibase 10.1103/PhysRevD.69.025001} {\bibfield
  {journal} {\bibinfo  {journal} {Phys.Rev.}\ }\textbf {\bibinfo {volume}
  {D69}},\ \bibinfo {pages} {025001} (\bibinfo {year} {2004})},\ \Eprint
  {http://arxiv.org/abs/hep-th/0209183} {arXiv:hep-th/0209183 [hep-th]}
  \BibitemShut {NoStop}%
\bibitem [{\citenamefont {Floerchinger}\ and\ \citenamefont
  {Wetterich}(2009)}]{Floerchinger:2009uf}%
  \BibitemOpen
  \bibfield  {author} {\bibinfo {author} {\bibfnamefont {S.}~\bibnamefont
  {Floerchinger}}\ and\ \bibinfo {author} {\bibfnamefont {C.}~\bibnamefont
  {Wetterich}},\ }\href {\doibase 10.1016/j.physletb.2009.09.014} {\bibfield
  {journal} {\bibinfo  {journal} {Phys.Lett.}\ }\textbf {\bibinfo {volume}
  {B680}},\ \bibinfo {pages} {371} (\bibinfo {year} {2009})},\ \Eprint
  {http://arxiv.org/abs/0905.0915} {arXiv:0905.0915 [hep-th]} \BibitemShut
  {NoStop}%
\bibitem [{\citenamefont {Litim}(2000)}]{Litim:2000ci}%
  \BibitemOpen
  \bibfield  {author} {\bibinfo {author} {\bibfnamefont {D.~F.}\ \bibnamefont
  {Litim}},\ }\href {\doibase 10.1016/S0370-2693(00)00748-6} {\bibfield
  {journal} {\bibinfo  {journal} {Phys.Lett.}\ }\textbf {\bibinfo {volume}
  {B486}},\ \bibinfo {pages} {92} (\bibinfo {year} {2000})},\ \Eprint
  {http://arxiv.org/abs/hep-th/0005245} {arXiv:hep-th/0005245 [hep-th]}
  \BibitemShut {NoStop}%
\bibitem [{\citenamefont {Litim}(2001)}]{Litim:2001up}%
  \BibitemOpen
  \bibfield  {author} {\bibinfo {author} {\bibfnamefont {D.~F.}\ \bibnamefont
  {Litim}},\ }\href {\doibase 10.1103/PhysRevD.64.105007} {\bibfield  {journal}
  {\bibinfo  {journal} {Phys.Rev.}\ }\textbf {\bibinfo {volume} {D64}},\
  \bibinfo {pages} {105007} (\bibinfo {year} {2001})},\ \Eprint
  {http://arxiv.org/abs/hep-th/0103195} {arXiv:hep-th/0103195 [hep-th]}
  \BibitemShut {NoStop}%
\bibitem [{\citenamefont {Papp}\ \emph {et~al.}(2000)\citenamefont {Papp},
  \citenamefont {Schaefer}, \citenamefont {Pirner},\ and\ \citenamefont
  {Wambach}}]{Papp:1999he}%
  \BibitemOpen
  \bibfield  {author} {\bibinfo {author} {\bibfnamefont {G.}~\bibnamefont
  {Papp}}, \bibinfo {author} {\bibfnamefont {B.-J.}\ \bibnamefont {Schaefer}},
  \bibinfo {author} {\bibfnamefont {H.}~\bibnamefont {Pirner}}, \ and\ \bibinfo
  {author} {\bibfnamefont {J.}~\bibnamefont {Wambach}},\ }\href {\doibase
  10.1103/PhysRevD.61.096002} {\bibfield  {journal} {\bibinfo  {journal}
  {Phys.Rev.}\ }\textbf {\bibinfo {volume} {D61}},\ \bibinfo {pages} {096002}
  (\bibinfo {year} {2000})},\ \Eprint {http://arxiv.org/abs/hep-ph/9909246}
  {arXiv:hep-ph/9909246 [hep-ph]} \BibitemShut {NoStop}%
\bibitem [{\citenamefont {Mueller}\ and\ \citenamefont
  {Pawlowski}(2015)}]{Mueller:2015fka}%
  \BibitemOpen
  \bibfield  {author} {\bibinfo {author} {\bibfnamefont {N.}~\bibnamefont
  {Mueller}}\ and\ \bibinfo {author} {\bibfnamefont {J.~M.}\ \bibnamefont
  {Pawlowski}},\ }\href {\doibase 10.1103/PhysRevD.91.116010} {\bibfield
  {journal} {\bibinfo  {journal} {Phys. Rev.}\ }\textbf {\bibinfo {volume}
  {D91}},\ \bibinfo {pages} {116010} (\bibinfo {year} {2015})},\ \Eprint
  {http://arxiv.org/abs/1502.08011} {arXiv:1502.08011 [hep-ph]} \BibitemShut
  {NoStop}%
\bibitem [{\citenamefont {Fukushima}(2004)}]{Fukushima:2003fw}%
  \BibitemOpen
  \bibfield  {author} {\bibinfo {author} {\bibfnamefont {K.}~\bibnamefont
  {Fukushima}},\ }\href {\doibase 10.1016/j.physletb.2004.04.027} {\bibfield
  {journal} {\bibinfo  {journal} {Phys.Lett.}\ }\textbf {\bibinfo {volume}
  {B591}},\ \bibinfo {pages} {277} (\bibinfo {year} {2004})},\ \Eprint
  {http://arxiv.org/abs/hep-ph/0310121} {arXiv:hep-ph/0310121 [hep-ph]}
  \BibitemShut {NoStop}%
\bibitem [{\citenamefont {Ratti}\ \emph {et~al.}(2006)\citenamefont {Ratti},
  \citenamefont {Thaler},\ and\ \citenamefont {Weise}}]{Ratti:2005jh}%
  \BibitemOpen
  \bibfield  {author} {\bibinfo {author} {\bibfnamefont {C.}~\bibnamefont
  {Ratti}}, \bibinfo {author} {\bibfnamefont {M.~A.}\ \bibnamefont {Thaler}}, \
  and\ \bibinfo {author} {\bibfnamefont {W.}~\bibnamefont {Weise}},\ }\href
  {\doibase 10.1103/PhysRevD.73.014019} {\bibfield  {journal} {\bibinfo
  {journal} {Phys.Rev.}\ }\textbf {\bibinfo {volume} {D73}},\ \bibinfo {pages}
  {014019} (\bibinfo {year} {2006})},\ \Eprint
  {http://arxiv.org/abs/hep-ph/0506234} {arXiv:hep-ph/0506234 [hep-ph]}
  \BibitemShut {NoStop}%
\bibitem [{\citenamefont {Fu}\ \emph {et~al.}(2008)\citenamefont {Fu},
  \citenamefont {Zhang},\ and\ \citenamefont {Liu}}]{Fu:2007xc}%
  \BibitemOpen
  \bibfield  {author} {\bibinfo {author} {\bibfnamefont {W.-j.}\ \bibnamefont
  {Fu}}, \bibinfo {author} {\bibfnamefont {Z.}~\bibnamefont {Zhang}}, \ and\
  \bibinfo {author} {\bibfnamefont {Y.-x.}\ \bibnamefont {Liu}},\ }\href
  {\doibase 10.1103/PhysRevD.77.014006} {\bibfield  {journal} {\bibinfo
  {journal} {Phys.Rev.}\ }\textbf {\bibinfo {volume} {D77}},\ \bibinfo {pages}
  {014006} (\bibinfo {year} {2008})},\ \Eprint {http://arxiv.org/abs/0711.0154}
  {arXiv:0711.0154 [hep-ph]} \BibitemShut {NoStop}%
\bibitem [{\citenamefont {Schaefer}\ \emph {et~al.}(2007)\citenamefont
  {Schaefer}, \citenamefont {Pawlowski},\ and\ \citenamefont
  {Wambach}}]{Schaefer:2007pw}%
  \BibitemOpen
  \bibfield  {author} {\bibinfo {author} {\bibfnamefont {B.-J.}\ \bibnamefont
  {Schaefer}}, \bibinfo {author} {\bibfnamefont {J.~M.}\ \bibnamefont
  {Pawlowski}}, \ and\ \bibinfo {author} {\bibfnamefont {J.}~\bibnamefont
  {Wambach}},\ }\href {\doibase 10.1103/PhysRevD.76.074023} {\bibfield
  {journal} {\bibinfo  {journal} {Phys.Rev.}\ }\textbf {\bibinfo {volume}
  {D76}},\ \bibinfo {pages} {074023} (\bibinfo {year} {2007})},\ \Eprint
  {http://arxiv.org/abs/0704.3234} {arXiv:0704.3234 [hep-ph]} \BibitemShut
  {NoStop}%
\bibitem [{\citenamefont {Fischer}\ \emph
  {et~al.}(2014{\natexlab{a}})\citenamefont {Fischer}, \citenamefont
  {Luecker},\ and\ \citenamefont {Welzbacher}}]{Fischer:2014ata}%
  \BibitemOpen
  \bibfield  {author} {\bibinfo {author} {\bibfnamefont {C.~S.}\ \bibnamefont
  {Fischer}}, \bibinfo {author} {\bibfnamefont {J.}~\bibnamefont {Luecker}}, \
  and\ \bibinfo {author} {\bibfnamefont {C.~A.}\ \bibnamefont {Welzbacher}},\
  }\href@noop {} {\  (\bibinfo {year} {2014}{\natexlab{a}})},\ \Eprint
  {http://arxiv.org/abs/1405.4762} {arXiv:1405.4762 [hep-ph]} \BibitemShut
  {NoStop}%
\bibitem [{\citenamefont {Lo}\ \emph {et~al.}(2013{\natexlab{a}})\citenamefont
  {Lo}, \citenamefont {Friman}, \citenamefont {Kaczmarek}, \citenamefont
  {Redlich},\ and\ \citenamefont {Sasaki}}]{Lo:2013etb}%
  \BibitemOpen
  \bibfield  {author} {\bibinfo {author} {\bibfnamefont {P.~M.}\ \bibnamefont
  {Lo}}, \bibinfo {author} {\bibfnamefont {B.}~\bibnamefont {Friman}}, \bibinfo
  {author} {\bibfnamefont {O.}~\bibnamefont {Kaczmarek}}, \bibinfo {author}
  {\bibfnamefont {K.}~\bibnamefont {Redlich}}, \ and\ \bibinfo {author}
  {\bibfnamefont {C.}~\bibnamefont {Sasaki}},\ }\href {\doibase
  10.1103/PhysRevD.88.014506} {\bibfield  {journal} {\bibinfo  {journal}
  {Phys.Rev.}\ }\textbf {\bibinfo {volume} {D88}},\ \bibinfo {pages} {014506}
  (\bibinfo {year} {2013}{\natexlab{a}})},\ \Eprint
  {http://arxiv.org/abs/1306.5094} {arXiv:1306.5094 [hep-lat]} \BibitemShut
  {NoStop}%
\bibitem [{\citenamefont {Lo}\ \emph {et~al.}(2013{\natexlab{b}})\citenamefont
  {Lo}, \citenamefont {Friman}, \citenamefont {Kaczmarek}, \citenamefont
  {Redlich},\ and\ \citenamefont {Sasaki}}]{Lo:2013hla}%
  \BibitemOpen
  \bibfield  {author} {\bibinfo {author} {\bibfnamefont {P.~M.}\ \bibnamefont
  {Lo}}, \bibinfo {author} {\bibfnamefont {B.}~\bibnamefont {Friman}}, \bibinfo
  {author} {\bibfnamefont {O.}~\bibnamefont {Kaczmarek}}, \bibinfo {author}
  {\bibfnamefont {K.}~\bibnamefont {Redlich}}, \ and\ \bibinfo {author}
  {\bibfnamefont {C.}~\bibnamefont {Sasaki}},\ }\href {\doibase
  10.1103/PhysRevD.88.074502} {\bibfield  {journal} {\bibinfo  {journal}
  {Phys.Rev.}\ }\textbf {\bibinfo {volume} {D88}},\ \bibinfo {pages} {074502}
  (\bibinfo {year} {2013}{\natexlab{b}})},\ \Eprint
  {http://arxiv.org/abs/1307.5958} {arXiv:1307.5958 [hep-lat]} \BibitemShut
  {NoStop}%
\bibitem [{\citenamefont {Braun}\ \emph {et~al.}(2010)\citenamefont {Braun},
  \citenamefont {Gies},\ and\ \citenamefont {Pawlowski}}]{Braun:2007bx}%
  \BibitemOpen
  \bibfield  {author} {\bibinfo {author} {\bibfnamefont {J.}~\bibnamefont
  {Braun}}, \bibinfo {author} {\bibfnamefont {H.}~\bibnamefont {Gies}}, \ and\
  \bibinfo {author} {\bibfnamefont {J.~M.}\ \bibnamefont {Pawlowski}},\ }\href
  {\doibase 10.1016/j.physletb.2010.01.009} {\bibfield  {journal} {\bibinfo
  {journal} {Phys.Lett.}\ }\textbf {\bibinfo {volume} {B684}},\ \bibinfo
  {pages} {262} (\bibinfo {year} {2010})},\ \Eprint
  {http://arxiv.org/abs/0708.2413} {arXiv:0708.2413 [hep-th]} \BibitemShut
  {NoStop}%
\bibitem [{\citenamefont {Fister}\ and\ \citenamefont
  {Pawlowski}(2013)}]{Fister:2013bh}%
  \BibitemOpen
  \bibfield  {author} {\bibinfo {author} {\bibfnamefont {L.}~\bibnamefont
  {Fister}}\ and\ \bibinfo {author} {\bibfnamefont {J.~M.}\ \bibnamefont
  {Pawlowski}},\ }\href {\doibase 10.1103/PhysRevD.88.045010} {\bibfield
  {journal} {\bibinfo  {journal} {Phys.Rev.}\ }\textbf {\bibinfo {volume}
  {D88}},\ \bibinfo {pages} {045010} (\bibinfo {year} {2013})},\ \Eprint
  {http://arxiv.org/abs/1301.4163} {arXiv:1301.4163 [hep-ph]} \BibitemShut
  {NoStop}%
\bibitem [{\citenamefont {Fischer}\ \emph
  {et~al.}(2014{\natexlab{b}})\citenamefont {Fischer}, \citenamefont {Fister},
  \citenamefont {Luecker},\ and\ \citenamefont {Pawlowski}}]{Fischer:2013eca}%
  \BibitemOpen
  \bibfield  {author} {\bibinfo {author} {\bibfnamefont {C.~S.}\ \bibnamefont
  {Fischer}}, \bibinfo {author} {\bibfnamefont {L.}~\bibnamefont {Fister}},
  \bibinfo {author} {\bibfnamefont {J.}~\bibnamefont {Luecker}}, \ and\
  \bibinfo {author} {\bibfnamefont {J.~M.}\ \bibnamefont {Pawlowski}},\ }\href
  {\doibase 10.1016/j.physletb.2014.03.057} {\bibfield  {journal} {\bibinfo
  {journal} {Phys.Lett.}\ }\textbf {\bibinfo {volume} {B732}},\ \bibinfo
  {pages} {273} (\bibinfo {year} {2014}{\natexlab{b}})},\ \Eprint
  {http://arxiv.org/abs/1306.6022} {arXiv:1306.6022 [hep-ph]} \BibitemShut
  {NoStop}%
\bibitem [{\citenamefont {Greensite}(2012)}]{Greensite:2012dy}%
  \BibitemOpen
  \bibfield  {author} {\bibinfo {author} {\bibfnamefont {J.}~\bibnamefont
  {Greensite}},\ }\href {\doibase 10.1103/PhysRevD.86.114507} {\bibfield
  {journal} {\bibinfo  {journal} {Phys.Rev.}\ }\textbf {\bibinfo {volume}
  {D86}},\ \bibinfo {pages} {114507} (\bibinfo {year} {2012})},\ \Eprint
  {http://arxiv.org/abs/1209.5697} {arXiv:1209.5697 [hep-lat]} \BibitemShut
  {NoStop}%
\bibitem [{\citenamefont {Langfeld}\ and\ \citenamefont
  {Pawlowski}(2013)}]{Langfeld:2013xbf}%
  \BibitemOpen
  \bibfield  {author} {\bibinfo {author} {\bibfnamefont {K.}~\bibnamefont
  {Langfeld}}\ and\ \bibinfo {author} {\bibfnamefont {J.~M.}\ \bibnamefont
  {Pawlowski}},\ }\href@noop {} {\  (\bibinfo {year} {2013})},\ \Eprint
  {http://arxiv.org/abs/1307.0455} {arXiv:1307.0455 [hep-lat]} \BibitemShut
  {NoStop}%
\bibitem [{\citenamefont {Smith}\ \emph {et~al.}(2013)\citenamefont {Smith},
  \citenamefont {Dumitru}, \citenamefont {Pisarski},\ and\ \citenamefont {von
  Smekal}}]{Smith:2013msa}%
  \BibitemOpen
  \bibfield  {author} {\bibinfo {author} {\bibfnamefont {D.}~\bibnamefont
  {Smith}}, \bibinfo {author} {\bibfnamefont {A.}~\bibnamefont {Dumitru}},
  \bibinfo {author} {\bibfnamefont {R.}~\bibnamefont {Pisarski}}, \ and\
  \bibinfo {author} {\bibfnamefont {L.}~\bibnamefont {von Smekal}},\ }\href
  {\doibase 10.1103/PhysRevD.88.054020} {\bibfield  {journal} {\bibinfo
  {journal} {Phys.Rev.}\ }\textbf {\bibinfo {volume} {D88}},\ \bibinfo {pages}
  {054020} (\bibinfo {year} {2013})},\ \Eprint {http://arxiv.org/abs/1307.6339}
  {arXiv:1307.6339 [hep-lat]} \BibitemShut {NoStop}%
\bibitem [{\citenamefont {Diakonov}\ \emph {et~al.}(2013)\citenamefont
  {Diakonov}, \citenamefont {Petrov}, \citenamefont {Schadler},\ and\
  \citenamefont {Gattringer}}]{Diakonov:2013lja}%
  \BibitemOpen
  \bibfield  {author} {\bibinfo {author} {\bibfnamefont {D.}~\bibnamefont
  {Diakonov}}, \bibinfo {author} {\bibfnamefont {V.}~\bibnamefont {Petrov}},
  \bibinfo {author} {\bibfnamefont {H.-P.}\ \bibnamefont {Schadler}}, \ and\
  \bibinfo {author} {\bibfnamefont {C.}~\bibnamefont {Gattringer}},\ }\href
  {\doibase 10.1007/JHEP11(2013)207} {\bibfield  {journal} {\bibinfo  {journal}
  {JHEP}\ }\textbf {\bibinfo {volume} {1311}},\ \bibinfo {pages} {207}
  (\bibinfo {year} {2013})},\ \Eprint {http://arxiv.org/abs/1308.2328}
  {arXiv:1308.2328 [hep-lat]} \BibitemShut {NoStop}%
\bibitem [{\citenamefont {Greensite}\ and\ \citenamefont
  {Langfeld}(2014)}]{Greensite:2014isa}%
  \BibitemOpen
  \bibfield  {author} {\bibinfo {author} {\bibfnamefont {J.}~\bibnamefont
  {Greensite}}\ and\ \bibinfo {author} {\bibfnamefont {K.}~\bibnamefont
  {Langfeld}},\ }\href {\doibase 10.1103/PhysRevD.90.014507} {\bibfield
  {journal} {\bibinfo  {journal} {Phys. Rev.}\ }\textbf {\bibinfo {volume}
  {D90}},\ \bibinfo {pages} {014507} (\bibinfo {year} {2014})},\ \Eprint
  {http://arxiv.org/abs/1403.5844} {arXiv:1403.5844 [hep-lat]} \BibitemShut
  {NoStop}%
\bibitem [{\citenamefont {Fu}\ \emph {et~al.}(2015)\citenamefont {Fu},
  \citenamefont {Herbst}, \citenamefont {Pawlowski}, \citenamefont {Rennecke},\
  and\ \citenamefont {Schaefer}}]{FHPR}%
  \BibitemOpen
  \bibfield  {author} {\bibinfo {author} {\bibfnamefont {W.-j.}\ \bibnamefont
  {Fu}}, \bibinfo {author} {\bibfnamefont {T.~K.}\ \bibnamefont {Herbst}},
  \bibinfo {author} {\bibfnamefont {J.~M.}\ \bibnamefont {Pawlowski}}, \bibinfo
  {author} {\bibfnamefont {F.}~\bibnamefont {Rennecke}}, \ and\ \bibinfo
  {author} {\bibfnamefont {B.-J.}\ \bibnamefont {Schaefer}},\ }\href@noop {}
  {\bibfield  {journal} {\bibinfo  {journal} {in progress}\ } (\bibinfo {year}
  {2015})}\BibitemShut {NoStop}%
\bibitem [{\citenamefont {Borsanyi}\ \emph
  {et~al.}(2010{\natexlab{a}})\citenamefont {Borsanyi}, \citenamefont
  {Endrodi}, \citenamefont {Fodor}, \citenamefont {Jakovac}, \citenamefont
  {Katz} \emph {et~al.}}]{Borsanyi:2010cj}%
  \BibitemOpen
  \bibfield  {author} {\bibinfo {author} {\bibfnamefont {S.}~\bibnamefont
  {Borsanyi}}, \bibinfo {author} {\bibfnamefont {G.}~\bibnamefont {Endrodi}},
  \bibinfo {author} {\bibfnamefont {Z.}~\bibnamefont {Fodor}}, \bibinfo
  {author} {\bibfnamefont {A.}~\bibnamefont {Jakovac}}, \bibinfo {author}
  {\bibfnamefont {S.~D.}\ \bibnamefont {Katz}},  \emph {et~al.},\ }\href
  {\doibase 10.1007/JHEP11(2010)077} {\bibfield  {journal} {\bibinfo  {journal}
  {JHEP}\ }\textbf {\bibinfo {volume} {1011}},\ \bibinfo {pages} {077}
  (\bibinfo {year} {2010}{\natexlab{a}})},\ \Eprint
  {http://arxiv.org/abs/1007.2580} {arXiv:1007.2580 [hep-lat]} \BibitemShut
  {NoStop}%
\bibitem [{\citenamefont {Borsanyi}\ \emph
  {et~al.}(2010{\natexlab{b}})\citenamefont {Borsanyi} \emph
  {et~al.}}]{Borsanyi:2010bp}%
  \BibitemOpen
  \bibfield  {author} {\bibinfo {author} {\bibfnamefont {S.}~\bibnamefont
  {Borsanyi}} \emph {et~al.} (\bibinfo {collaboration} {Wuppertal-Budapest
  Collaboration}),\ }\href {\doibase 10.1007/JHEP09(2010)073} {\bibfield
  {journal} {\bibinfo  {journal} {JHEP}\ }\textbf {\bibinfo {volume} {1009}},\
  \bibinfo {pages} {073} (\bibinfo {year} {2010}{\natexlab{b}})},\ \Eprint
  {http://arxiv.org/abs/1005.3508} {arXiv:1005.3508 [hep-lat]} \BibitemShut
  {NoStop}%
\bibitem [{\citenamefont {Adamczyk}\ \emph {et~al.}(2014)\citenamefont
  {Adamczyk} \emph {et~al.}}]{Adamczyk:2013dal}%
  \BibitemOpen
  \bibfield  {author} {\bibinfo {author} {\bibfnamefont {L.}~\bibnamefont
  {Adamczyk}} \emph {et~al.} (\bibinfo {collaboration} {STAR}),\ }\href
  {\doibase 10.1103/PhysRevLett.112.032302} {\bibfield  {journal} {\bibinfo
  {journal} {Phys.Rev.Lett.}\ }\textbf {\bibinfo {volume} {112}},\ \bibinfo
  {pages} {032302} (\bibinfo {year} {2014})},\ \Eprint
  {http://arxiv.org/abs/1309.5681} {arXiv:1309.5681 [nucl-ex]} \BibitemShut
  {NoStop}%
\bibitem [{fQC()}]{fQCD}%
  \BibitemOpen
  \href@noop {} {}\bibinfo {note} {FQCD Collaboration, J. Braun, A.K. Cyrol, L.
  Fister, W.-J. Fu, T.K. Herbst, M. Mitter, J.M. Pawlowski, F. Rennecke, and N.
  Strodthoff}\BibitemShut {NoStop}%
\bibitem [{\citenamefont {Christiansen}\ \emph {et~al.}(2014)\citenamefont
  {Christiansen}, \citenamefont {Knorr}, \citenamefont {Pawlowski},\ and\
  \citenamefont {Rodigast}}]{Christiansen:2014raa}%
  \BibitemOpen
  \bibfield  {author} {\bibinfo {author} {\bibfnamefont {N.}~\bibnamefont
  {Christiansen}}, \bibinfo {author} {\bibfnamefont {B.}~\bibnamefont {Knorr}},
  \bibinfo {author} {\bibfnamefont {J.~M.}\ \bibnamefont {Pawlowski}}, \ and\
  \bibinfo {author} {\bibfnamefont {A.}~\bibnamefont {Rodigast}},\ }\href@noop
  {} {\  (\bibinfo {year} {2014})},\ \Eprint {http://arxiv.org/abs/1403.1232}
  {arXiv:1403.1232 [hep-th]} \BibitemShut {NoStop}%
\bibitem [{\citenamefont {Christiansen}\ \emph {et~al.}(2015)\citenamefont
  {Christiansen}, \citenamefont {Knorr}, \citenamefont {Meibohm}, \citenamefont
  {Pawlowski},\ and\ \citenamefont {Reichert}}]{Christiansen:2015rva}%
  \BibitemOpen
  \bibfield  {author} {\bibinfo {author} {\bibfnamefont {N.}~\bibnamefont
  {Christiansen}}, \bibinfo {author} {\bibfnamefont {B.}~\bibnamefont {Knorr}},
  \bibinfo {author} {\bibfnamefont {J.}~\bibnamefont {Meibohm}}, \bibinfo
  {author} {\bibfnamefont {J.~M.}\ \bibnamefont {Pawlowski}}, \ and\ \bibinfo
  {author} {\bibfnamefont {M.}~\bibnamefont {Reichert}},\ }\href@noop {} {\
  (\bibinfo {year} {2015})},\ \Eprint {http://arxiv.org/abs/1506.07016}
  {arXiv:1506.07016 [hep-th]} \BibitemShut {NoStop}%
\bibitem [{\citenamefont {Fister}\ and\ \citenamefont
  {Pawlowski}(2015)}]{Fister:2015eca}%
  \BibitemOpen
  \bibfield  {author} {\bibinfo {author} {\bibfnamefont {L.}~\bibnamefont
  {Fister}}\ and\ \bibinfo {author} {\bibfnamefont {J.~M.}\ \bibnamefont
  {Pawlowski}},\ }\href@noop {} {\  (\bibinfo {year} {2015})},\ \Eprint
  {http://arxiv.org/abs/1504.05166} {arXiv:1504.05166 [hep-ph]} \BibitemShut
  {NoStop}%
\end{thebibliography}%

\end{document}